\documentclass{elsarticle}

\usepackage{algorithm}
\usepackage{algorithmic}
\usepackage{amsfonts}
\usepackage{amsmath}
\usepackage{amsthm}
\usepackage{color}
\usepackage{graphicx}
\usepackage{lineno, hyperref}
\usepackage{epstopdf}

\biboptions{authoryear}

\newcommand{\given}{\,|\,}

\theoremstyle{plain}

\theoremstyle{definition}

\theoremstyle{remark}

\begin{document}

\begin{frontmatter}

\title{{Nonparametric Bayesian label prediction on a graph}}
\author{Jarno Hartog\fnref{jarno}}
\author{Harry van Zanten\fnref{harry}}
\fntext[jarno]{University of Amsterdam, Science Park 107, 1098 XG Amsterdam, {\tt j.hartog2@uva.nl}}
\fntext[harry]{University of Amsterdam, Science Park 107, 1098 XG Amsterdam, {\tt hvzanten@uva.nl}}

\begin{abstract}
\noindent  An implementation of a nonparametric Bayesian approach to solving binary classification problems on graphs is described. A hierarchical Bayesian approach with a randomly scaled Gaussian prior is considered. The prior uses the graph Laplacian to take into account the underlying geometry of the graph. A method based on a theoretically optimal prior and a more flexible variant using partial conjugacy are proposed. Two simulated data examples and two examples using real data are used in order to illustrate the proposed methods.
\end{abstract}

\begin{keyword}
Binary classification; graph; Bayesian nonparametrics
\end{keyword}

\end{frontmatter}

%\maketitle

\section{Introduction}

In this article we consider prediction problems that can be seen as classification
problems on graphs. These arise in several applied settings, for instance in the 
prediction of the biological function of a protein in a protein-protein interaction graph
(e.g. \cite{kolaczyk2009, nariai2007, sharan2007}), or in graph-based semi-supervised learning 
(e.g.\ \cite{belkin2004, sindhwani2007}). 
We have problems in mind in which the graph is given by the application context and the graph has vertices
of different types, coded by vertex labels that can have two possible values, say. 
The available data are noisy observations of some of the labels. The goal of the statistical
procedure is to classify the vertices correctly, including those for which there is no 
observation available. The idea is that typically, the location of a given vertex in the graph, 
in combination with (noisy) information about the labels of vertices close to it, should have 
predictive power for the label of the vertex of interest. Hence, successful prediction of 
labels should be possible to some degree.

Several approaches for graph-based prediction have been considered in the literature. 
In this paper we investigate a nonparametric full Bayes procedure that was recently considered 
 in \cite{kirichenko2017} and that has so far only been studied theoretically. 
Concretely, we consider a connected, simple graph $G = (V, E)$, with $\# V = n$ vertices which 
we denote simply by $V = \{1, 2, \ldots, n\}$. 
We formalize the notion of nonparametric prediction in this setting by postulating the existence
of a {\em soft label function} 
\[
\ell : V \to (0,1)
\]
that determines the hard labels $y_1, y_2, \ldots, y_n$ that we (partly) observe. 
The  label $y_i$ corresponding to vertex $i$ is a Bernoulli random variable with $P(y_i=1)
=1-P(y_i=0) = \ell(i)$ and the $y_i$'s are assumed to be independent.
The underlying idea is that the ``real'', hard labels $h(i)$ of the vertices are obtained 
by thresholding the soft labels at level $1/2$, i.e.\ $h(i) = 1_{\ell(i) > 1/2}$. 
The $y_i$ are noisy versions of the real hard labels $h(i)$, in the sense 
that they can be wrong with some positive probability. Specifically, it holds in this setup that
$P(y_i \not = h(i)) = |h(i)-\ell(i)|$. 
The Bayesian approach proposed in \cite{kirichenko2017} consists in endowing the soft label 
function $\ell$ with a prior distribution and determining the corresponding posterior. 
 The priors we consider are described in 
detail in the next section.

The posterior distribution for $\ell$ that results from a Bayesian analysis can be used for instance 
for prediction. For the priors we consider, the computation of the posterior mode  is 
closely related to the computation of a kernel-based regression estimate
with a kernel based on the Laplacian matrix associated with the graph (e.g. \cite{ando2007, belkin2004, kolaczyk2009, smola2003, zhu2005}). In that sense the method we consider is 
close to those used in the cited papers. A benefit of the full 
Bayesian framework is that the spread of the posterior may be used to produce a quantification of the uncertainty  in the predictions.
Moreover, we specifically consider a full, hierarchical Bayes procedure because of 
its capability, when properly designed at least, to automatically let the data 
determine the appropriate value of crucial tuning parameters.

As is well known,  the choice of bandwidths, smoothness, or regularization parameters 
in nonparametric methods is a delicate issue in general. The graph context is no exception
in this regard and  it is recognized  that it would be beneficial to have a better understanding 
of how to choose the regularization parameters (e.g. \cite{belkin2006}).
The theoretical results in \cite{kirichenko2017} indicate how the performance of 
nonparametric Bayesian prediction on graphs depends on both the geometry 
of the underlying graph and on the  ``smoothness'' of the (unobserved) soft label function $\ell$.
Moreover, for a certain family of Gaussian process priors \cite{kirichenko2017} 
gives guidelines for
choosing the hyperparameters in such a way that asymptotically optimal performance is obtained.

The aim of the present paper is to provide an implementation of nonparametric 
Bayesian prediction on graphs using Gaussian priors based on the  Laplacian on the graph.
Moreover, we investigate numerically  the influence of the geometry of the 
graph and the smoothness of the soft label function, motivated by the asymptotics 
given in \cite{kirichenko2017}. In this manner we arrive at recommended 
choices for tuning parameters and hyper priors that are in line with 
the theoretical guarantees and that are also computationally convenient.

The rest of this paper is organized as follows. In the next section a more precise description of the problem setting and of the priors we consider are given. 
Algorithms to sample from the posterior are given in Section 3 and computational
aspects are discussed in Section 4. In Section 5 we present numerical experiments. 
We first apply and test the procedure on two simulated datasets, one involving the 
path graph and one a small-world graph, respectively. Next we study the performance
on real data, considering the problems of predicting the function of a protein in a 
protein-protein interaction graph, and classifying hand-written digits using a nearest-neighbour graph.
Concluding remarks are given in  Section 6.

\section{Observation model and priors}

\subsection{Observation model}

Again, we start with a connected, simple undirected graph $G = (V, E)$, with $\# V = n$ vertices denoted
 by $V = \{1, 2, \ldots, n\}$. Associated to every vertex $i$ is a noisy hard label $y_i$. 
We assume the $y_i$'s are independent Bernoulli variables,  with 
\[
P(y_i = 1) = 1-P(y_i=0) = \ell(i), 
\]
where $\ell: V \to (0,1)$ is an unobserved function on the vertices of the graph, the so-called
soft label function. 
We observe only a subset $Y^\text{obs} \subset \{y_1, \ldots, y_n\}$ of all the noisy labels.
This can be a random subset of all the $\{y_1, \ldots, y_n\}$, generated in an arbitrary way, 
but independent of the values of the labels. {Note that in this setup we either observe the label of a vertex or not, so multiple observations of the same vertex are not possible.}

\subsection{Prior on the soft label function}

Our prediction method consists in first inferring the 
 soft label function $\ell$  from $Y^\text{obs}$ and 
subsequently predicting the hard labels by thresholding. 
We take a Bayesian approach which is nonparametric, in the sense 
that we do not assume that $\ell$ belongs to some low-dimensional, for instance
generalized linear family of functions.

\subsubsection{Prior on $\ell$}

To put a prior on $\ell$ we first use the  probit link $\Phi$ (i.e.\ the cdf of the standard normal
distribution) to write $\ell = \Phi(f)$ for some  function $f: V \to \mathbb{R}$ and then put a prior on $f$. 
To achieve a form of Laplacian regularization, which takes the geometry of the graph into account (e.g. \cite{belkin2004, kirichenko2017, kolaczyk2009}) we employ a Gaussian prior 
with a covariance structure that is based on the Laplacian $L$ of the graph $G$. 
Recall that this is the matrix defined as $L = D - A$, where $D$ is the diagonal matrix of vertex degrees and $A$ is the adjacency matrix
of the graph. (See for instance \cite{cvetkovic2010} for background information.)  

We want to consider a Gaussian prior on $f$ with  a fixed power of the Laplacian matrix as precision (inverse covariance) matrix. As the Laplacian matrix has eigenvalue $0$ however, it is not invertible. 
Therefore we add a small number $1/n^2$ to all eigenvalues of $L$ to overcome this problem. By Theorem 4.2 of \cite{mohar1991}, we know that the smallest positive eigenvalue $\lambda_1$ of 
{$L$} satisfies $\lambda_1 \geq 4/n^2$, which motivates this choice. 
To make the prior flexible enough we add  a multiplicative scale parameter $c > 0$ as well. 
Together, this results  in a Gaussian prior for $f$ with zero mean and precision matrix 
$c(L + n^{-2}I)^q$, where $q, c > 0$. We then have
\begin{align*}
\ell & = \Phi(f),\\
f|c &  \sim N(0, (c(L + n^{-2}I)^q)^{-1}).
\end{align*}

To make the connection with kernel-based learning we note that in the corresponding regularized 
kernel-based regression model, the matrix $(L + n^{-2}I)^q$ corresponds to the kernel and the scale 
parameter $c$ to the regularization parameter controlling the trade-off between fitting the observed 
data  and the ``smoothness'' of the function estimate, as measured by the squared ``smoothness'' 
norm $f^T(L + n^{-2}I)^qf$.

\subsubsection{Prior on $c$}

As with all nonparametric methods, the performance of our procedure will depend
crucially on the choice of the hyperparameters $c$ and $q$, which control 
the bias-variance trade-off. The ``correct'' choices of these parameters depends
in principle on  properties of the unobserved function $f$ (or equivalently, the function $\ell$). 
Theoretical considerations in \cite{kirichenko2017} have shown that good 
performance can {be} obtained across a range of regularities of $f$ by fixing $q$ at an appropriate level, and putting
a prior on the hyperparameter {$c$}, so that we obtain a hierarchical Bayes procedure. 
The choices for the prior on $c$ and for $q$ that were shown to work well in 
\cite{kirichenko2017} depend on the geometry of the graph $G$. A main goal
of the present paper is to investigate numerically whether this dependence is 
indeed visible when the method is implemented and to investigate choices for $q$
and $c$ that yield good performance and are computationally convenient as well.

The geometry of the graph $G$ enters through the eigenvalues of the Laplacian, 
which we denote by 
 $0 = \lambda_0 < \lambda_1 \leq \cdots \leq \lambda_{n-1}$.
 (See, e.g., Chapter 7 of \cite{cvetkovic2010} for the main properties of the spectrum of $L$.)
In \cite{kirichenko2017} the theoretical performance of nonparametric Bayes methods
on graphs is studied under the assumption that for some parameter $r \geq 1$,  there exist $i_0 \in \mathbb{N}$, 
$\kappa \in (0, 1]$ and $C_1, C_2 > 0$ such that for all $n$ large enough, 
\begin{equation}\label{eq: geom}
C_1 \left( \frac{i}{n} \right)^{2/r} \leq \lambda_i \leq C_2 \left( \frac{i}{n} \right)^{2/r}, \quad \forall i \in \{ i_0, \ldots, \kappa n \}.
\end{equation}
This condition can be verified numerically for a given graph (as is done for instance
for certain propotein-protein interaction and small world graphs in \cite{kirichenko2017}) 
and can be shown to hold theoretically for instance for graphs that look like regular grids or {tori} of
arbitrary dimensions. 

In our notation, the hyper prior for the scaling parameter $c$ 
that was shown to have good theoretical properties in \cite{kirichenko2017} (under the geometry condition \eqref{eq: geom}) 
is the prior with density $p$ given by 
\begin{equation}\label{eq: c1}
p(c)  \propto c^{-r/(2q)-1}e^{-nc^{-r/(2q)}}, \qquad c > 0. 
\end{equation}
If the true (unknown) soft label function $\ell$ has regularity $\beta > 0$, defined in an 
appropriate, Sobolev-type sense, then this choice for $c$ guarantees that the posterior 
contracts around the truth at the optimal rate, provided the hyper parameter $q$ 
has been chosen such that $q \ge \beta$. Below we investigate the effect 
of choosing $q$ or $r$ too high or too low relative to these ``optimal'' choices. 

To understand better how crucial it is to use the prior \eqref{eq: c1} and to set 
its hyperparameters just right,  we compare it to  a slightly simpler choice that is natural here, 
which is a gamma prior on $c$ with density 
\begin{equation}\label{eq: c2}
p(c)  \propto c^{a-1}e^{-bc}, \qquad c > 0, 
\end{equation}
for certain $a,b > 0$. This choice is computationally convenient due to the usual 
normal-inverse gamma partial conjugacy (see e.g.\ \cite{choudhuri2007, liang2007}
in the context of our setting).
It introduces two more hyperparameters $a$ and $b$. The authors of \cite{choudhuri2007} and \cite{liang2007} mention $a=b=0$, corresponding to Jeffreys prior, which is improper, but does not result in any computational restrictions. 
We will see in the numerical experiments  that this choice is a reasonable one in our setting as well.

To distinguish between the two priors \eqref{eq: c1} and \eqref{eq: c2} in the paper 
we always call \eqref{eq: c2} the {\em ordinary gamma} prior for $c$, and 
\eqref{eq: c1} the {\em generalized gamma} prior for $c$.

We remark that since we are considering two competing priors on $c$, we could in principle 
consider some form of (Bayesian) model selection. We will however see in the numerical experiments
that the ordinary gamma prior generally performs better than the generalized gamma. Since the ordinary 
gamma prior is also preferable from the computational perspective, 
we would recommend to use the ordinary gamma instead of a combined method.

\subsection{Latent variables and missing labels}
\label{sec: latent}

Combining what we have so far, we obtain  a hierarchical  model that can be described as follows:
\begin{align*}
y_i \given f, c & \sim \text{indep.\ Bernoulli}(\Phi(f_i)),  \qquad i =1, \ldots, n,\\
f \given c & \sim N(0, (c(L + n^{-2}I)^q)^{-1}),\\
c & \sim p \quad \text{given by \eqref{eq: c1} or \eqref{eq: c2}}. 
\end{align*}
As is well known (cf.\ \cite{albert1993}) an equivalent formulation using an additional 
layer of  latent variables $z = (z_1, \ldots, z_n)$ is given by 
\begin{align*}
y_i  & = 1_{z_i > 0},  \qquad i = 1, \ldots, n,\\
z \given f, c & \sim   N(f, I),\\
f \given c & \sim N(0, (c(L + n^{-2}I)^q)^{-1}),\\
c & \sim p  \quad \text{given by \eqref{eq: c1} or \eqref{eq: c2}}. 
\end{align*}
This is a more convenient representation which we will use in  our computations.

We consider the situation in which we do not observe all the labels 
$y_i$, but only a certain subset $Y^\text{obs}$. {The labels that we observe, are only observed once.} The precise mechanism that determines 
which $y_i$'s we observe and which ones are missing is not important for 
the algorithm we propose. We
only assume that it is independent of the other elements of the model. 
Specifically, we assume that for some arbitrary distribution $\mu$ on the collection $2^V$ of 
subsets vertices, a set of vertices $I^\text{obs} \subset V$ is drawn and that 
we see which vertices were selected and what the corresponding noisy labels 
are. In other words, the  observed data is
$D = \{(i, y_i) : i \in I^\text{obs}\}$.  

All in all, the full hierarchical scheme we will work with is the following:
\begin{equation}
\begin{split}\label{eq: fullmodel}
D & = \{(i, y_i) : i \in I^\text{obs}\},\\
I^\text{obs} & \sim \mu,\\
y_i  & = 1_{z_i > 0},  \qquad i = 1, \ldots, n, \\
z \given f, c & \sim   N(f, I),\\
f \given c & \sim N(0, (c(L + n^{-2}I)^q)^{-1}),\\
c & \sim p \quad \text{given by \eqref{eq: c1} or \eqref{eq: c2}}. 
\end{split}
\end{equation}
Our goal is to compute the posterior $f \given D$ and to use it to predict the 
unobserved labels.

\section{Sampling scheme}

We use the latent variable approach as in \cite{albert1993, choudhuri2007}, for instance,
and implement a Gibbs sampler to sample from the posterior $f \given D$ in the setup 
\eqref{eq: fullmodel}. This involves sampling repeatedly from the conditionals
$p(z\given c, f, D)$, $p(f\given c, z, D)$ and $p(c\given z, f, D)$.
We detail these three steps in the following subsections.

\subsection{Sampling from $p(z\given c, f, D)$}

By construction, we have given $D$ that 
the latent variables $z_i$, $i \not \in I^\text{obs}$ 
corresponding to the missing observations are independent of those 
corresponding to the observed labels. We simply have  
that given $c, f$ and $D$, the variables  $z_i$, $i \not \in I^\text{obs}$, 
are independent $N(f_i, 1)$-variables.

As for the latent variables corresponding to the observed labels, we have 
by \eqref{eq: fullmodel} that the $z_i$ are independent given $c, f$ and $D$ and
\[
p(z_i\given c, f, D) = p(z_i\given c, f, y_i) \propto p(y_i\given z_i) p(z_i \given f, c)
\propto 1_{z_i\text{ matches } y_i} e^{-\frac12(z_i-f_i)^2},
\]
where we say that $z_i$ matches $y_i$ if $y_i = 1_{z_i > 0}$. For $y_i= 1$, 
this describes the $N(f_i, 1)$-distribution, conditioned to be positive.
We denote this distribution by $N_+(f_i, 1)$.  
For $y_i=0$ it corresponds to the $N(f_i, 1)$-distribution, conditioned to be
negative. We denote this distribution by $N_-(f_i, 1)$.  

Put together, we see that given $c, f$, and $D$, the $z_i$ are independent and 
\[
z_i \given c,f, D \sim
\begin{cases}
N(f_i, 1), & \text{if $i \not \in I^\text{obs}$}, \\
N_+(f_i, 1), & \text{if $i  \in I^\text{obs}$ and $y_i = 1$},\\
N_-(f_i, 1), & \text{if $i  \in I^\text{obs}$ and $y_i = 0$}.
\end{cases}
\]
We note that generating normal random variables conditioned to be positive or negative can for example be done by a simple rejection algorithm or inversion (e.g. \cite{devroye1986}).

\subsection{Sampling from $p(f\given c, z, D)$}

Since given $z$ we know all the $y_i$'s, and $I^\text{obs}$ 
is independent of all other elements of the model,  we have $p(f\given c, z, D)  = p(f\given c, z)$.
Next, we have
\[
p(f \given c, z) \propto p(z\given f, c)p(f\given c). 
 \]
By plugging in what we know from \eqref{eq: fullmodel} we get 
\[
p(f \given c, z) \propto e^{-\frac12f^T(I + c(L+n^{-2}I)^q )f + z^Tf}. 
 \]
Completing the square, it follows that 
\begin{equation}\label{eq: f}
f \given c, z \sim N\Big(((I + c(L+n^{-2}I)^q)^{-1}z, (I + c(L+n^{-2}I)^q)^{-1}\Big).
\end{equation}

\subsection{Sampling from $p(c\given z, f, D)$}

Again we use that given $z$ we know all the $y_i$'s, and that $I^\text{obs}$ 
is independent of everything else, which gives  $p(c\given  z, f, D)  = p(c\given  z, f)$.
Since given $f$, $z$ is independent of $c$, we have
\begin{equation}\label{eq: piet}
p(c\given  z, f) \propto p(f \given c)p(c). 
\end{equation}
Now the method for (approximate) sampling from $c \given z, f$ depends 
on the choice of the prior for $c$.

\subsubsection{Ordinary gamma prior for $c$}

If the prior density for $c$ is the ordinary gamma density given by \eqref{eq: c2}
we have the usual Gaussian-inverse gamma conjugacy. Indeed,
then  we have
\[
p(f \given c)p(c) \propto c^{n/2}e^{-\frac12 cf^T(L+n^{-2}I)^qf} c^{a-1}e^{-bc}
\propto c^{a+ n/2 -1}e^{-c(b+ \frac12f^T(L+n^{-2}I)^qf)}.
\]
In other words, in this case we have 
\[
c\given  z, f \sim \Gamma\Big(a+n/2, b+ \frac12f^T(L+n^{-2}I)^qf\Big). 
\]

\subsubsection{Generalized gamma prior for $c$}

If the prior density for $c$ is the generalized gamma density given by \eqref{eq: c1}
we do not have conjugacy and we replace drawing from the 
exact conditional $c \given z,f $, as done in the preceding subsection, by a Metropolis-Hastings
step. To this end we choose  a proposal density $s(c'\given c)$. 
To generate a new draw for $c$ we follow the usual steps:
\begin{itemize}
\item
draw a proposal $c' \sim s(\cdot\given c)$;
\item
draw an independent uniform variable $V$ on $[0,1]$;
\item
if
\[
V \le \frac{t(c') s(c\given c')}{t(c)s(c'\given c)},
\]
where 
\[
t(c) = c^{n/2-r/(2q)-1}e^{-\frac12 c f^T(L+n^{-2}I)^q)f-nc^{-r/(2q)}},
\]
then accept the proposal $c'$ as new draw, else retain the old draw $c$. 
\end{itemize}
Note that  $t(c)$ is indeed proportional to $p(f\given c)p(c)$, 
as required (cf.\ \eqref{eq: piet}).

We have considered different proposal distributions $s(c'\given c)$ 
in our experiments. Our experiments indicate that a random walk
proposal works well.

\subsection{Overview of the sampling schemes}

For convenience we summarize our sampling scheme, which depends  on 
the prior for $c$ that we use. 

For the generalised gamma prior 
\eqref{eq: c1} we have the following:

\bigskip

\begin{algorithm}[H]
\caption{Sampling scheme when using the generalized  gamma prior on $c$.}
\label{alg: mcmc1}
\begin{algorithmic}[1]
\REQUIRE Data $D= \{(i, y_i) : i \in I^\text{obs}\}$, initial values $f = f^{(0)}$ and $c = c^{(0)}$.
\ENSURE MCMC sample from the joint posterior $p(z, f, c| D)$.
\REPEAT
\STATE 
For $i =1, \ldots, n$, draw independent
\[
z_i \sim \begin{cases}
N(f_i, 1), & \text{if $i \not \in I^\text{obs}$}, \\
N_+(f_i, 1), & \text{if $i  \in I^\text{obs}$ and $y_i = 1$},\\
N_-(f_i, 1), & \text{if $i  \in I^\text{obs}$ and $y_i = 0$}.
\end{cases}
\]
\STATE 
Draw
\[
f  \sim N\Big(((I + c(L+n^{-2}I)^q)^{-1}z, (I + c(L+n^{-2}I)^q)^{-1}\Big).
\]
\STATE 
Draw a proposal $c'  \sim s(\cdot\given c)$ and a uniform $V$ on $[0,1]$.  
\IF{
\[
V \le \Big(\frac{c'}{c}\Big)^{n/2-r/(2q)-1}e^{-\frac12 (c'-c) f^T(L+n^{-2}I)^q)f
-n((c')^{-r/(2q)}- c^{-r/(2q)})} \frac{s(c\given c')}{s(c'\given c)},
\]
}
\STATE
Set $c = c'$. 
\ELSE
\STATE
Retain $c$.
\ENDIF

\UNTIL{You have a large enough sample.}
\end{algorithmic}
\end{algorithm}

For the ordinary gamma prior \eqref{eq: c2} the algorithm looks as follows:

\begin{algorithm}[H]
\caption{Sampling scheme when using the ordinary gamma prior on $c$.}
\label{alg: mcmc2}
\begin{algorithmic}[1]
\REQUIRE Data $D= \{(i, y_i) : i \in I^\text{obs}\}$, initial values $f = f^{(0)}$ and $c = c^{(0)}$.
\ENSURE MCMC sample from the joint posterior $p(z, f, c| D)$.
\REPEAT
\STATE 
For $i =1, \ldots, n$, draw independent
\[
z_i \sim \begin{cases}
N(f_i, 1), & \text{if $i \not \in I^\text{obs}$}, \\
N_+(f_i, 1), & \text{if $i  \in I^\text{obs}$ and $y_i = 1$},\\
N_-(f_i, 1), & \text{if $i  \in I^\text{obs}$ and $y_i = 0$}.
\end{cases}
\]
\STATE 
Draw
\[
f  \sim N\Big(((I + c(L+n^{-2}I)^q)^{-1}z, (I + c(L+n^{-2}I)^q)^{-1}\Big).
\]
\STATE 
Draw
\[
c \sim \Gamma\Big(a+n/2, b+ \frac12f^T(L+n^{-2}I)^qf\Big). 
\]
\UNTIL{You have a large enough sample.}
\end{algorithmic}
\end{algorithm}

\section{Computational aspects}

\subsection{Using the eigendecomposition of the Laplacian}

In every iteration of either  Algorithm \ref{alg: mcmc1} or \ref{alg: mcmc2}
 the matrix $I+ c(L+n^{-2}I)^q$ has to be inverted.
Doing this naively  can in general be computationally demanding, taking $O(n^3)$ computations,
in particular if $L$ is not very sparse, 
i.e.\ if $G$ is a dense graph with  many vertices with relatively large degree. 
To relax the computational burden it can  be  advantageous to 
change coordinates and to work relative to a basis of 
eigenvectors of the graph Laplacian $L$. 

To make this concrete, suppose we have the eigendecomposition of the Laplacian matrix $L = U \Lambda U^T$, 
where $\Lambda$ is the diagonal matrix of eigenvalues of $L$ and $U$ is an orthogonal matrix
of eigenvectors. Computing this decomposition costs us one time $O(n^3)$ computations.
We can then parametrise the model by the vector $g = U^Tf$ instead of $f$. The corresponding equivalent 
formulation of \eqref{eq: fullmodel} is  given by 
\begin{equation}
\begin{split}\label{eq: fullmodel2}
D & = \{(i, y_i) : i \in I^\text{obs}\},\\
I^\text{obs} & \sim \mu,\\
y_i  & = 1_{z_i > 0},  \\
z \given g, c & \sim  \ N(Ug, I),\\
g \given c & \sim N(0, (c(\Lambda + n^{-2} I)^{q})^{-1}),\\
c & \sim p \quad \text{given by \eqref{eq: c1} or \eqref{eq: c2}}. 
\end{split}
\end{equation}

Making the appropriate, straightforward adaptations in the posterior computations, 
the sampling schemes take the following form in this parametrisation:

\begin{algorithm}[H]
\caption{Sampling scheme when using the generalized  gamma prior on $c$, using $L = U\Lambda U^T$.}
\label{alg: mcmc3}
\begin{algorithmic}[1]
\REQUIRE Data $D= \{(i, y_i) : i \in I^\text{obs}\}$, initial values $g = g^{(0)}$ and $c = c^{(0)}$.
\ENSURE MCMC sample from the joint posterior $p(z, g, c| D)$.
\REPEAT
\STATE 
Compute $f = Ug$ and for $i =1, \ldots, n$, draw independent
\[
z_i \sim \begin{cases}
N(f_i, 1), & \text{if $i \not \in I^\text{obs}$}, \\
N_+(f_i, 1), & \text{if $i  \in I^\text{obs}$ and $y_i = 1$},\\
N_-(f_i, 1), & \text{if $i  \in I^\text{obs}$ and $y_i = 0$}.
\end{cases}
\]
\STATE 
For $i =1 , \ldots, n$, draw independent
\[
g_i  \sim N\Big(\frac{{z^Tu^{(i)}}}{1+c(\lambda_{i-1} + 1/n^2)^q}, \frac{1}{1+c(\lambda_{i-1} + 1/n^2)^q}\Big).
\]
\STATE 
Draw a proposal $c'  \sim q(\cdot\given c)$ and a uniform $V$ on $[0,1]$.  
\IF{
\[
V \le \Big(\frac{c'}{c}\Big)^{n/2-r/(2q)-1}e^{-\frac12 (c'-c) \sum (\lambda_{i-1}+1/n^{2})^qg^2_i
-n((c')^{-r/(2q)}- c^{-r/(2q)})} \frac{q(c\given c')}{q(c'\given c)},
\]
}
\STATE
Set $c = c'$. 
\ELSE
\STATE
Retain $c$.
\ENDIF

\UNTIL{You have a large enough sample.}
\end{algorithmic}
\end{algorithm}

For the ordinary gamma prior \eqref{eq: c2} the algorithm looks as follows:

\begin{algorithm}[H]
\caption{Sampling scheme when using the ordinary gamma prior on $c$, using $L = U\Lambda U^T$.}
\label{alg: mcmc4}
\begin{algorithmic}[1]
\REQUIRE Data $D= \{(i, y_i) : i \in I^\text{obs}\}$, initial values $g = g^{(0)}$ and $c = c^{(0)}$.
\ENSURE MCMC sample from the joint posterior $p(z, g, c| D)$.
\REPEAT
\STATE 
Compute $f = Ug$ and for $i =1, \ldots, n$, draw independent
\[
z_i \sim \begin{cases}
N(f_i, 1), & \text{if $i \not \in I^\text{obs}$}, \\
N_+(f_i, 1), & \text{if $i  \in I^\text{obs}$ and $y_i = 1$},\\
N_-(f_i, 1), & \text{if $i  \in I^\text{obs}$ and $y_i = 0$}.
\end{cases}
\]
\STATE 
For $i =1 , \ldots, n$, draw independent
\[
g_i  \sim N\Big(\frac{{z^Tu^{(i)}}}{1+c(\lambda_{i-1} + 1/n^2)^q}, \frac{1}{1+c(\lambda_{i-1} + 1/n^2)^q}\Big).
\]
\STATE 
Draw
\[
c \sim \Gamma\Big(a+n/2, b+ \frac12\sum (\lambda_{i-1}+1/n^{2})^qg^2_i\Big). 
\]
\UNTIL{You have a large enough sample.}
\end{algorithmic}
\end{algorithm}

We note that in this approach we need to invest once in the computation 
of the spectral decomposition of the Laplacian $L$. This removes the 
$O(n^3)$ matrix inversions in each iteration of  the algorithms.  We do remark 
however that there is a matrix multiplication $Ug$ in line 2 of the algorithms.
This is in principle an $O(n^2)$ operation{, which is the most expensive operation in each iteration}. Moreover, 
Algorithms \ref{alg: mcmc3} and \ref{alg: mcmc4} produce 
samples of $g \given D$, that is, we obtain the posterior samples 
as vectors of coordinates relative to the eigenbasis of the Laplacian. If
we want the samples in the original basis, which is what we need for 
prediction, we need to multiply all these vectors by $U$.

\subsection{A strategy for sparse graphs}

Instead of sampling the Gaussian distribution in 
$ f\given c, z$ at once as in \eqref{eq: f}, it can be advantageous to sample 
this vector one coordinate at the time. 

Denoting by $v_{-i}$ the vector $v$ with coordinate $i$ removed, standard Gaussian
computations show that for every coordinate $i$, 
\[
f_i \given f_{-i}, c,z \sim N(\mu_i, \sigma^2_i), 
\]
where
\begin{equation*}
\mu_i =  \frac{z_i - c((L + n^{-2}I)^q)_{i, -i}f_{-i}}{c((L + n^{-2}I)^q)_{i, i} + 1},
\end{equation*}
and
\begin{equation*}
\sigma^2_i = \frac{1}{c((L+n^{-2}I)^q)_{i, i} + 1}.
\end{equation*} 
This method for  sampling from the conditional $f \given c, z$ does not require the eigendecomposition
of $L$. In case the power of the Laplacian matrix $q$ is an integer, the computations 
for a fixed $i$  only involve the $q$-step neighbors of vertex $i$, which might be computationally attractive in graphs where the number of $q$-step neighbors of each vertex is low. {For example, if the number of $q$-step neighbors is bounded by $K$, the complexity of each iteration is $O(Kn)$.}

\section{Numerical experiments}

\subsection{Path graph}

\begin{figure}[H]
\begin{minipage}{0.49\textwidth}
\includegraphics[width = \textwidth]{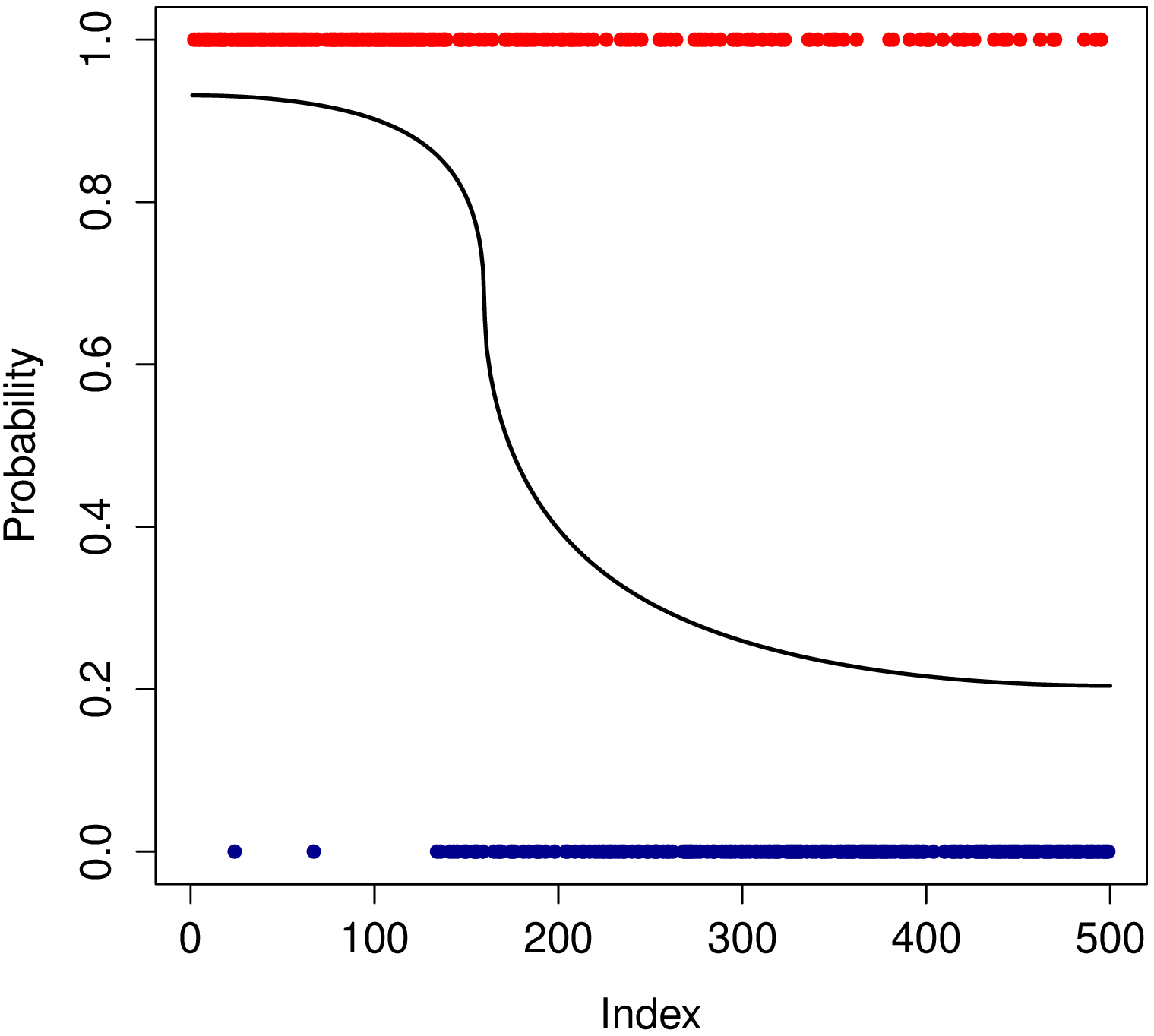}
\end{minipage}
\begin{minipage}{0.49\textwidth}
\includegraphics[width = \textwidth]{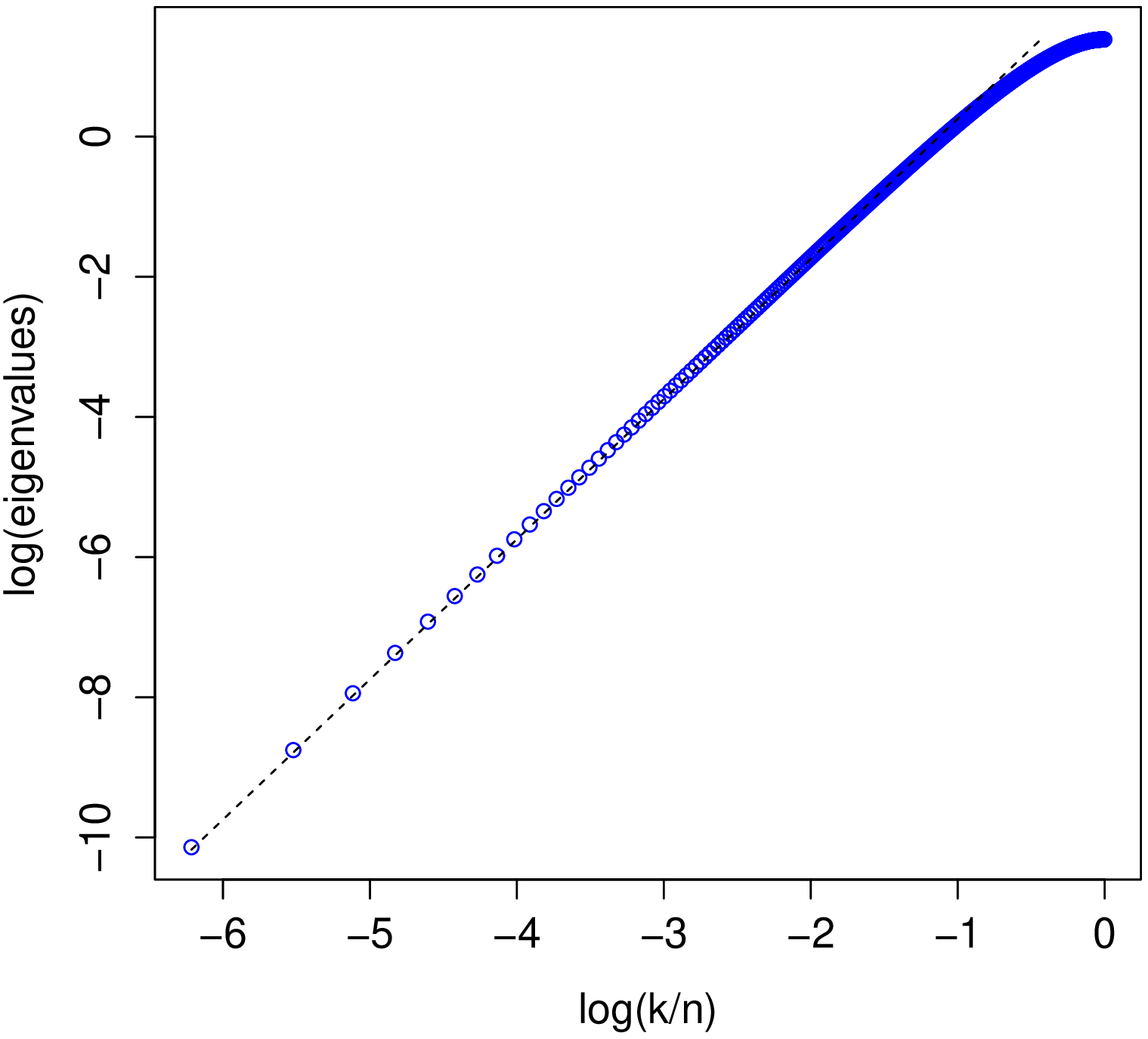}
\end{minipage}
\caption{Left: soft label function  and simulated noisy hard labels   
on a path graph with $500$ nodes. Right: spectrum of the Laplacian.}\label{fig: p1}
\end{figure}

To explore some of the issues involved in implementing Bayesian prediction prediction on graphs
we first consider the basic example of simulated data on  the path graph with $n=500$ 
vertices. In this case, it is  
known that the Laplacian eigenvalues are 
$\lambda_k = 4 \sin^2({\pi k}/({2 n}))$ for $k \geq 1$, with corresponding eigenvectors 
\begin{equation*}
u^{(k)}_i = \sqrt{2}\cos\left(\frac{\pi(i-\frac{1}{2})k}{n}\right), \quad i = 1, \ldots, n.
\end{equation*}
This graph satisfies the geometry condition \eqref{eq: geom}
 with $r = 1$. 
  To simulate data we construct a function $f_0$ on the graph by setting
\begin{equation*}
f_0 = \sum_{k = 1}^{n - 1} a_k u^{(k)},
\end{equation*}
where we choose  $a_k = k^{-1.5} \sin k$ for $k \geq 1$. 
This function has Sobolev-type smoothness $\beta = 1$, as defined precisely in 
\cite{kirichenko2017}. 
We simulate noisy labels $Y_i$ on the graph vertices satisfiying $P(Y_i = 1) = \ell_0(i) = \Phi(f_0(i))$, where $\Phi$ is the standard normal cdf. Finally we remove $20\%$ of the labels at 
random to generate the set of observed labels $Y^\text{obs}$.
The left panel of Figure \ref{fig: p1} shows the soft label function $\ell_0$ and simulated  noisy hard labels $Y_i$ on the path graph with $n=500$ vertices. 
In the right panel $\log\lambda_k$ is plotted against 
$\log (k/n)$ to illustrate that for this graph the geometry condition indeed holds with $r=1$.

\begin{figure}[H]
\begin{minipage}{0.32\textwidth}
\includegraphics[width = \textwidth]{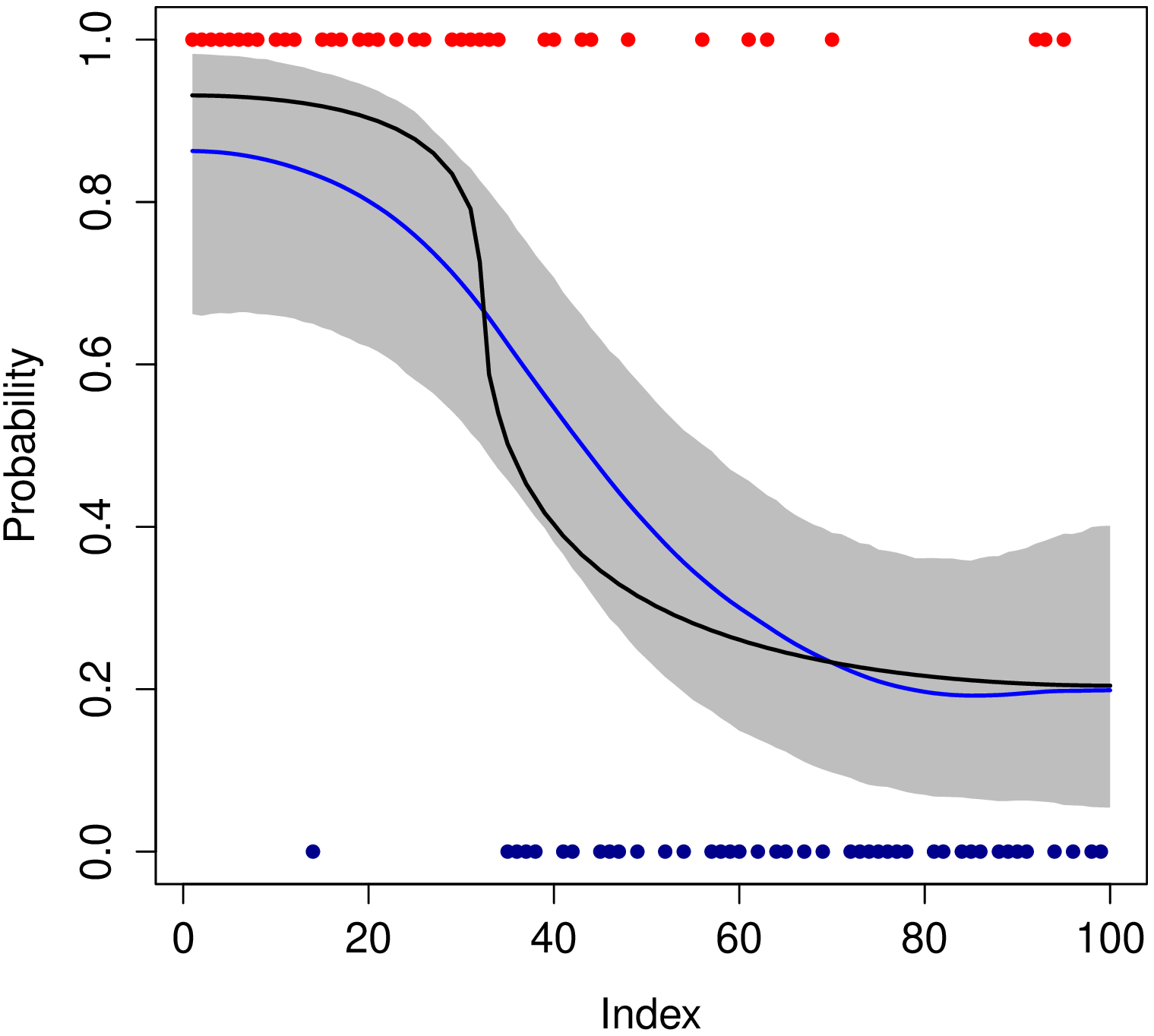}
\end{minipage}
\begin{minipage}{0.32\textwidth}
\includegraphics[width = \textwidth]{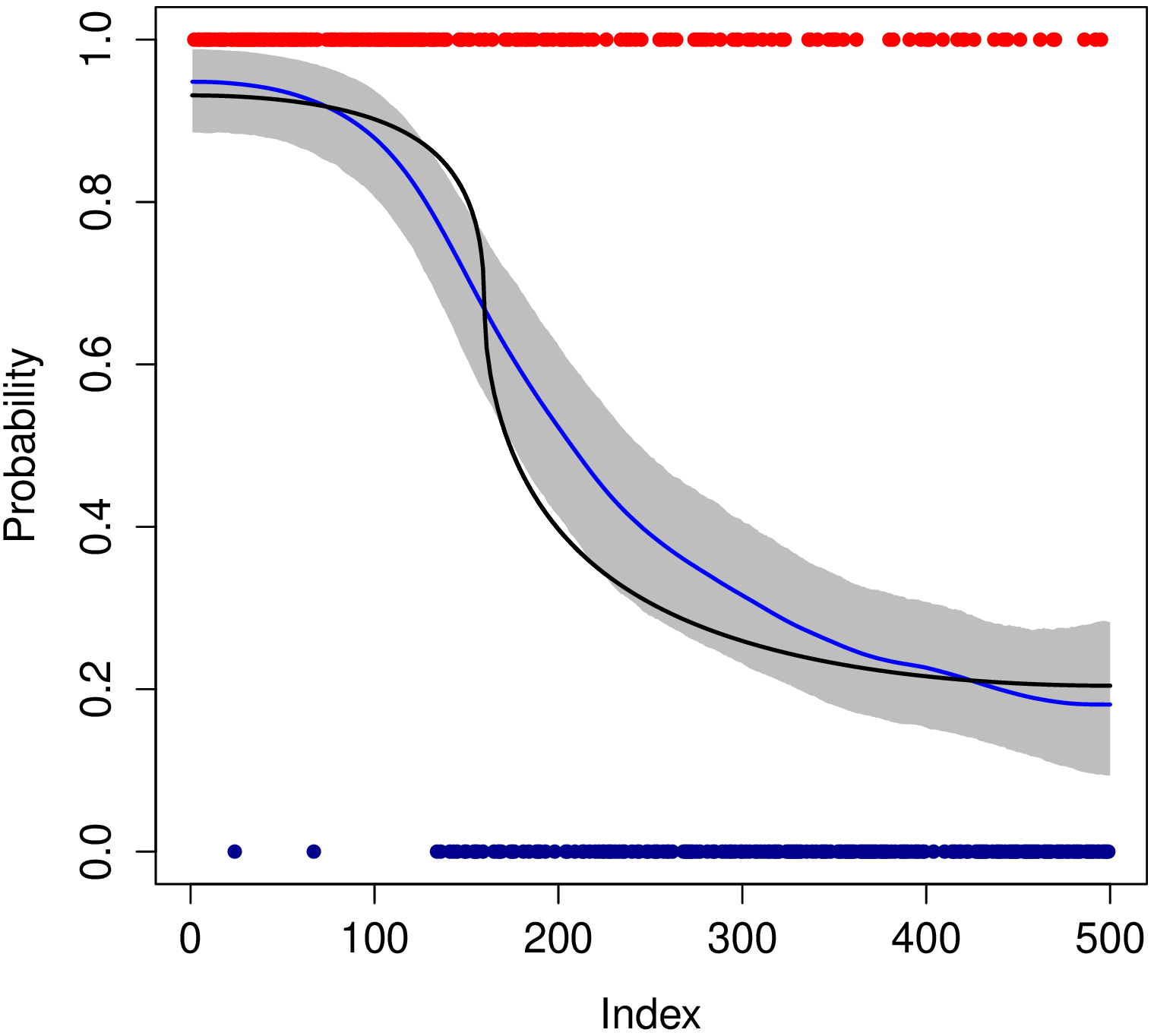}
\end{minipage}
\begin{minipage}{0.32\textwidth}
\includegraphics[width = \textwidth]{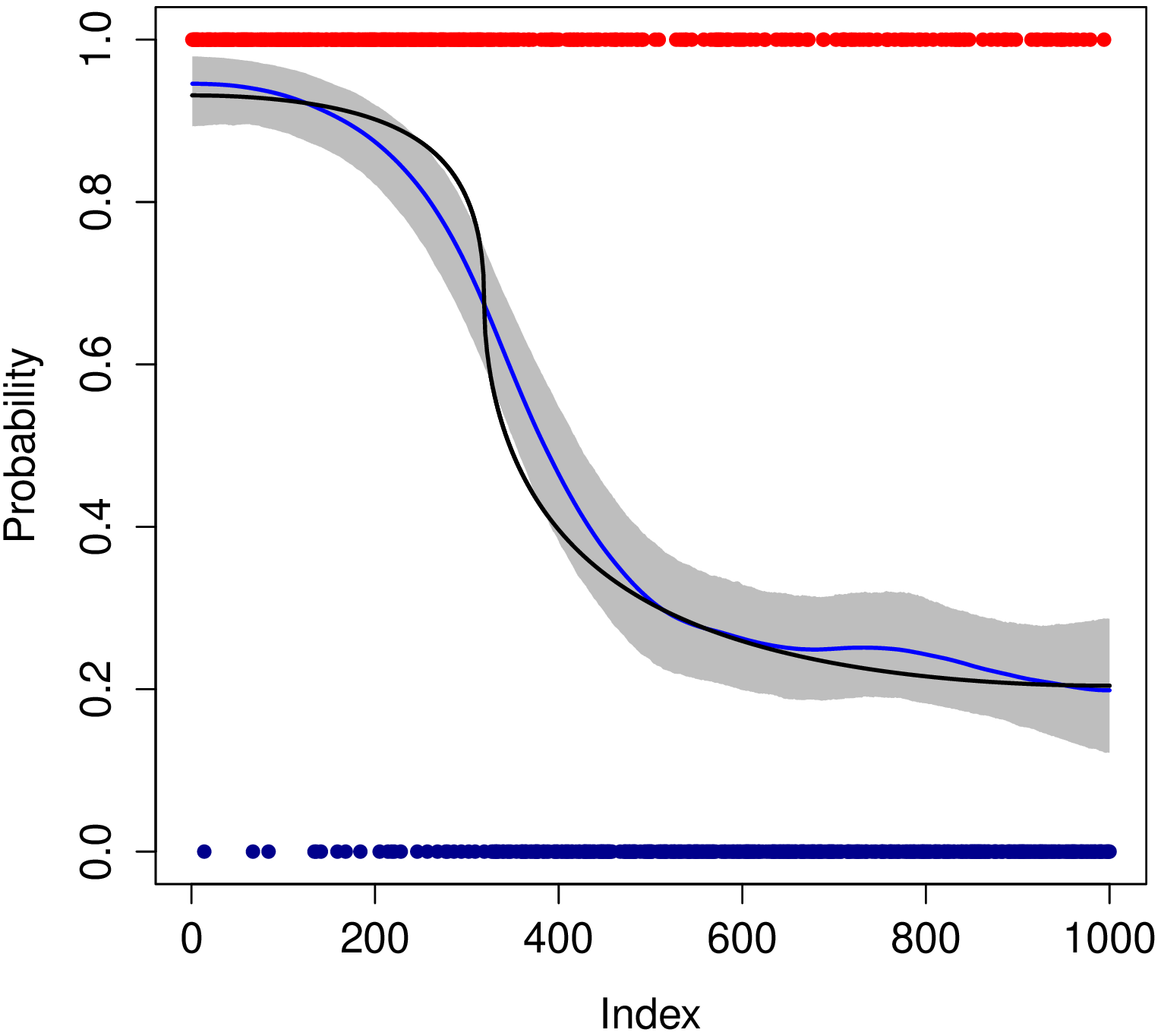}
\end{minipage}
\caption{Posteriors for the soft label function for  $n = 100, 500, 1000$.
Prior on $c$ is the generalised gamma.}\label{fig:pbasic}
\end{figure}

In Figure \ref{fig:pbasic} we visualise the posterior for the soft label function 
$\ell$ for various graph sizes $n$. Here we used the generalised gamma prior on $c$
with $r = 1$, $\alpha = \beta = 1.5$ and $q = \alpha + r/2$. 
These values are suggested by the theory in \cite{kirichenko2017}. 
The blue line is the posterior mean
and the gray area depicts  point-wise $95\%$ credible intervals.

\begin{figure}[H]
\begin{minipage}{0.32\textwidth}
\includegraphics[width = \textwidth]{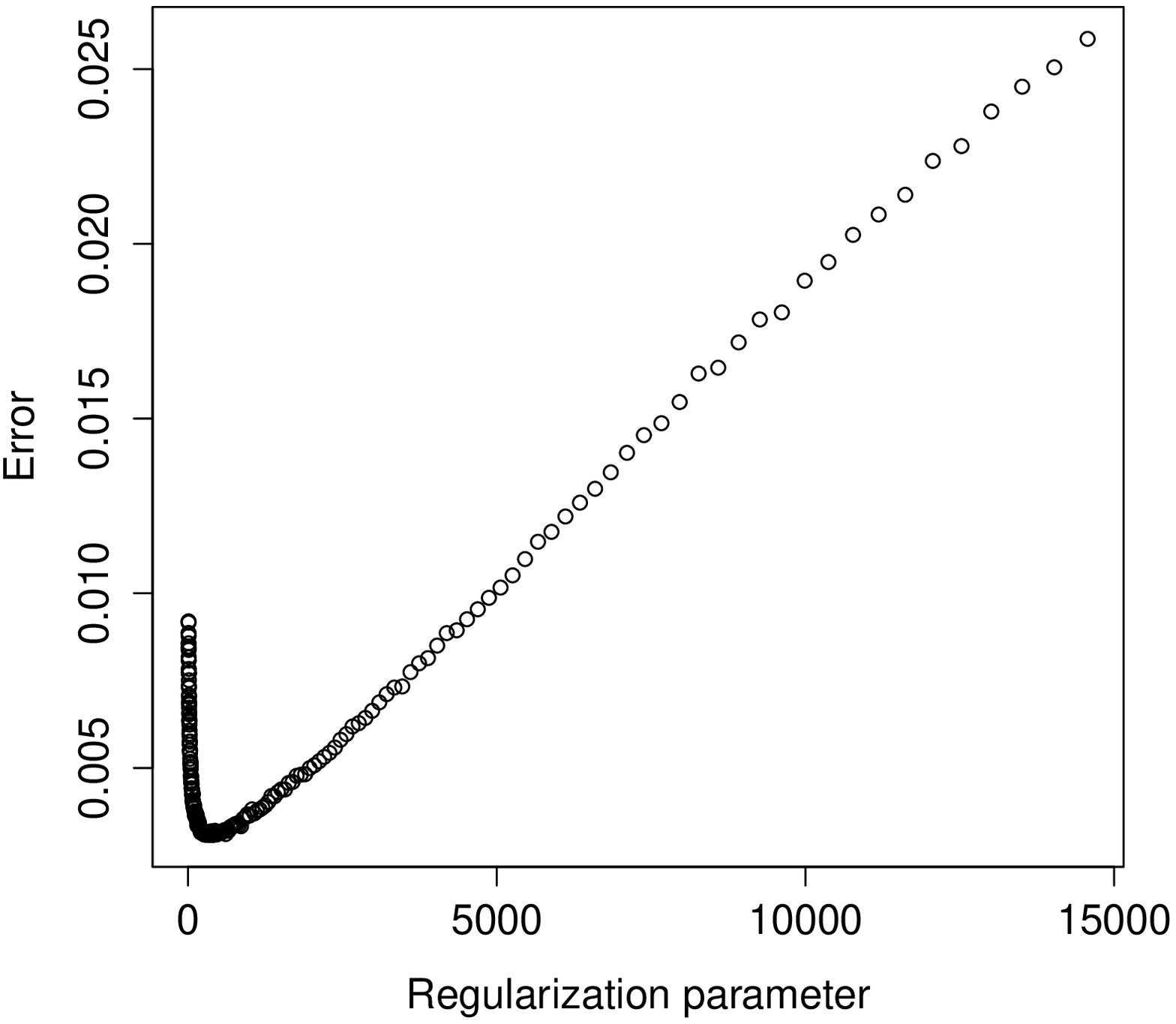}
\includegraphics[width = \textwidth]{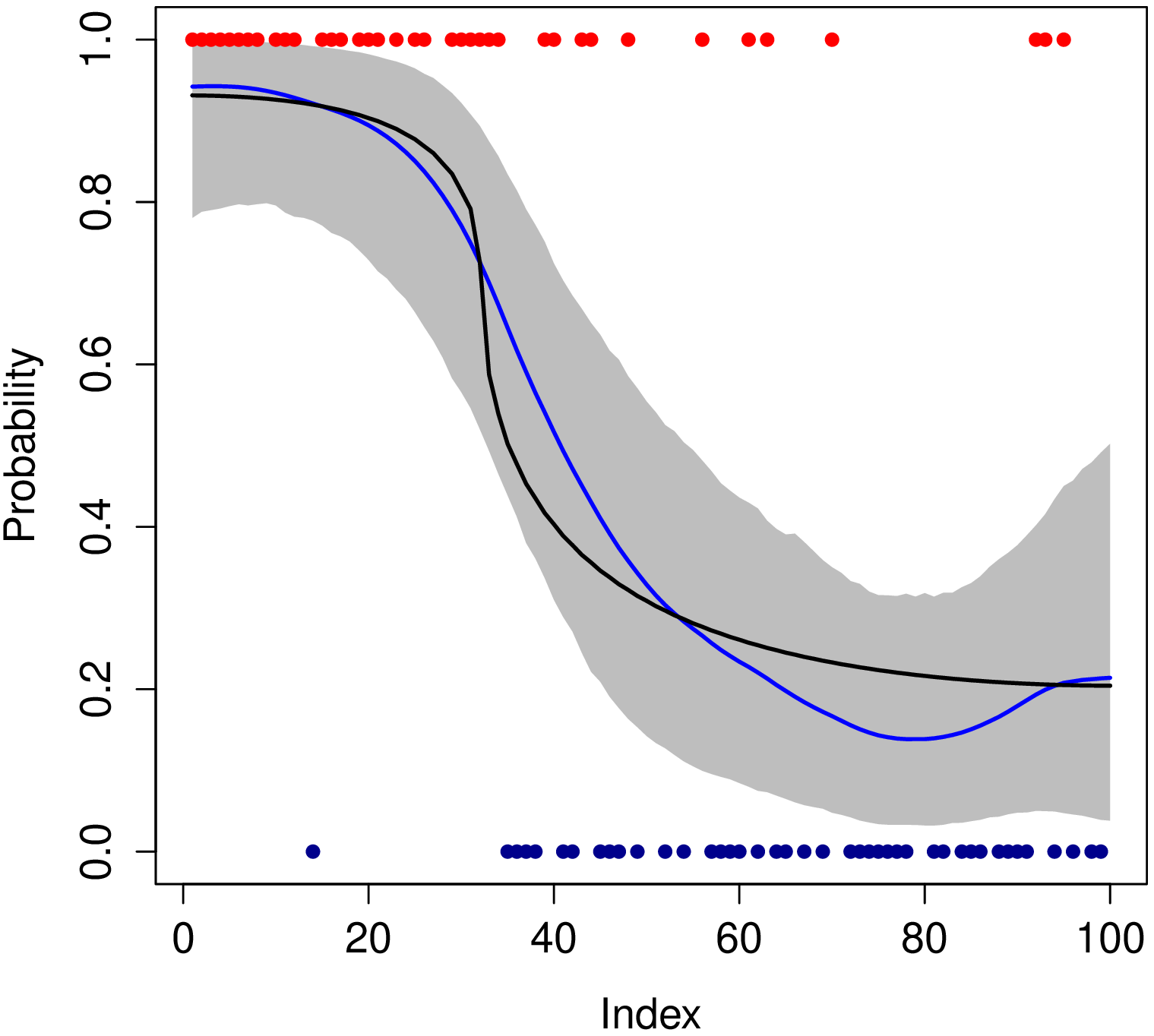}
\end{minipage}
\begin{minipage}{0.32\textwidth}
\includegraphics[width = \textwidth]{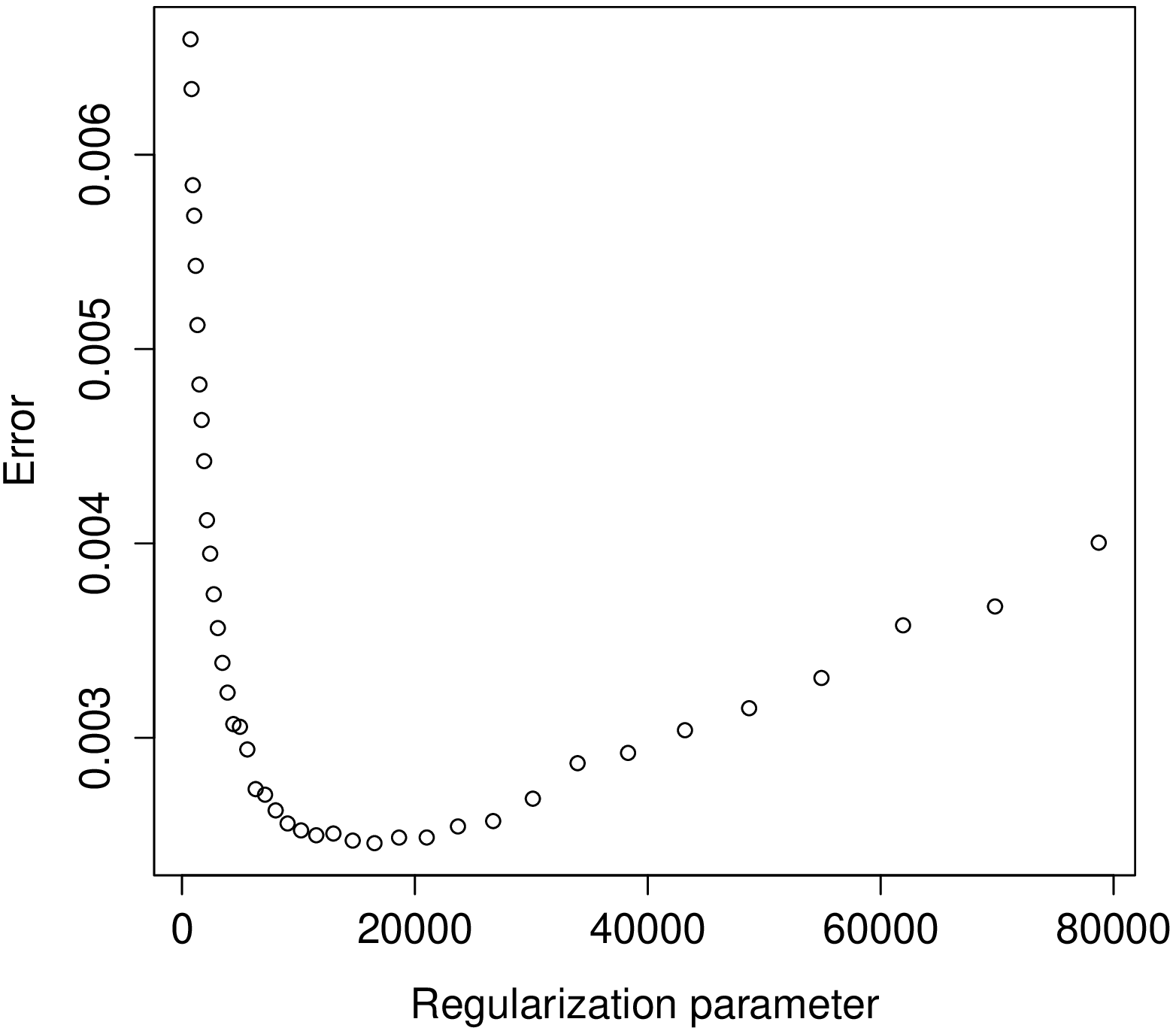}
\includegraphics[width = \textwidth]{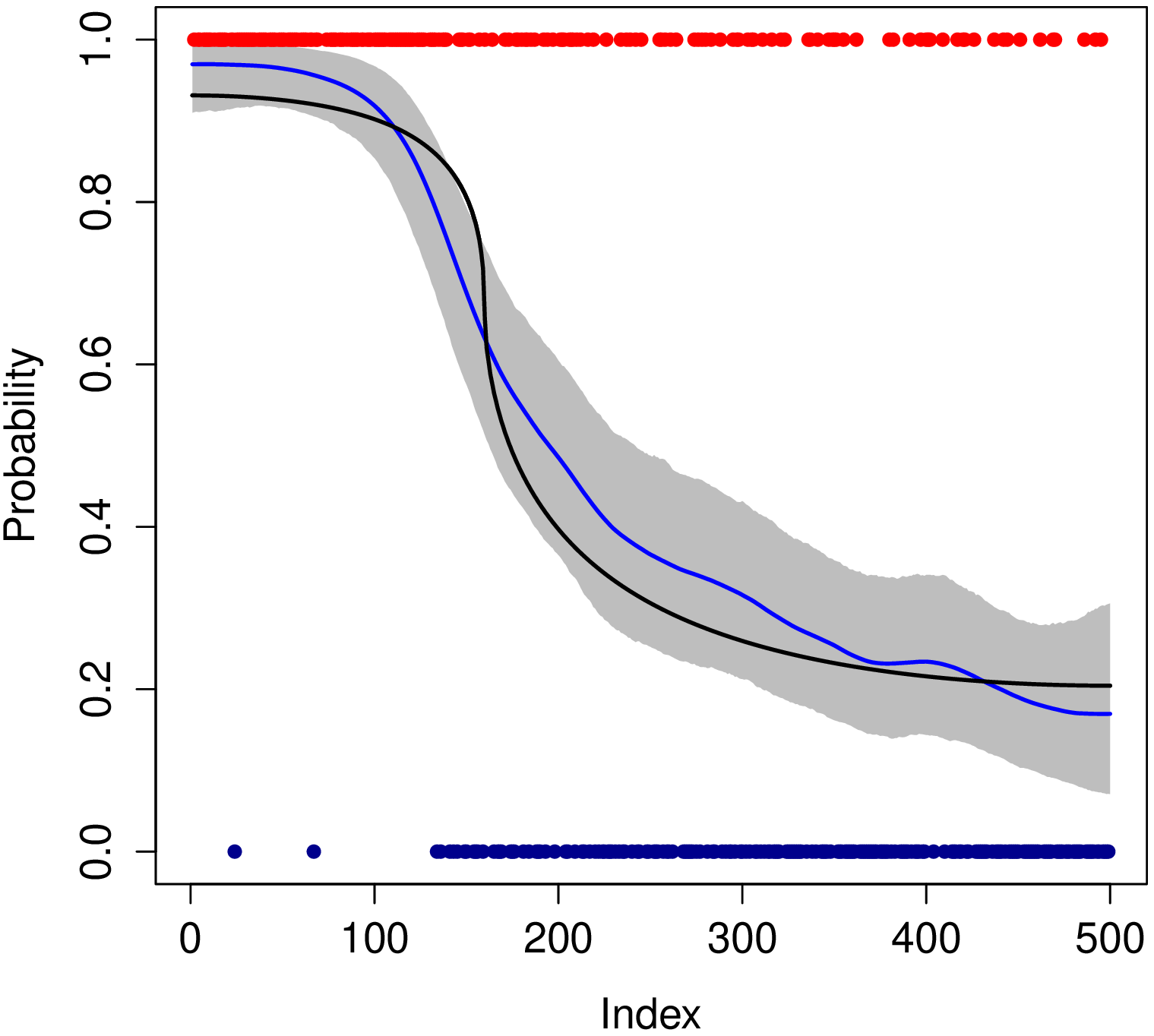}
\end{minipage}
\begin{minipage}{0.32\textwidth}
\includegraphics[width = \textwidth]{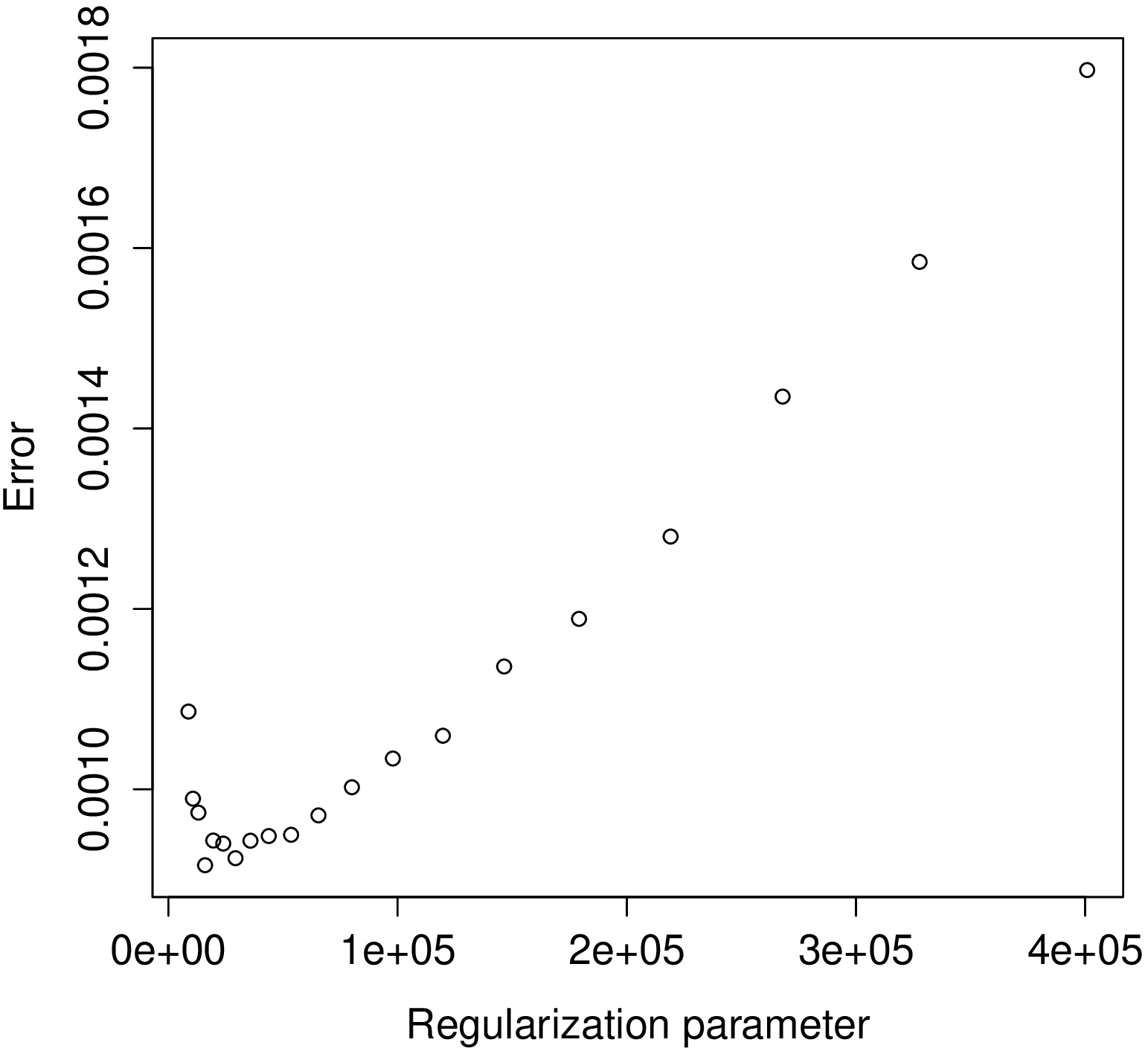}
\includegraphics[width = \textwidth]{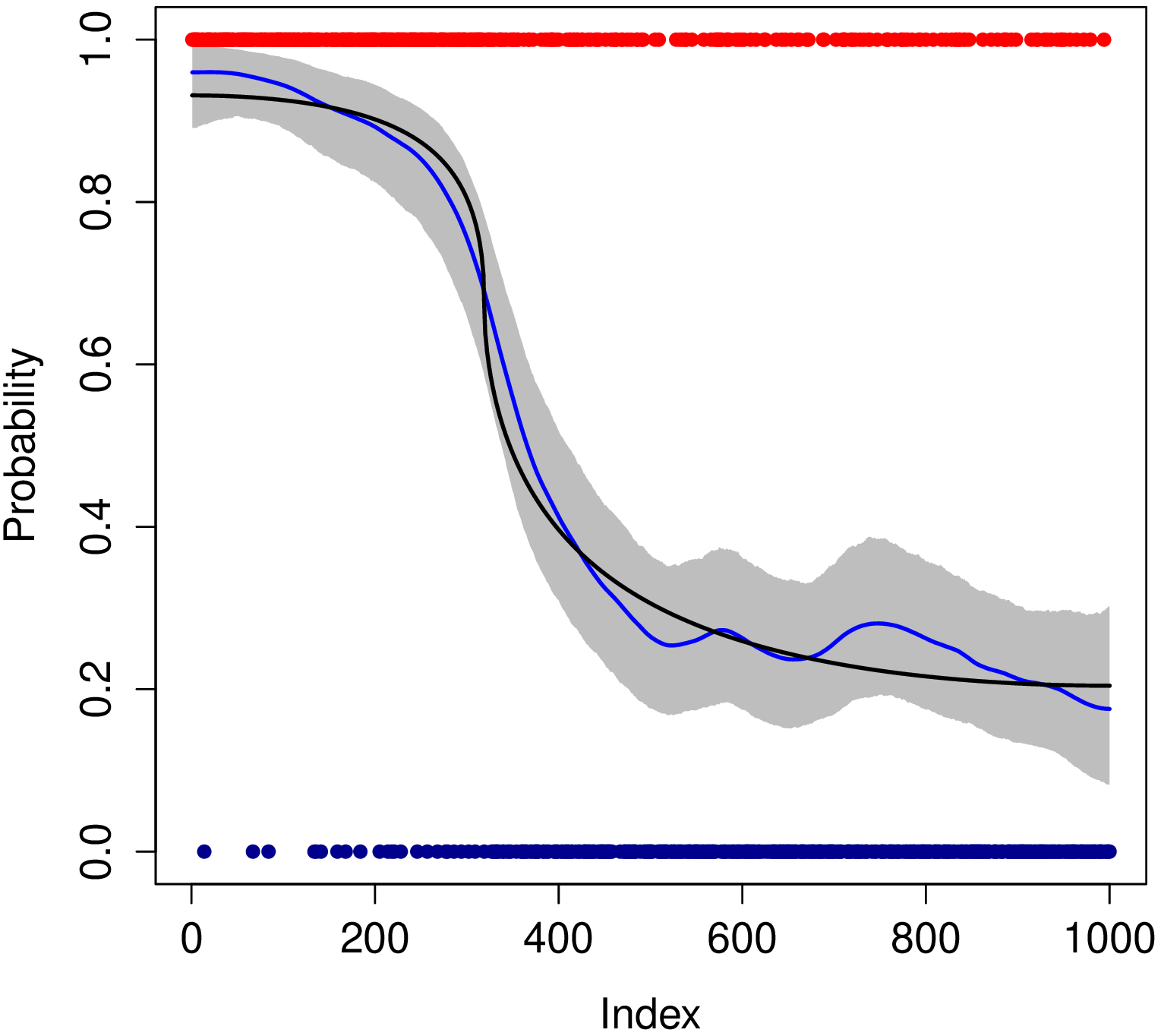}
\end{minipage}
\caption{Posteriors for fixed tuning parameter $c$ for $n = 100, 500, 1000$. 
Top panel: MSE of the posterior mean as function of $c$. 
Bottom panel: posterior for the soft label function for optimal choice of $c$.
Values of $c$ were $4.0\cdot10^2$, $1.6\cdot10^4$ and $2.9\cdot10^4$, 
respectively.}\label{fig:popt}
\end{figure}

At a first glance it appears that the procedure might be slightly oversmoothing, 
which could be due to the fact that the posterior for $c$ is concentrated at 
too large values.
To get more insight into this issue we compare to posteriors computed with 
a fixed tuning parameter $c$, set at the ``oracle value'' which minimises 
the MSE of the posterior mean, which we determined numerically.  
The results are given in Figure \ref{fig:popt}. The posteriors have slightly 
better coverage than those in Figure \ref{fig:pbasic}.

\begin{figure}[H]
\begin{minipage}{0.32\textwidth}
\includegraphics[width = \textwidth]{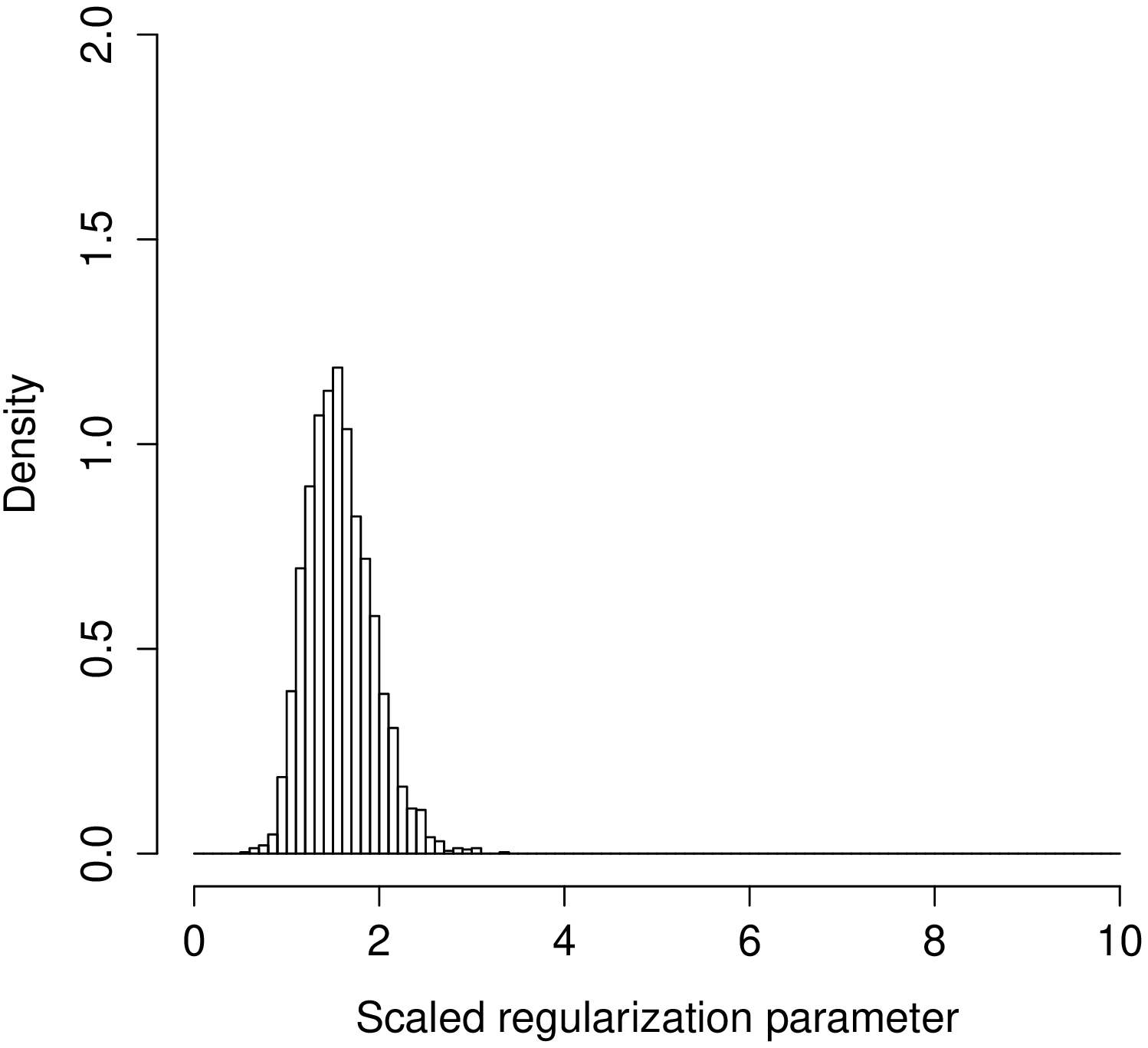}
\includegraphics[width = \textwidth]{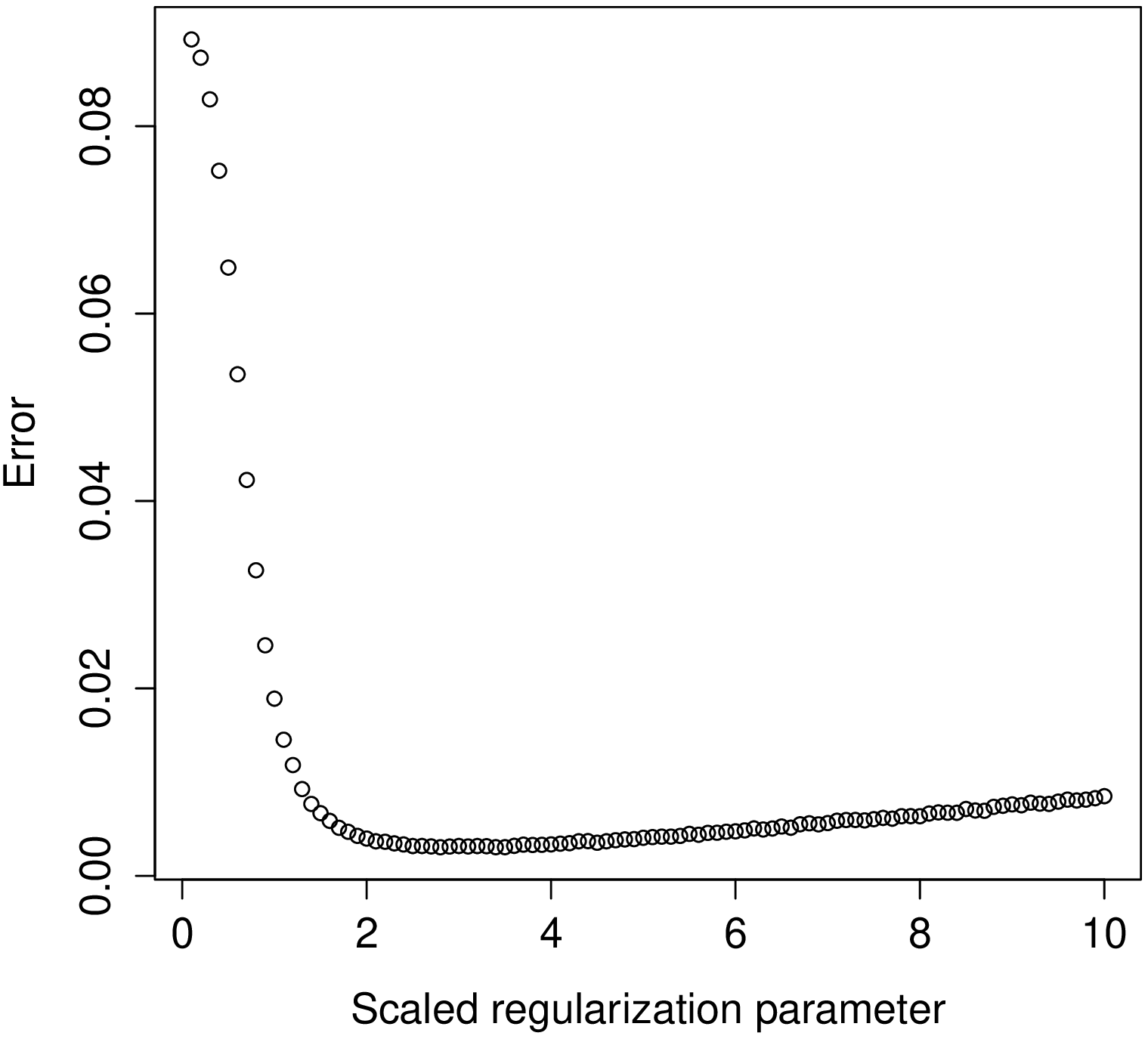}
\includegraphics[width = \textwidth]{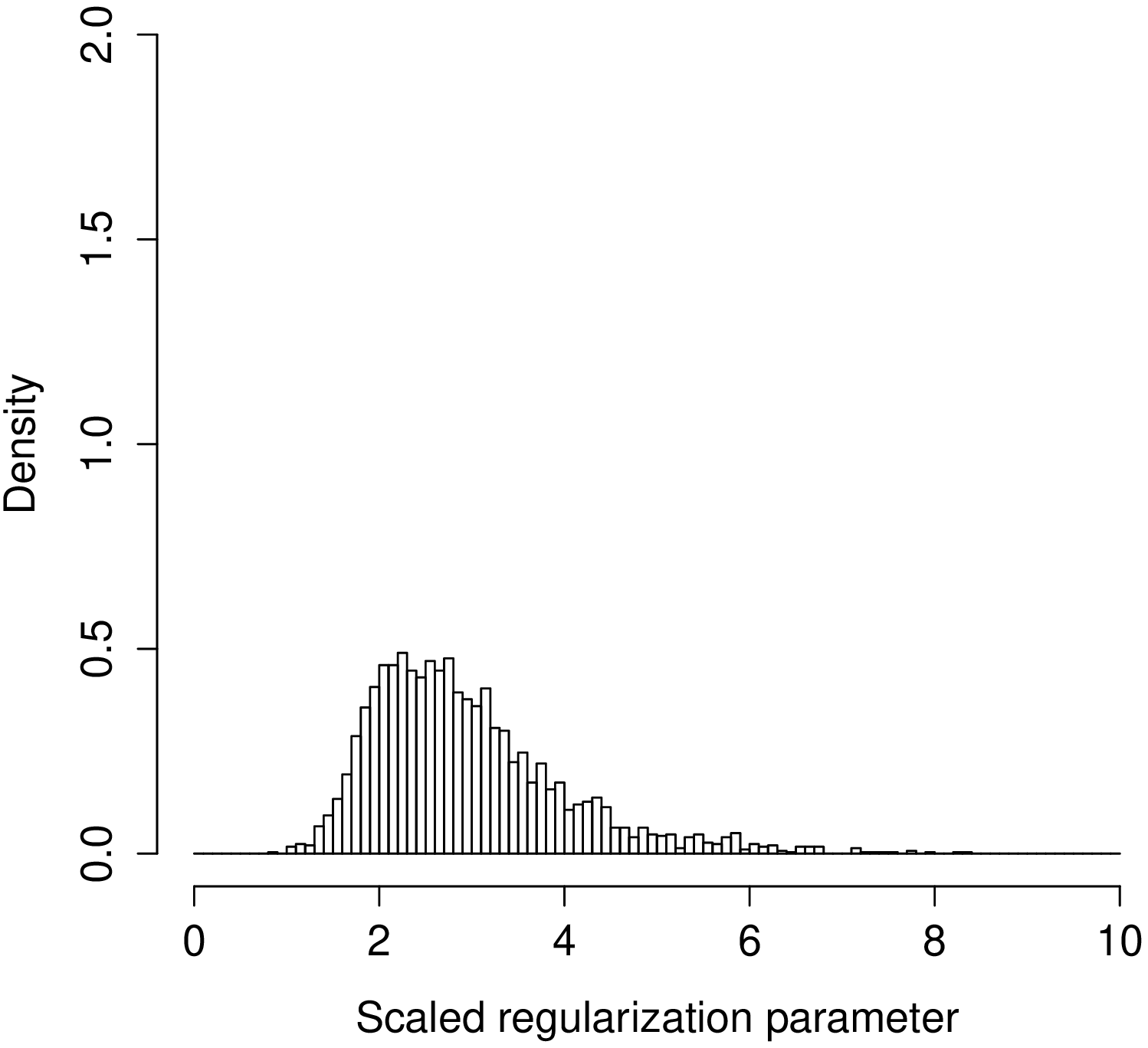}
\includegraphics[width = \textwidth]{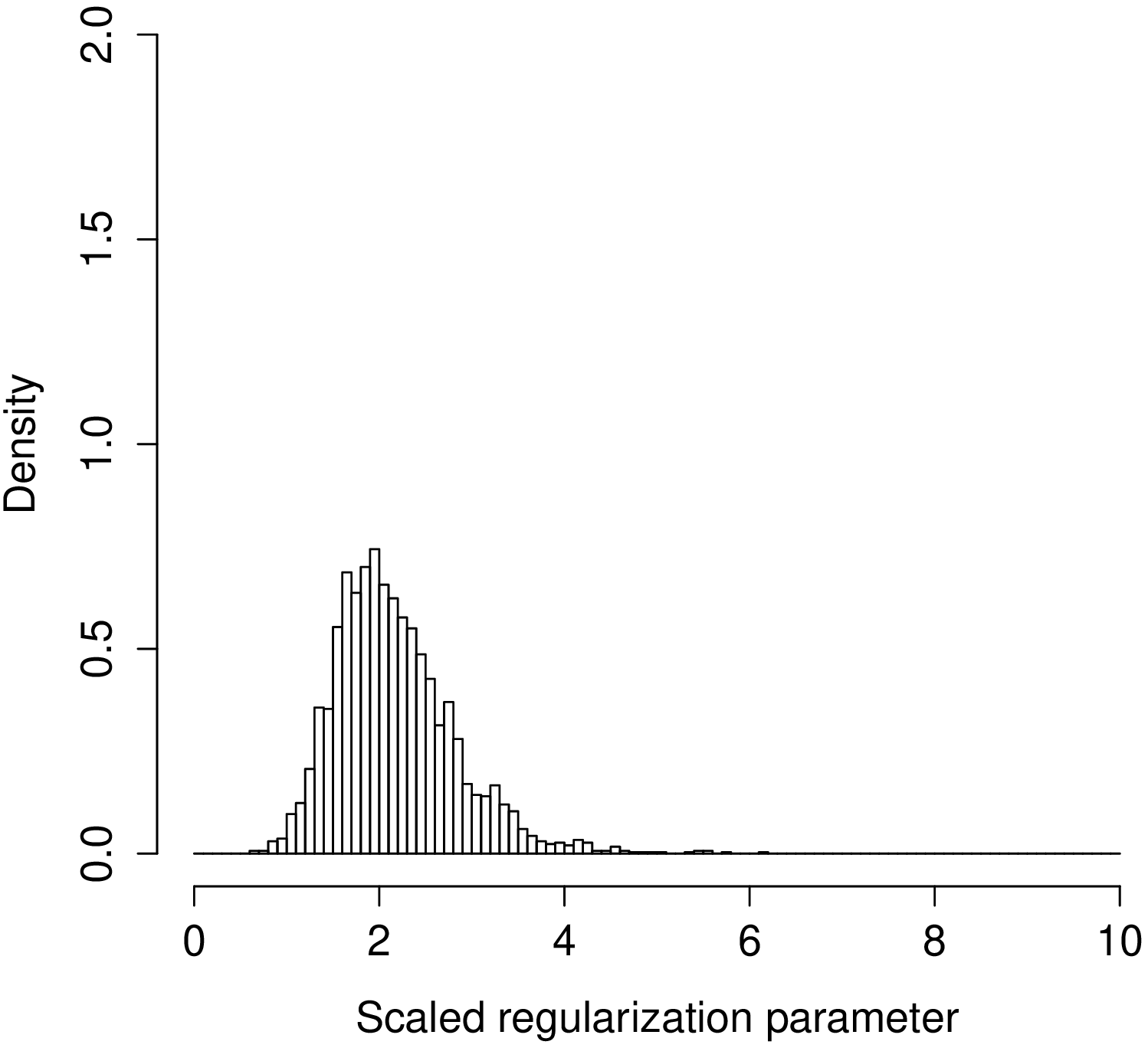}
\end{minipage}
\begin{minipage}{0.32\textwidth}
\includegraphics[width = \textwidth]{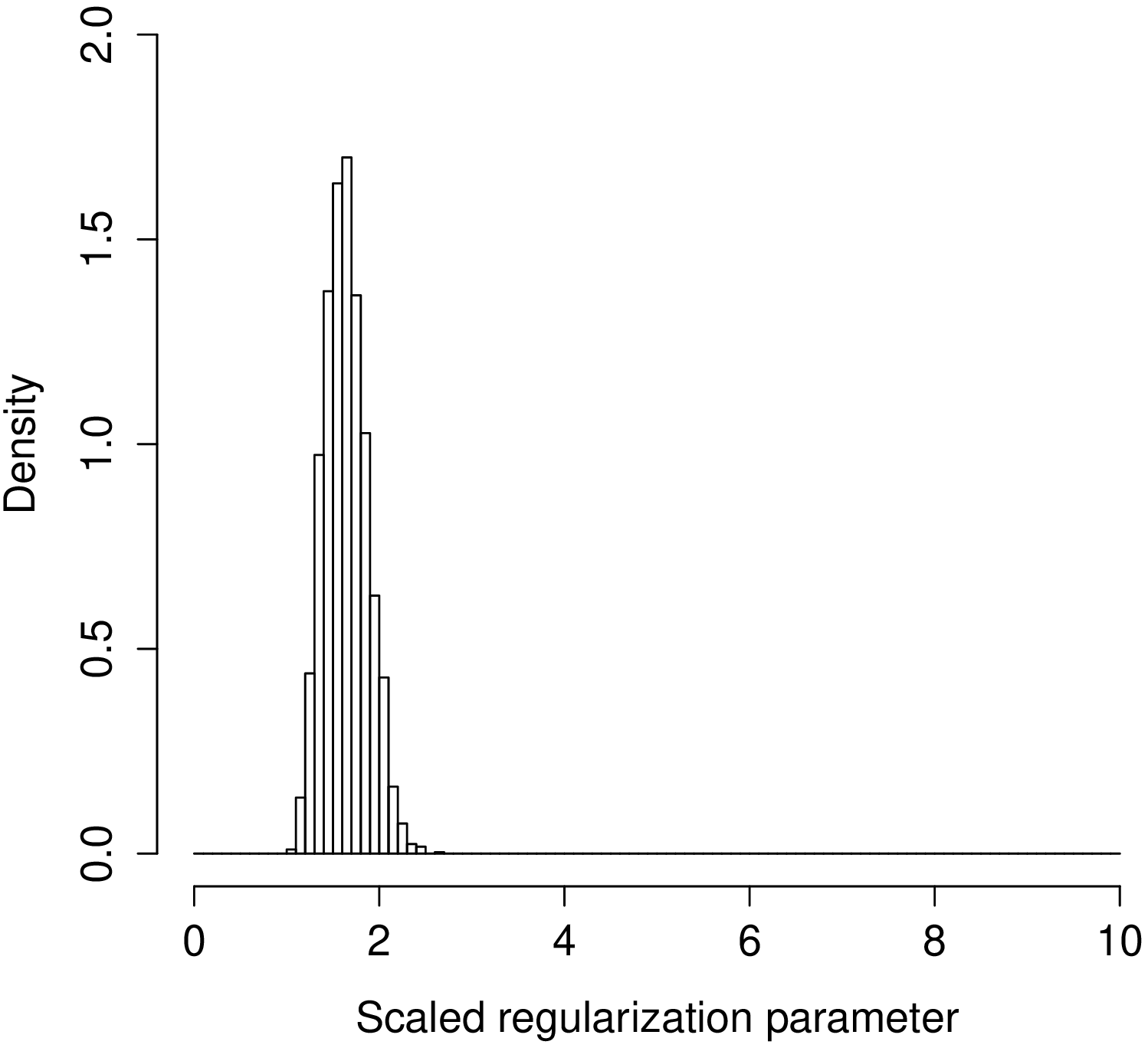}
\includegraphics[width = \textwidth]{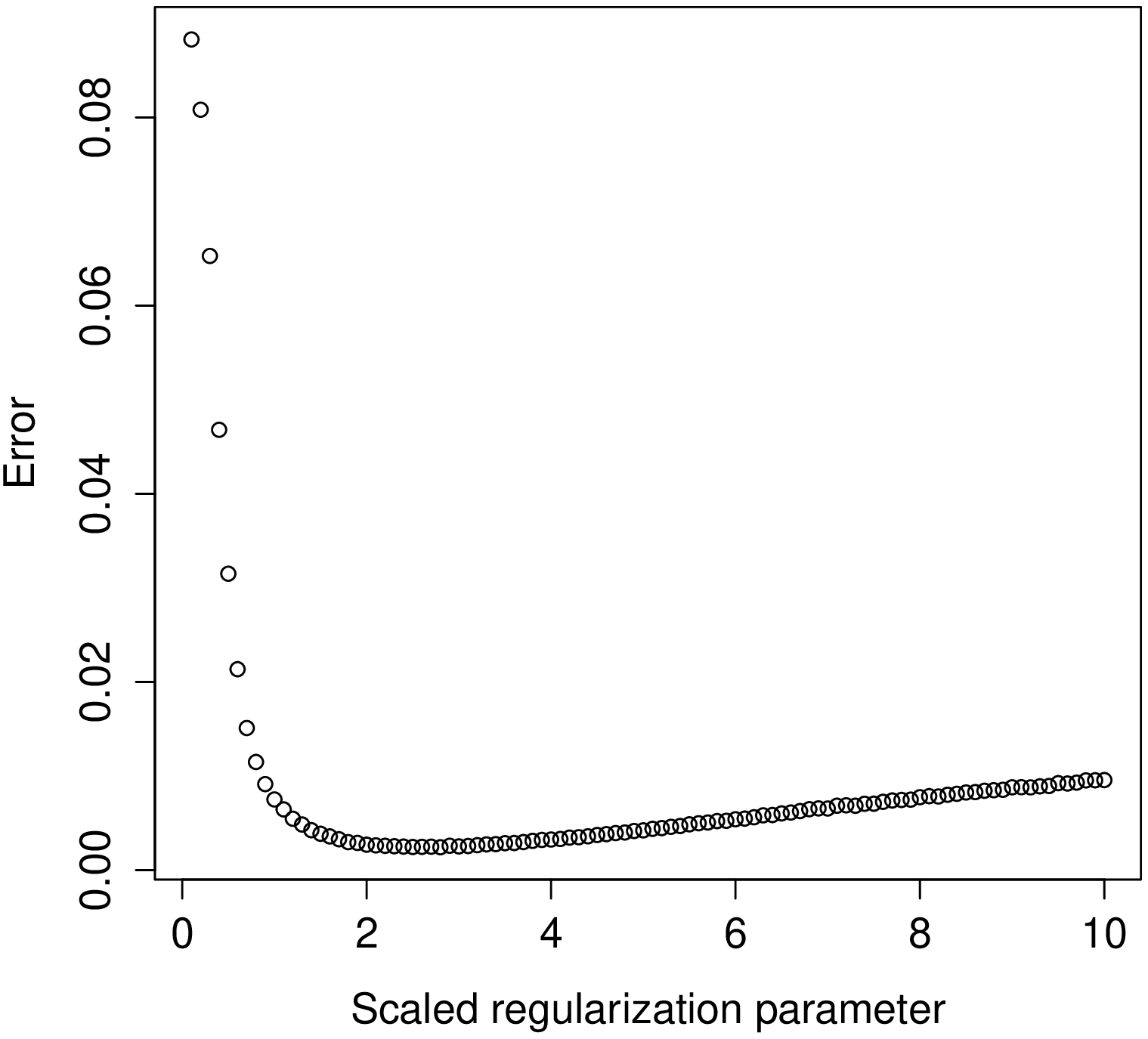}
\includegraphics[width = \textwidth]{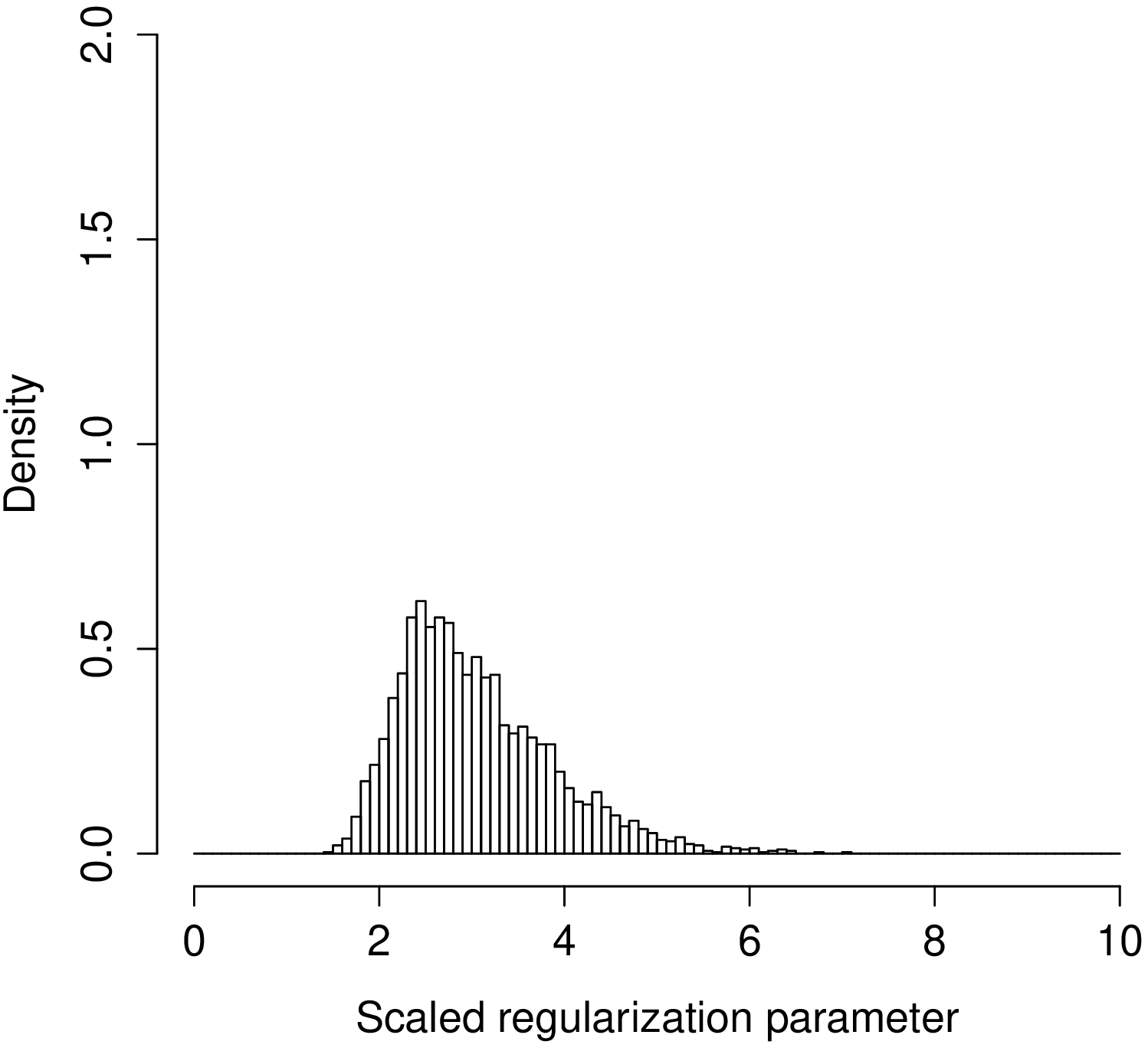}
\includegraphics[width = \textwidth]{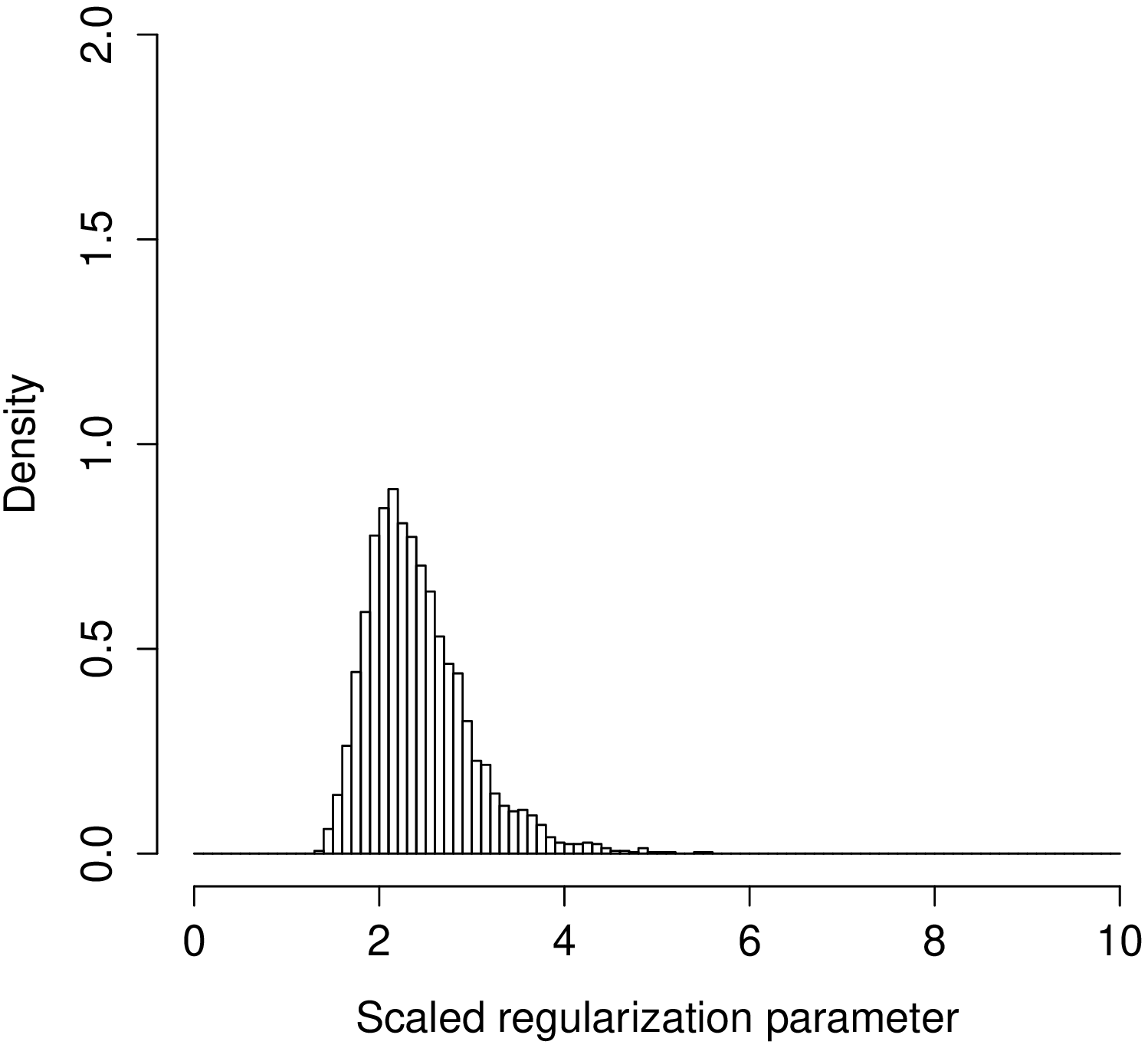}
\end{minipage}
\begin{minipage}{0.32\textwidth}
\includegraphics[width = \textwidth]{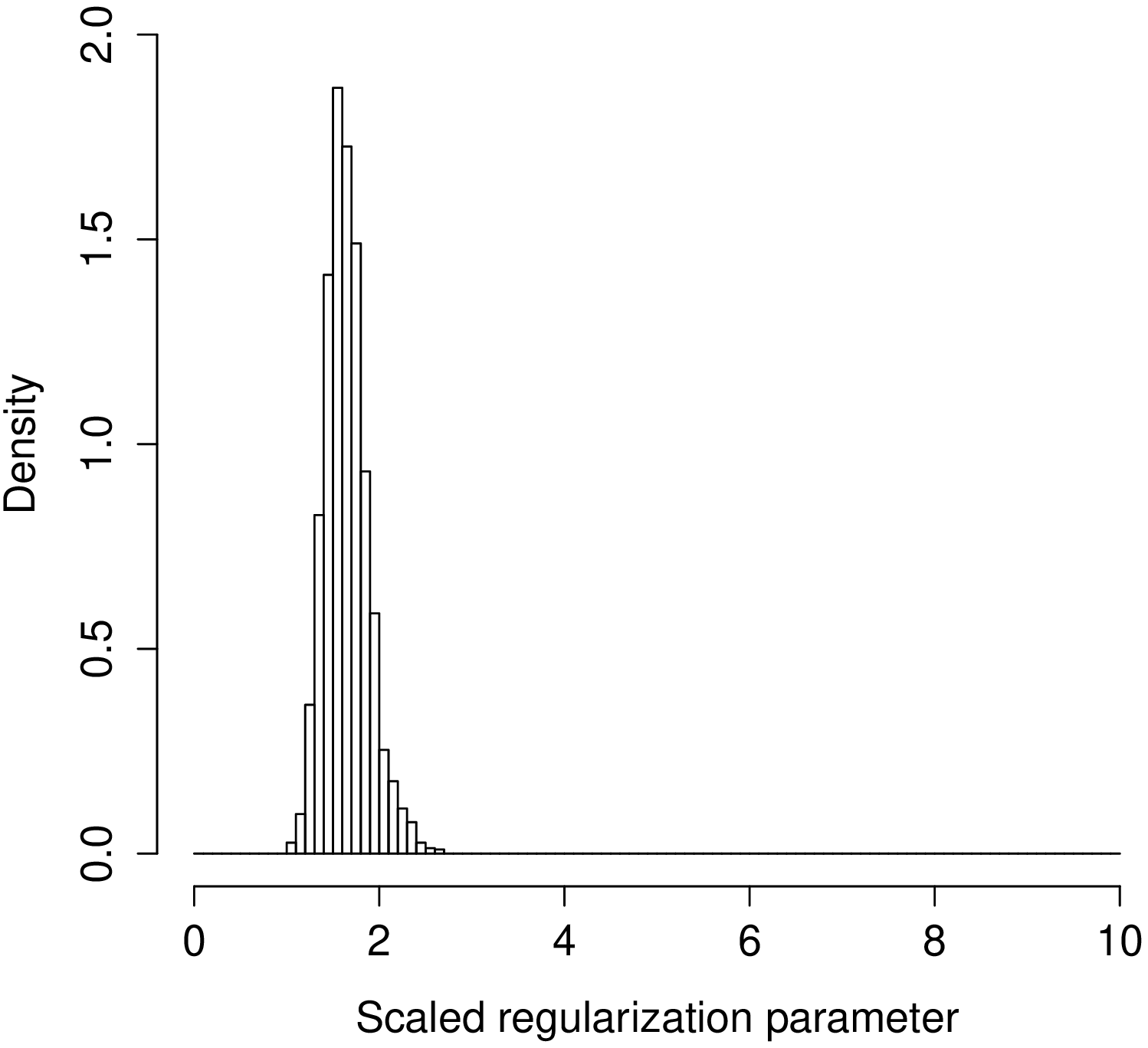}
\includegraphics[width = \textwidth]{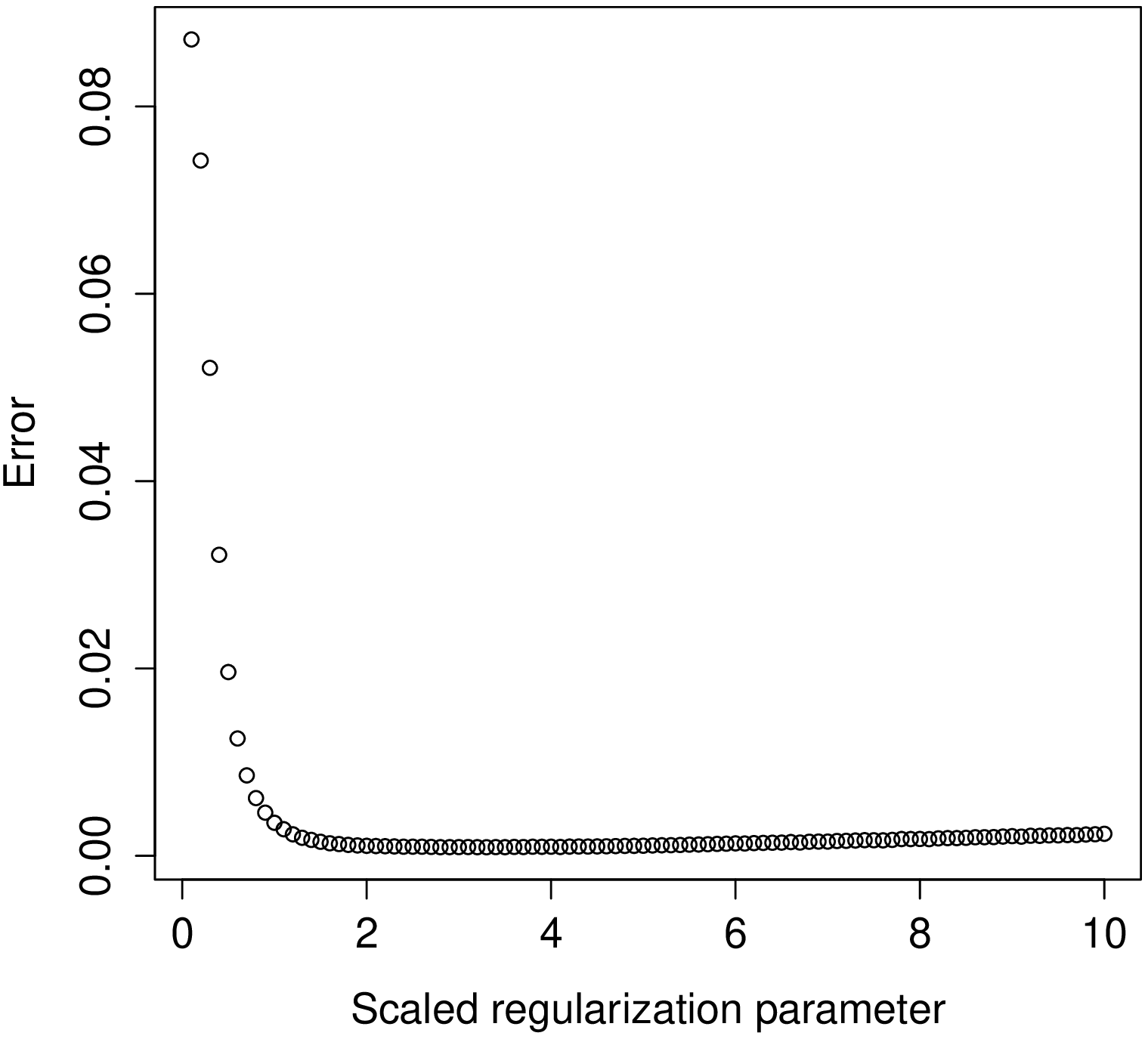}
\includegraphics[width = \textwidth]{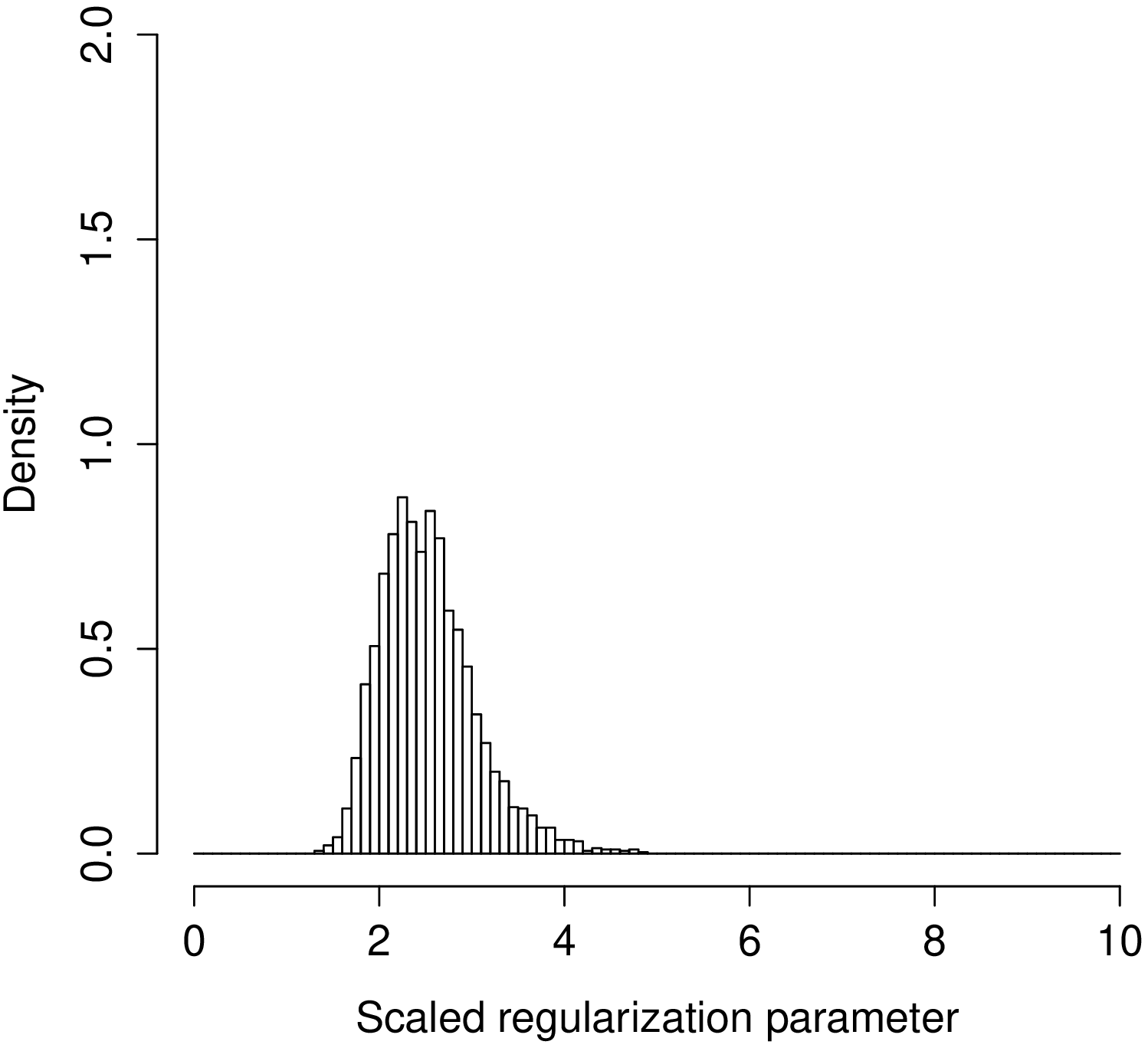}
\includegraphics[width = \textwidth]{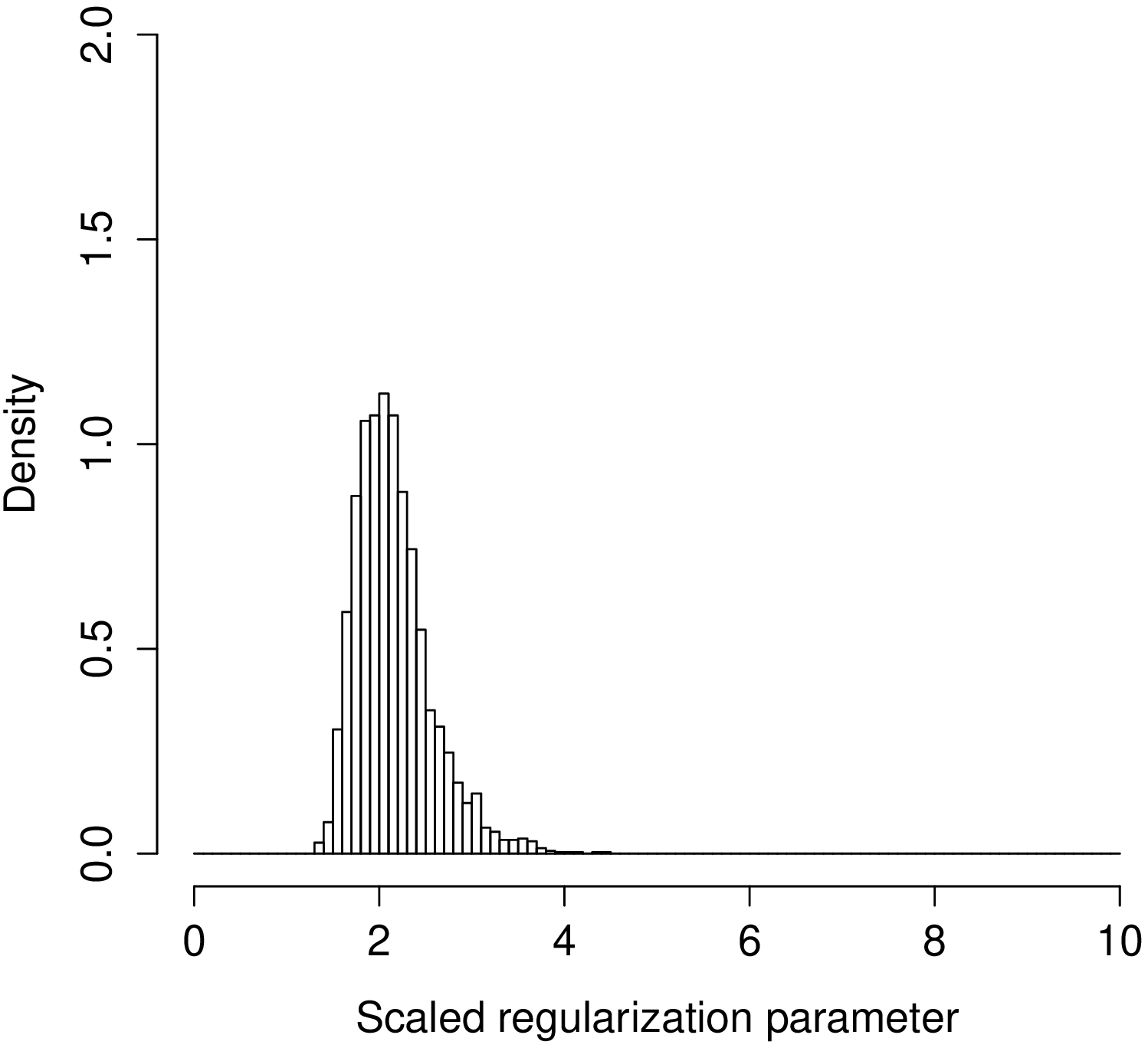}
\end{minipage}
\caption{Histograms of the scaled regularization parameter $n^{1-r/(2q)}c^{-r/(2q)}$ for $n = 100, 500, 1000$.  The top row: posterior corresponding to generalised gamma prior. 
Bottom rows: posterior corresponding to ordinary gamma prior with $a=b=0$ and $a=1$, $b=0$,
respectively. The second row shows the MSE of the posterior mean corresponding to fixed $c$ as function of $c$.}\label{fig:preg}
\end{figure}

Posterior histograms of the tuning parameter show if we use the 
generalised gamma prior for $c$, the posterior 
indeed favours too high values of  $c$, compared to the oracle choice. This 
results in the oversmoothing we observe in  Figure \ref{fig:pbasic}.
See the first two rows of Figure \ref{fig:preg}.

\begin{figure}[H]
\centering
\begin{minipage}{0.6\textwidth}
\includegraphics[width=\textwidth]{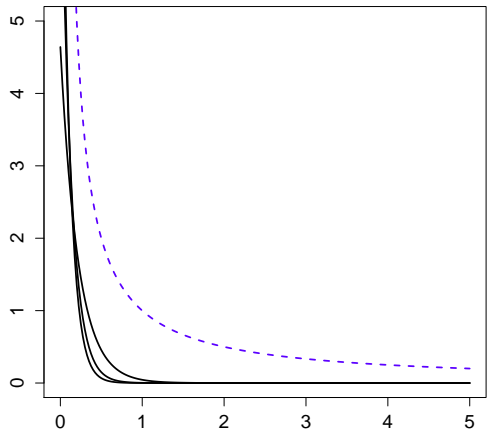}
\end{minipage}
\caption{Densities of the generalised gamma prior for $n=100, 500, 1000$ in black. Blue 
dashed line the ordinary gamma prior with $a=b=0$.}\label{fig: dens}
\end{figure}

When instead of the theoretically optimal generalised gamma prior on $c$ 
we use the ordinary gamma prior, we can use the hyper parameters $a$ and $b$ 
to ensure that the posterior for $c$ 
assigns more  mass close to  the oracle tuning  parameter. 
In practice, we do not know the true underlying function, so it is natural to 
spread the prior mass as much as possible. 
We can for example choose $a = b = 0$, corresponding to an improper prior $p(c) \propto 1/c$ (as in \cite{choudhuri2007}), or $a = 1$ and $b = 0$ such that $p(c) \propto 1$.
In Figure \ref{fig: dens} we plot the ordinary gamma prior density corresponding to $a=b=0$
(blue dashed line) and the generalised gamma prior densities for various $n$ (black lines). 
Since the ordinary gamma assigns more mass to smaller values of $c$, we might 
hope that if we use that prior on $c$, we get a posterior closer to the oracle 
and hence reduce the oversmoothing problem.

\begin{figure}[H]
\begin{minipage}{0.32\textwidth}
\includegraphics[width = \textwidth]{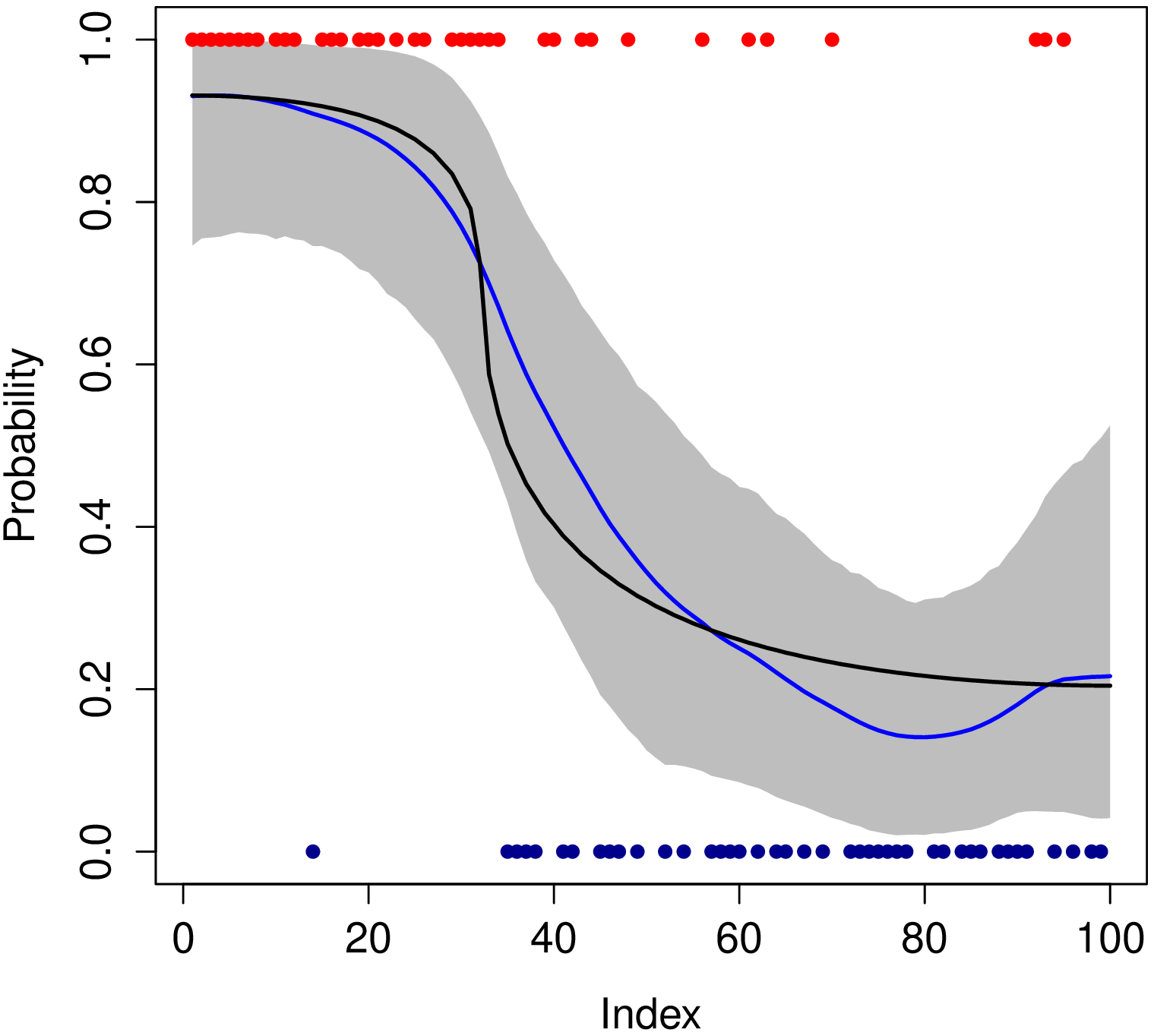}
\end{minipage}
\begin{minipage}{0.32\textwidth}
\includegraphics[width = \textwidth]{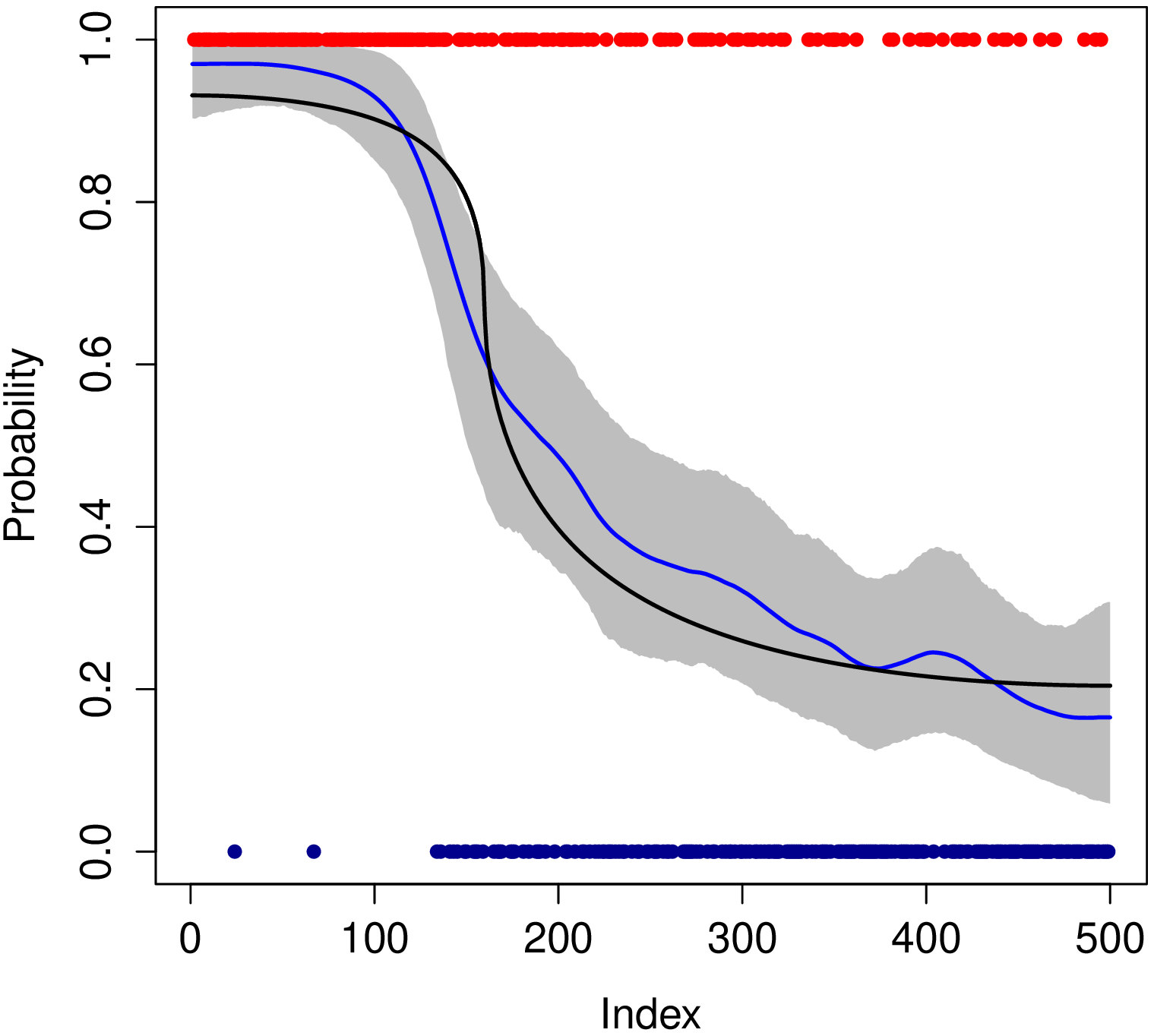}
\end{minipage}
\begin{minipage}{0.32\textwidth}
\includegraphics[width = \textwidth]{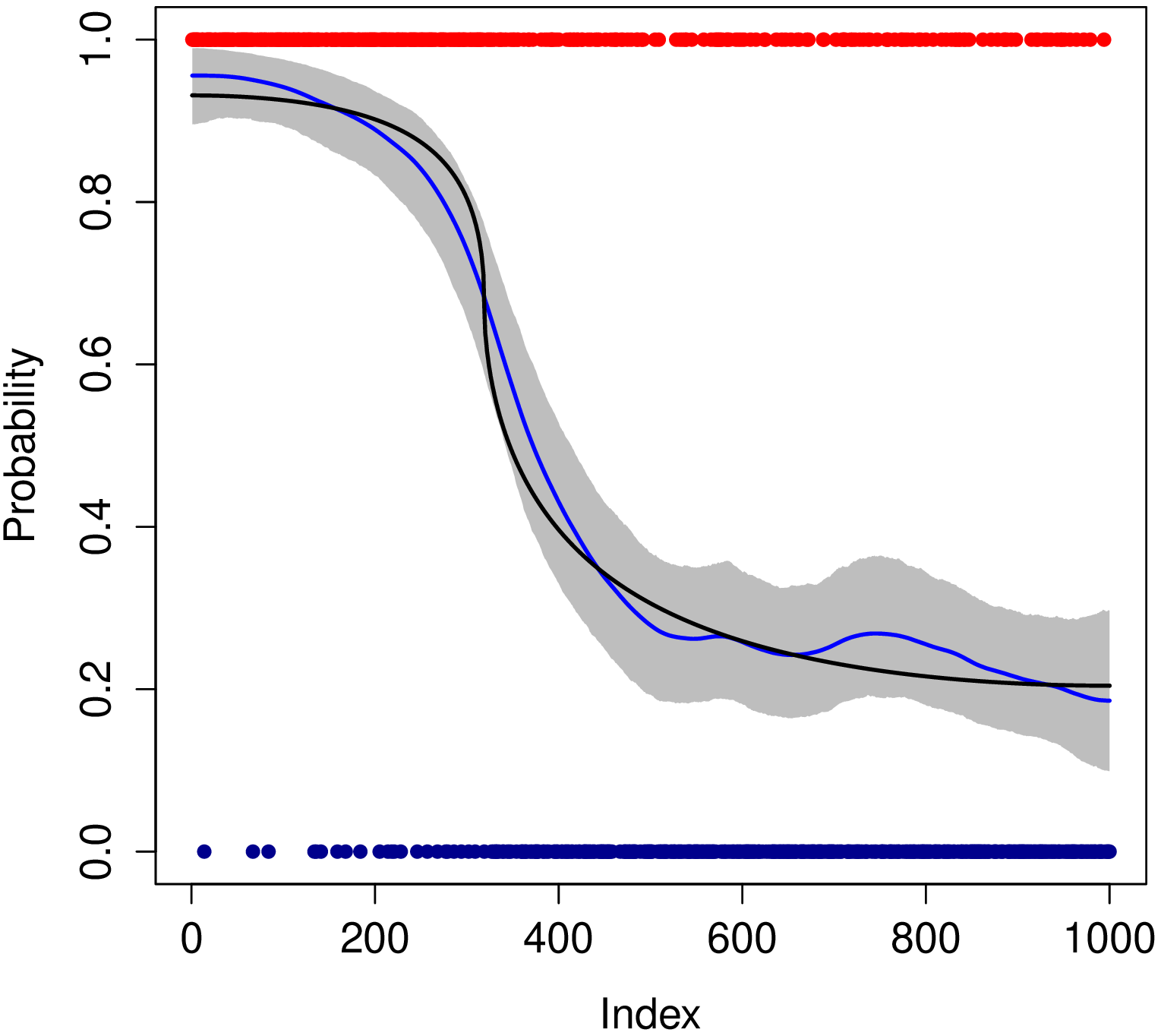}
\end{minipage}
\caption{Posteriors for the soft label function for  $n = 100, 500, 1000$.
Prior on $c$ is the ordinary gamma with $a=b=0$.}\label{fig:preg2}
\end{figure}

Figure \ref{fig:preg2} visualises  the posteriors that we get  for the soft label function 
when using the ordinary gamma prior with $a=b=0$. We see that indeed 
we get better posterior coverage than in Figure \ref{fig:pbasic}. 
The third row in Figure \ref{fig:preg} confirms that when using the ordinary gamma
prior on $c$, the posterior for $c$ puts more mass around the optimal value.

%\subsubsection{Impact of geometry condition number}
%
%
%
%
%
%
%
%\begin{figure}[H]
%\begin{minipage}{0.33\textwidth}
%\includegraphics[width = \textwidth]{Path100q6par10.eps}
%\includegraphics[width = \textwidth]{Path100q6par10-0-0.eps}
%\end{minipage}
%\begin{minipage}{0.33\textwidth}
%\includegraphics[width = \textwidth]{Path500q6par10.eps}
%\includegraphics[width = \textwidth]{Path500q6par10-0-0.eps}
%\end{minipage}
%\begin{minipage}{0.33\textwidth}
%\includegraphics[width = \textwidth]{Path1000q6par10.eps}
%\includegraphics[width = \textwidth]{Path1000q6par10-0-0.eps}
%\end{minipage}
%\caption{Path graph, $r = 10$, $\alpha = \beta$, $n = 100, 500, 1000$. Top is the theoretically optimal prior, bottom is the computationally attractive prior with $a=b=0$.}
%\end{figure}
%
%In case the geometry condition number is too large, the power of the Laplacian matrix $q=\alpha+r/2$ increases and therefore smooth function have more prior mass. In our  example we chose $r=10$ instead of the true $r=1$. This results in a smoother posterior estimate.

\subsubsection{Impact of prior smoothness}

In the paper \cite{kirichenko2017} it was 
suggested to take the power of the Laplacian equal to $q = \alpha + r/2$, 
where $r$ is the number appearing in the geometry condition \eqref{eq: geom}
and $\alpha$ is a tuning parameter that quantifies the smoothness 
of the prior in some sense. It was proved that when combining this 
with the generalised gamma prior \eqref{eq: c1} on $c$, we get 
good convergence rates if the Sobolev-type smoothness of the true soft-label 
function is less than $q$. This might suggest that it is advantageous to set $q$
high, since then the theory says that we get good rates across a large range of 
regularities of the true function. On the other hand, setting $q$ higher
means we favour smooth functions more. This could potentially lead to 
oversmoothing and hence to poor posterior coverage. In this section 
we investigate this issue numerically. 

\begin{figure}[H]
\begin{minipage}{0.32\textwidth}
\includegraphics[width = \textwidth]{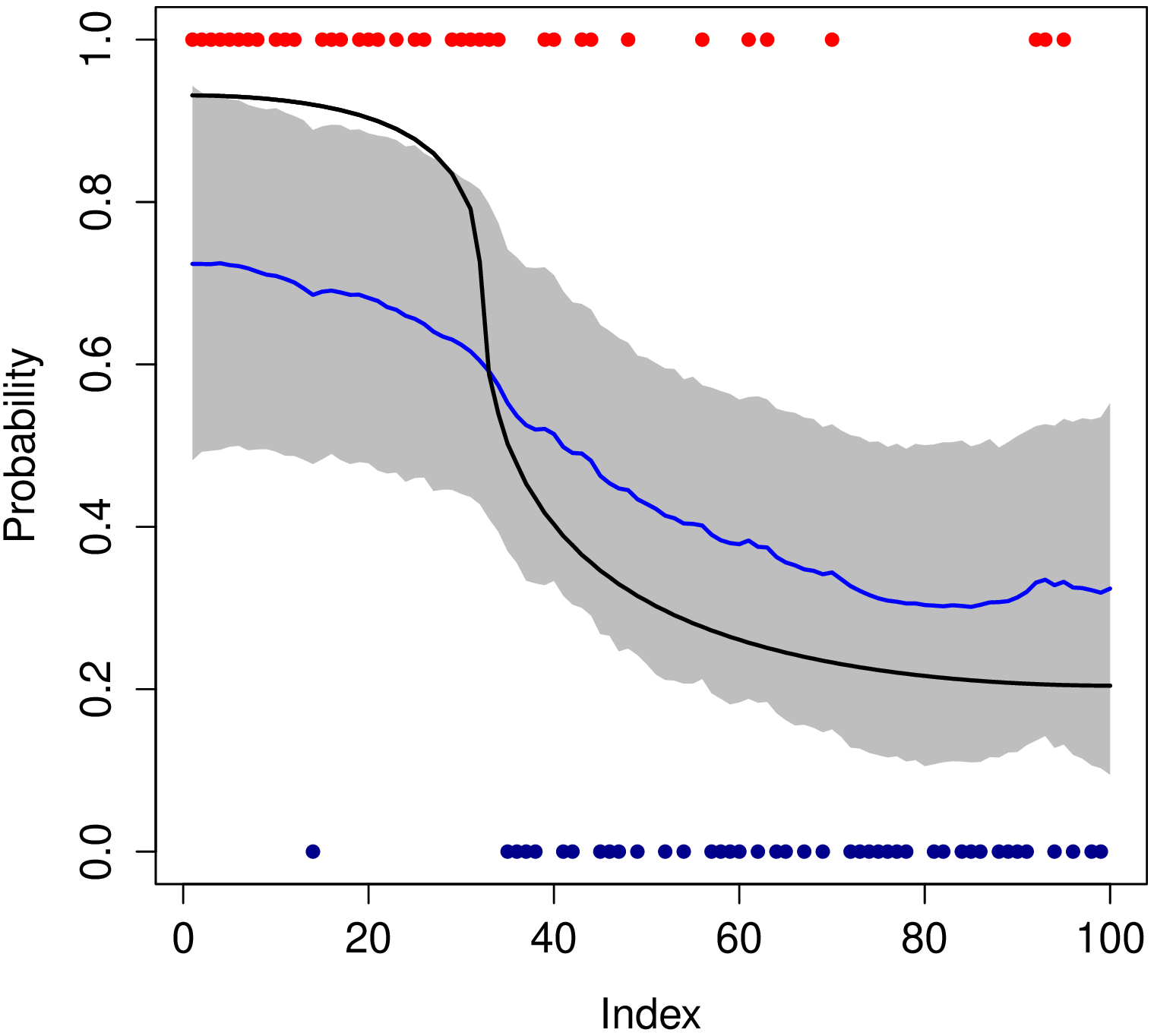}
\includegraphics[width = \textwidth]{Path100q1_5par1.eps}
\includegraphics[width = \textwidth]{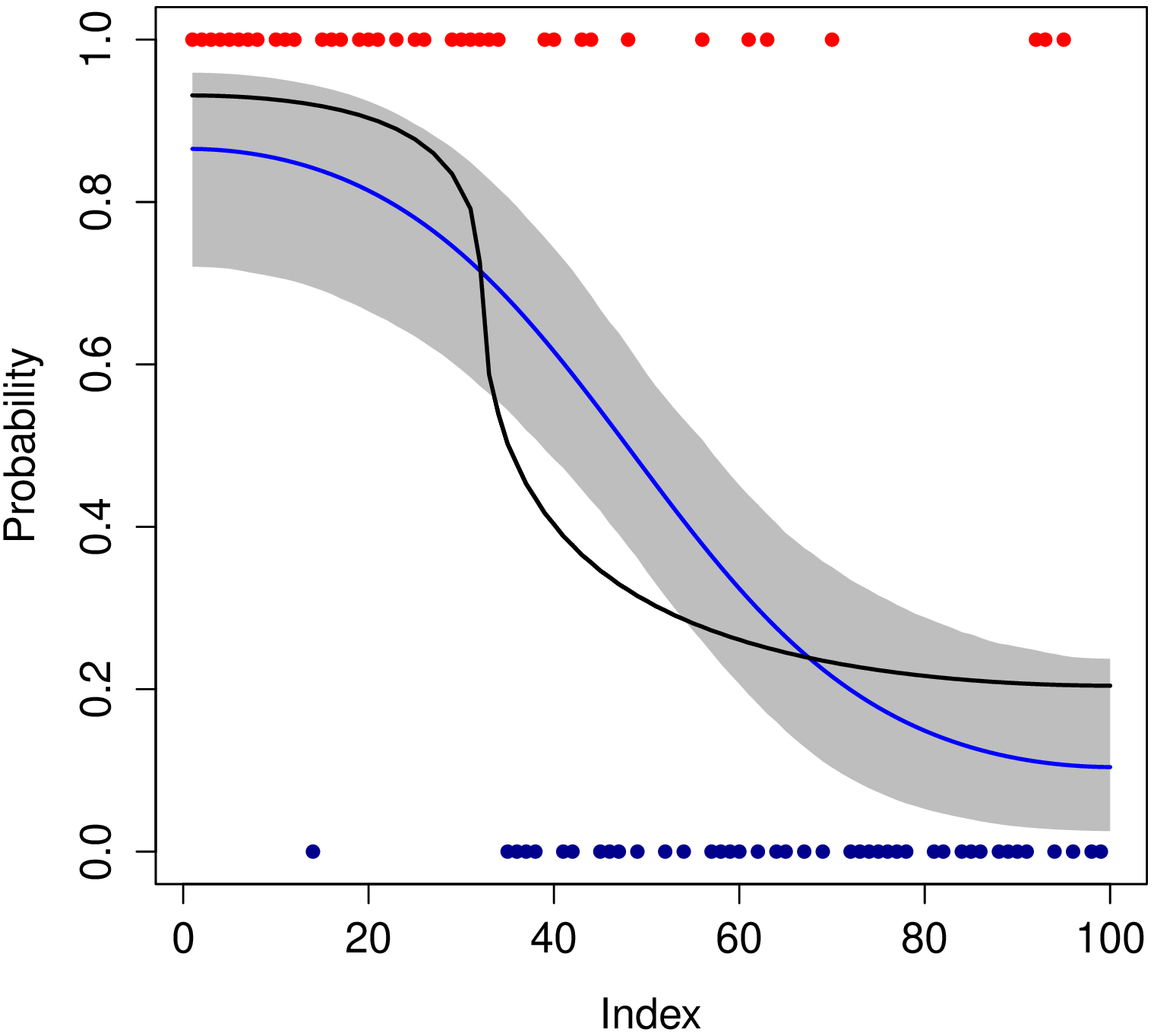}
\end{minipage}
\begin{minipage}{0.32\textwidth}
\includegraphics[width = \textwidth]{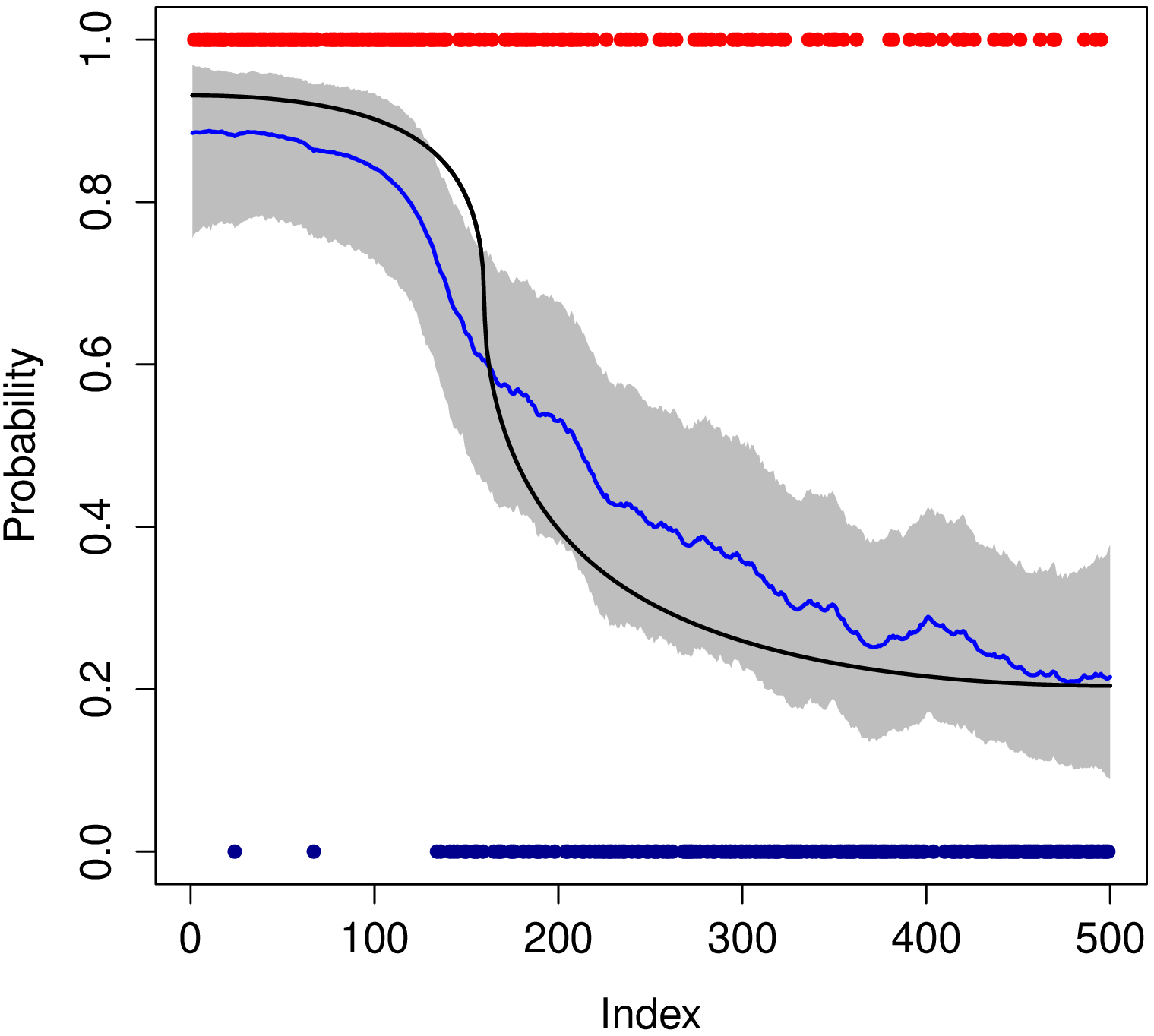}
\includegraphics[width = \textwidth]{Path500q1_5par1.eps}
\includegraphics[width = \textwidth]{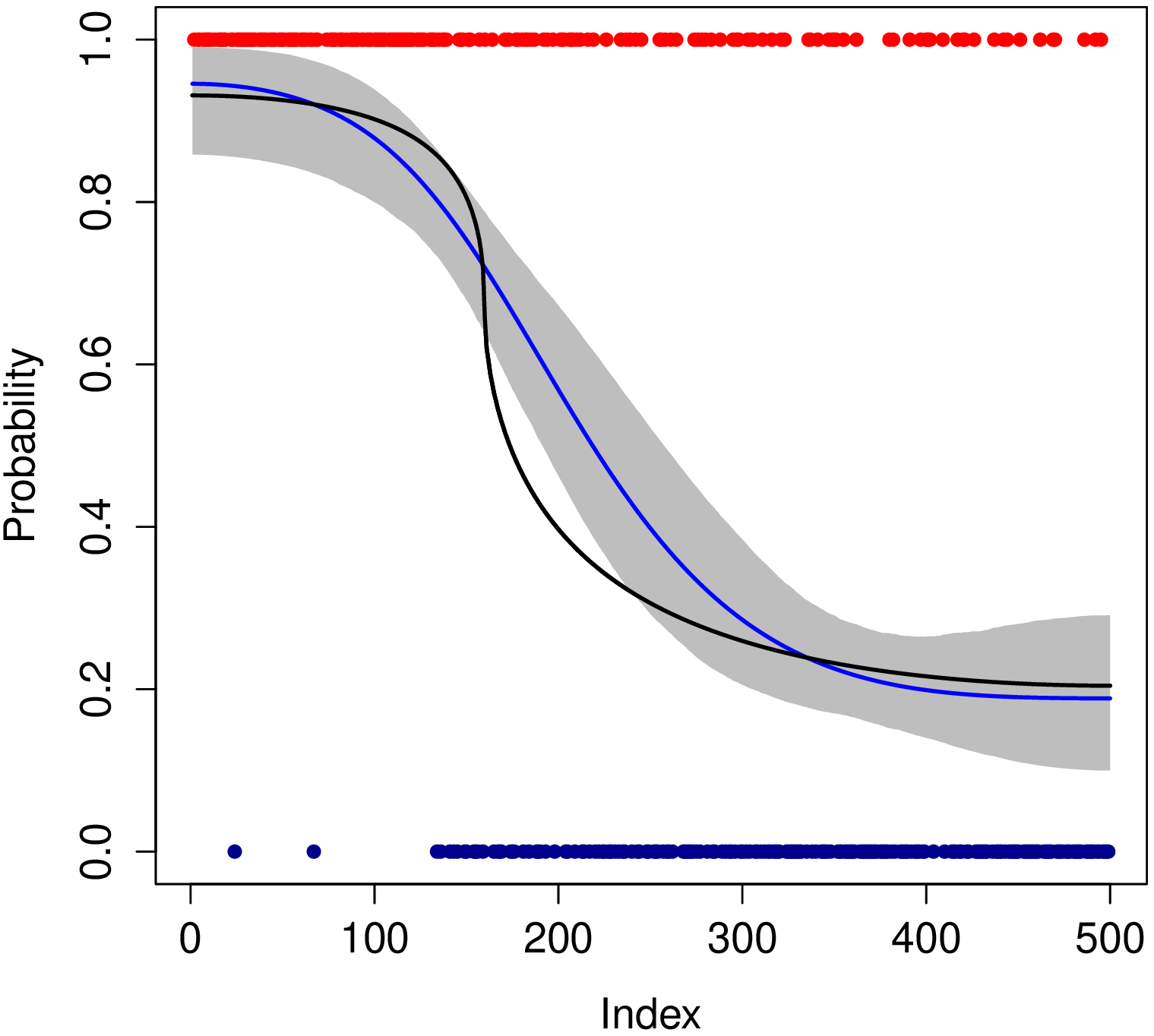}
\end{minipage}
\begin{minipage}{0.32\textwidth}
\includegraphics[width = \textwidth]{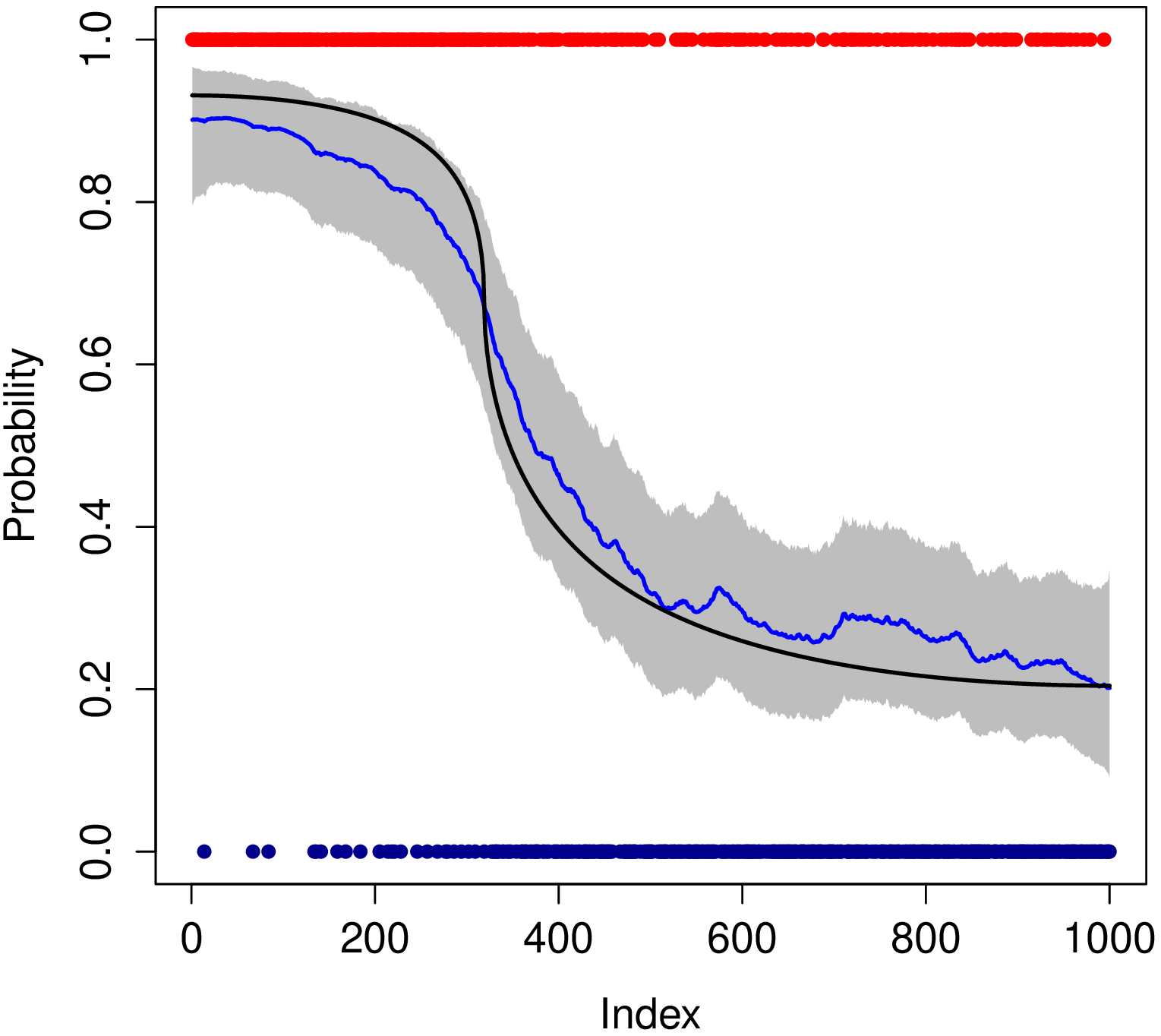}
\includegraphics[width = \textwidth]{Path1000q1_5par1.eps}
\includegraphics[width = \textwidth]{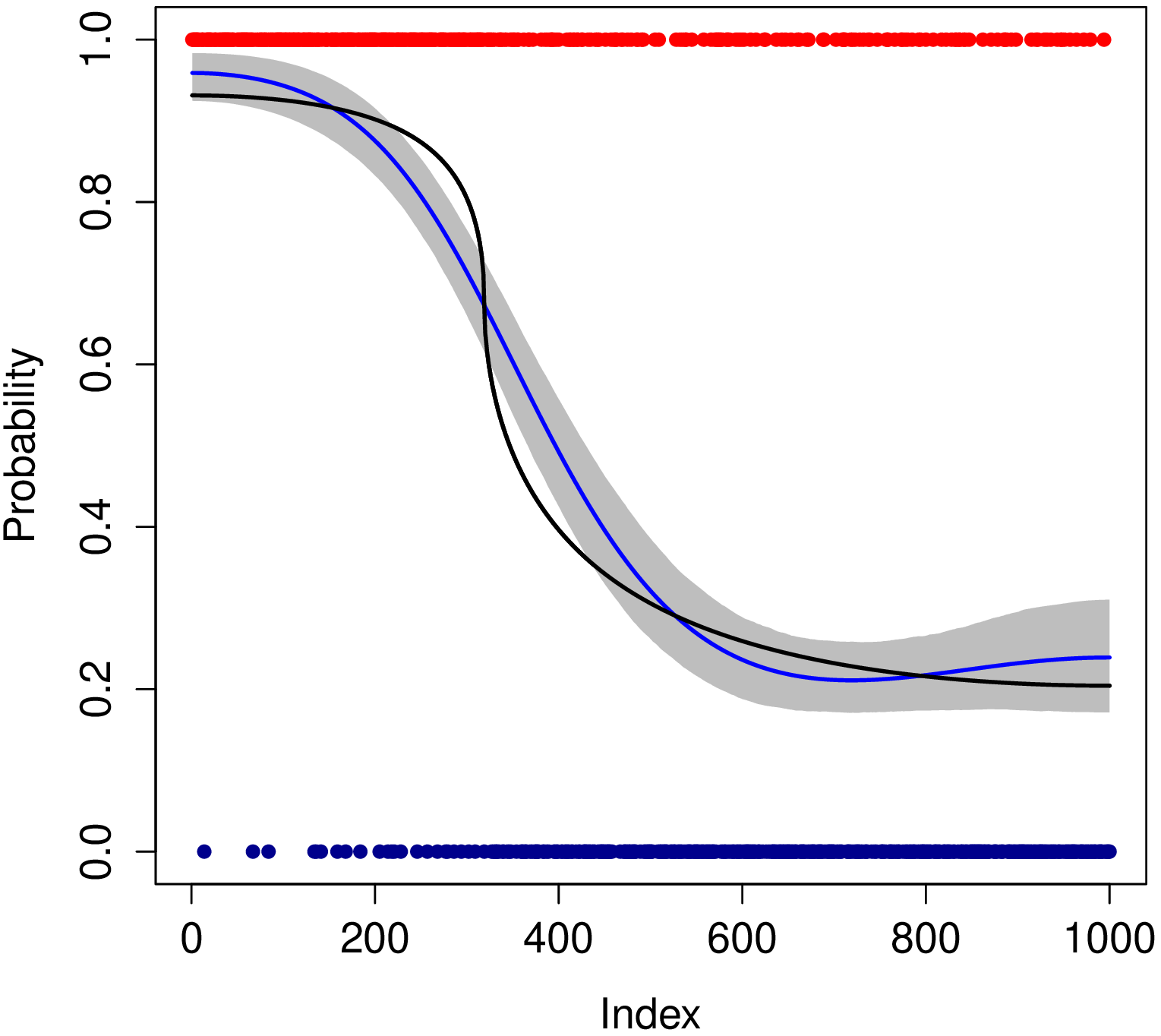}
\end{minipage}
\caption{Posteriors for $\ell$ when using the generalised gamma prior on $c$. From 
left two right we have $n = 100, 500, 1000$. From top to bottom we have
$q = \alpha + 1/2$, with $\alpha = 0.2, 1, 5$.}\label{fig: q1}
\end{figure}

In Figure \ref{fig: q1} we use the generalised gamma prior on $c$. 
We plot the posterior for $\ell$, varying $n$ from left to right  
and $q$ from top to bottom. In the top row $q = 0.2 + r/2$. Since this 
is less  than $\beta = 1$, the theory suggests that we are undersmoothing 
too much and will  get sub-optimal convergence rates. The figure seems to confirms this. 
In the middle row we have $q = 1+ r/2$. This means the prior smoothness matches
the true smoothness, which is asymptotically a good choice according to the theory.
This is the same picture as in Figure \ref{fig:pbasic}.
In the bottom row $q = 5+r/2$. In this case the prior smoothness is larger 
than the true smoothness $\beta=1$. However, the theory says that we should still 
get a good convergence rates, because to compensate for the smoothness mismatch
the posterior for $c$ will automatically charge smaller values of $c$ more.
In the simulations we see that the result is indeed not dramatic, but that 
the procedure is in actual fact oversmoothing somewhat, resulting in worse posterior coverage.

\begin{figure}[H]
\begin{minipage}{0.32\textwidth}
\includegraphics[width = \textwidth]{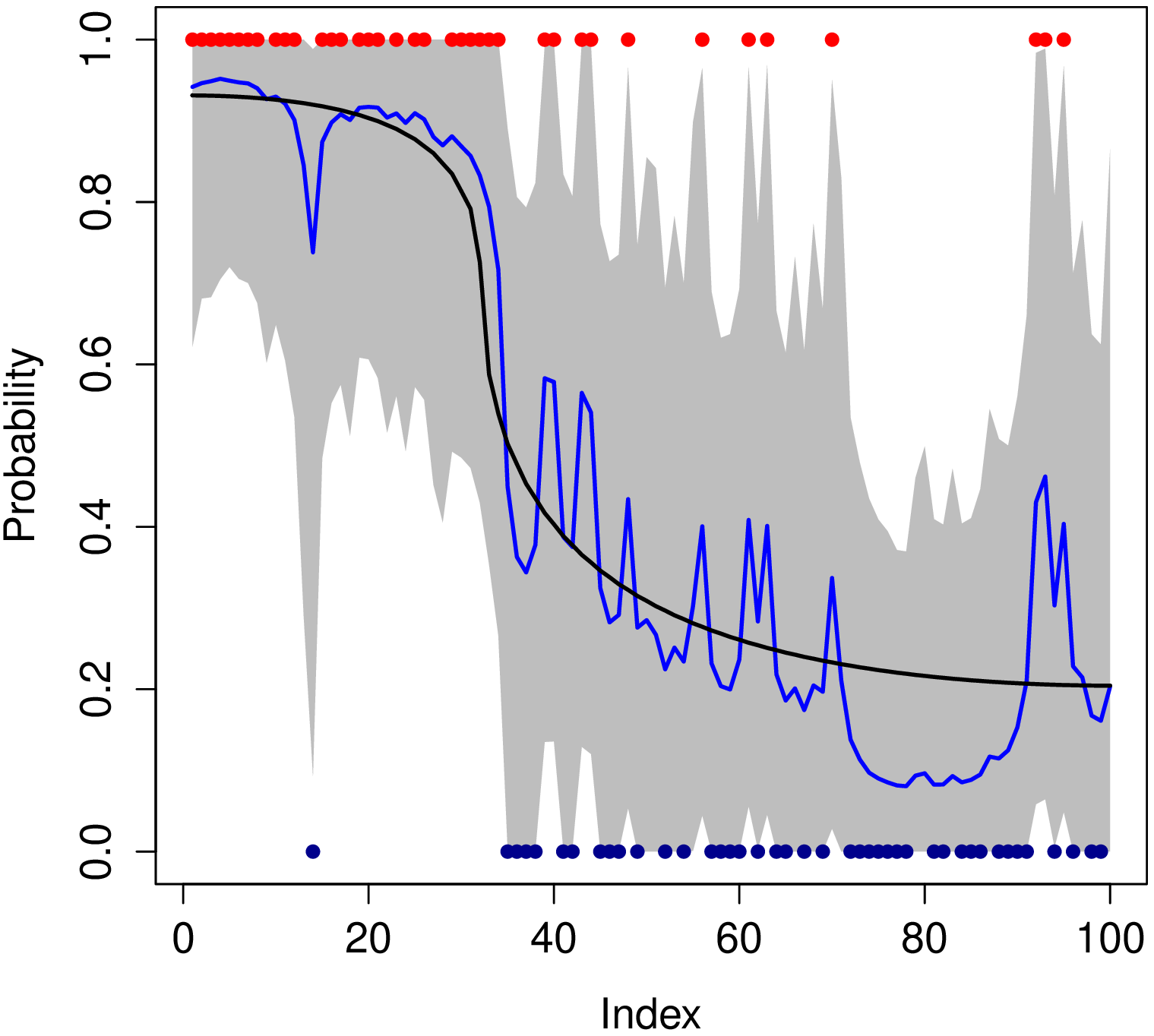}
\includegraphics[width = \textwidth]{Path100q1_5par1-0-0.eps}
\includegraphics[width = \textwidth]{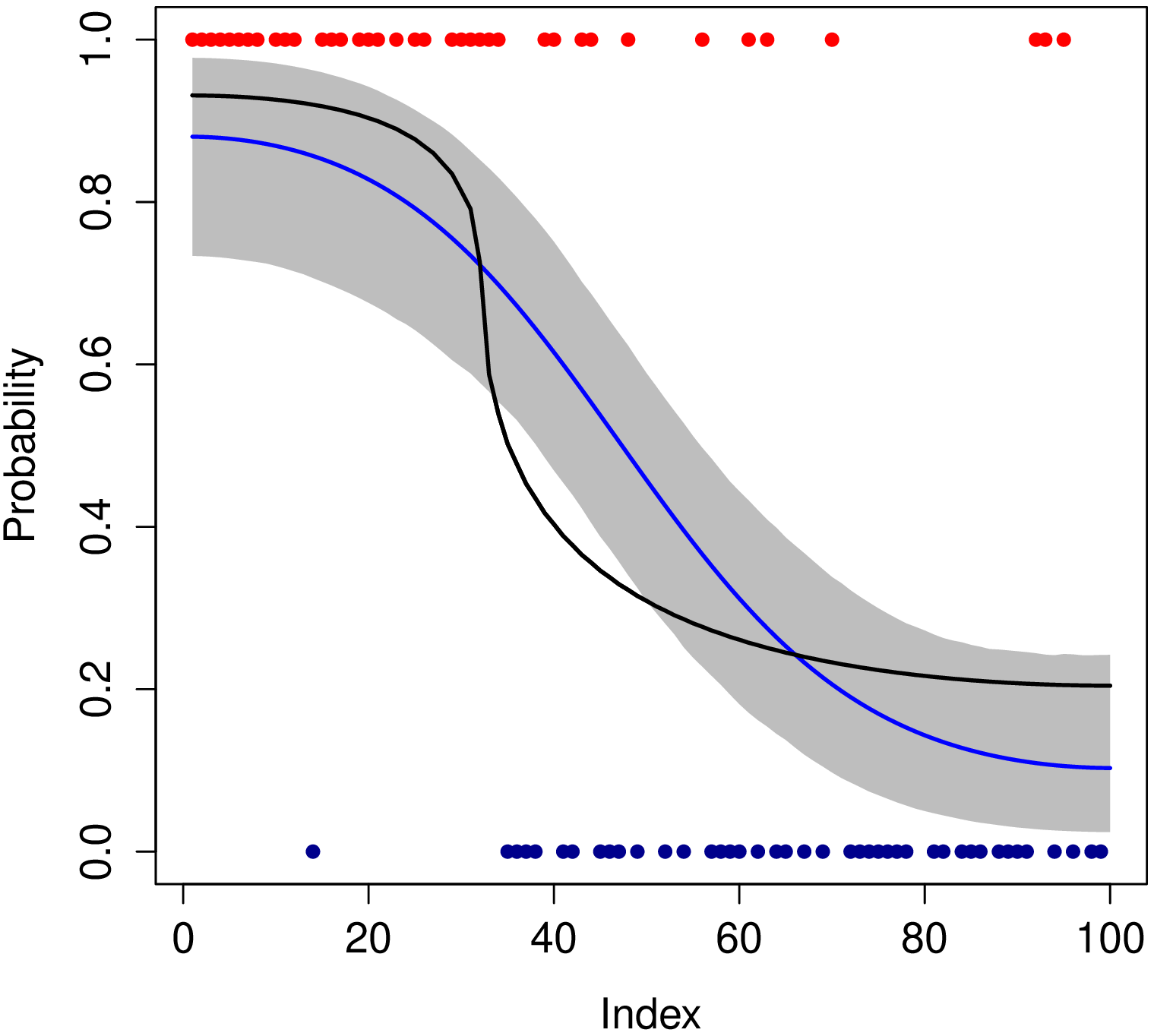}
\end{minipage}
\begin{minipage}{0.32\textwidth}
\includegraphics[width = \textwidth]{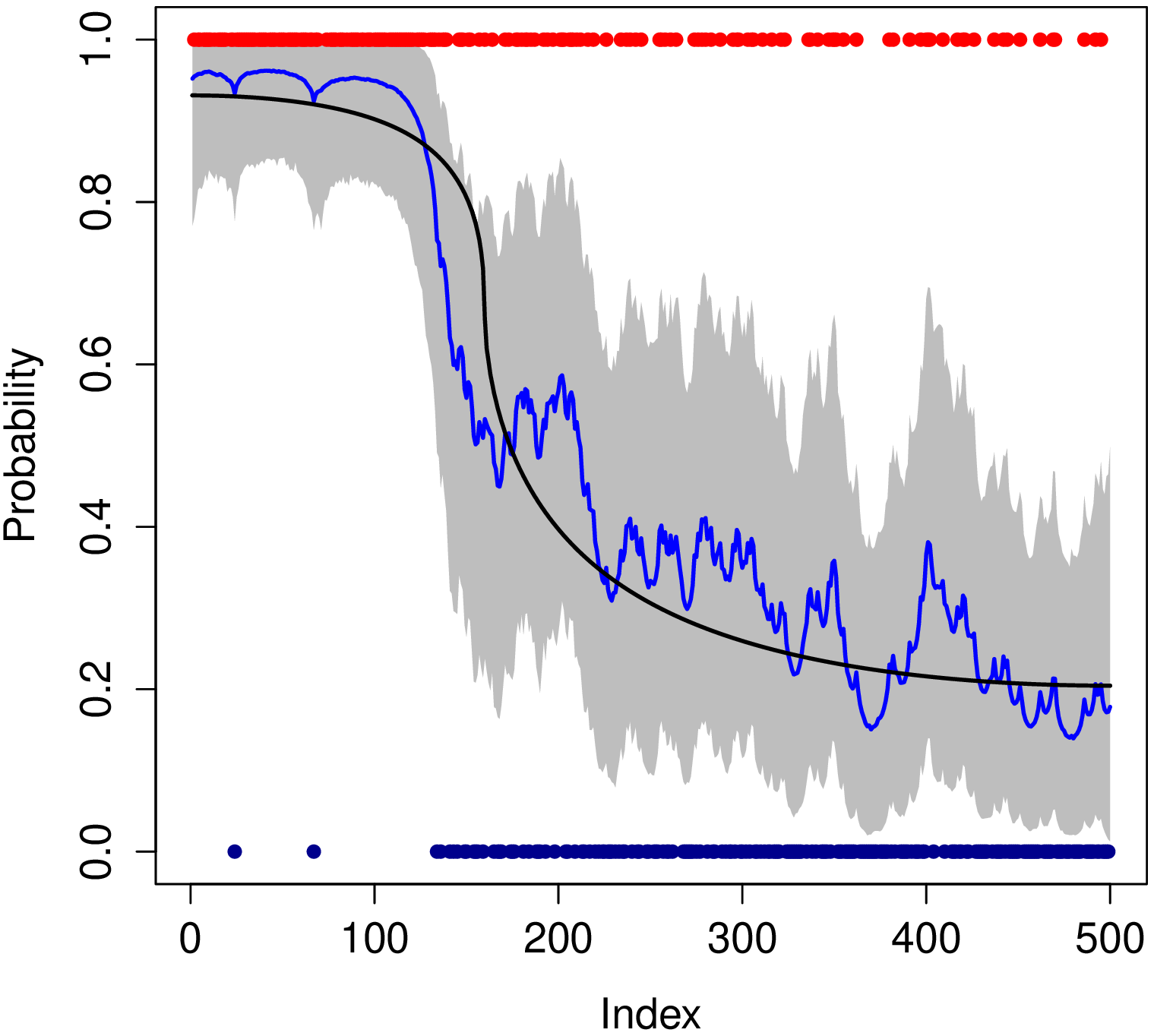}
\includegraphics[width = \textwidth]{Path500q1_5par1-0-0.eps}
\includegraphics[width = \textwidth]{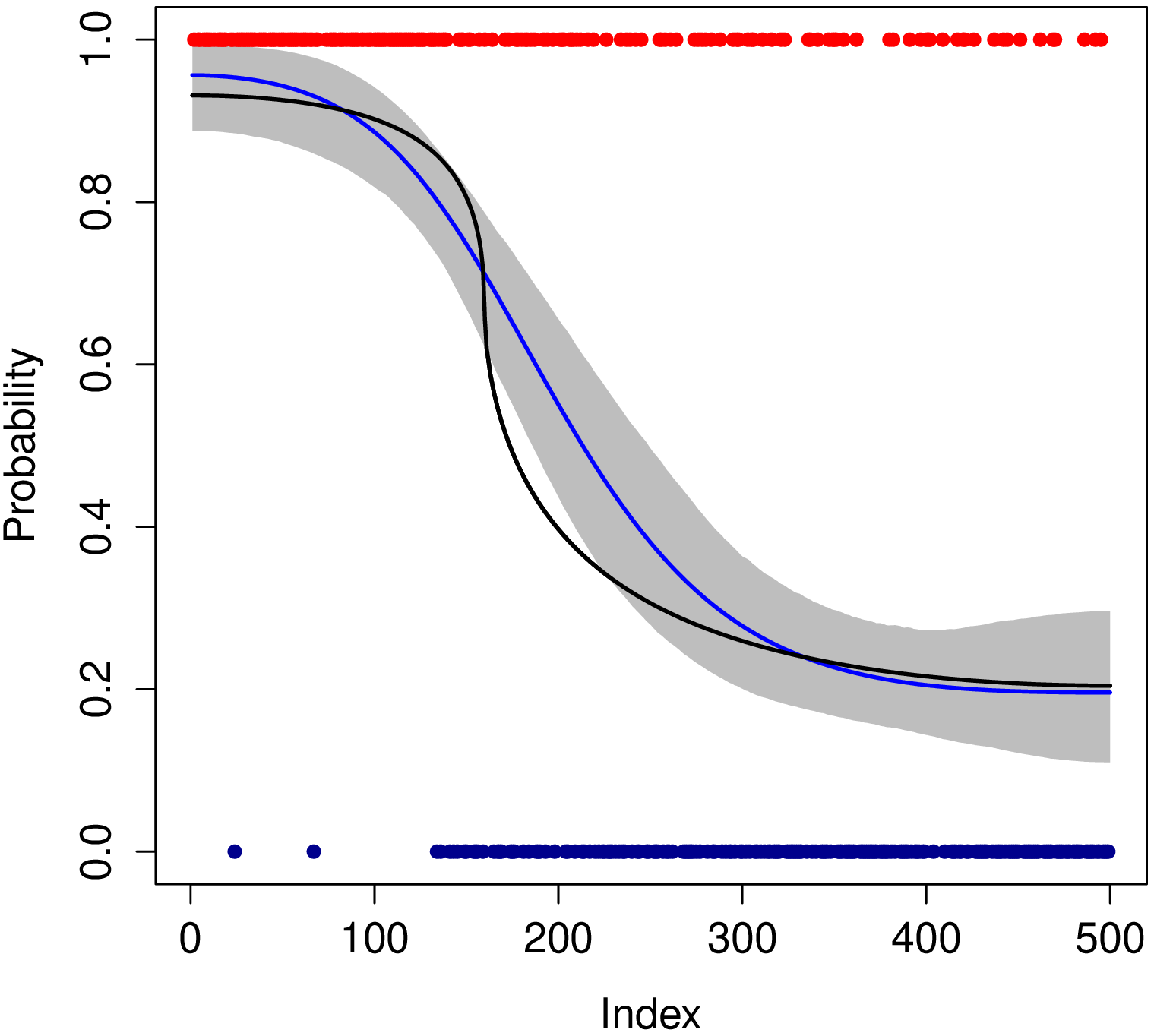}
\end{minipage}
\begin{minipage}{0.32\textwidth}
\includegraphics[width = \textwidth]{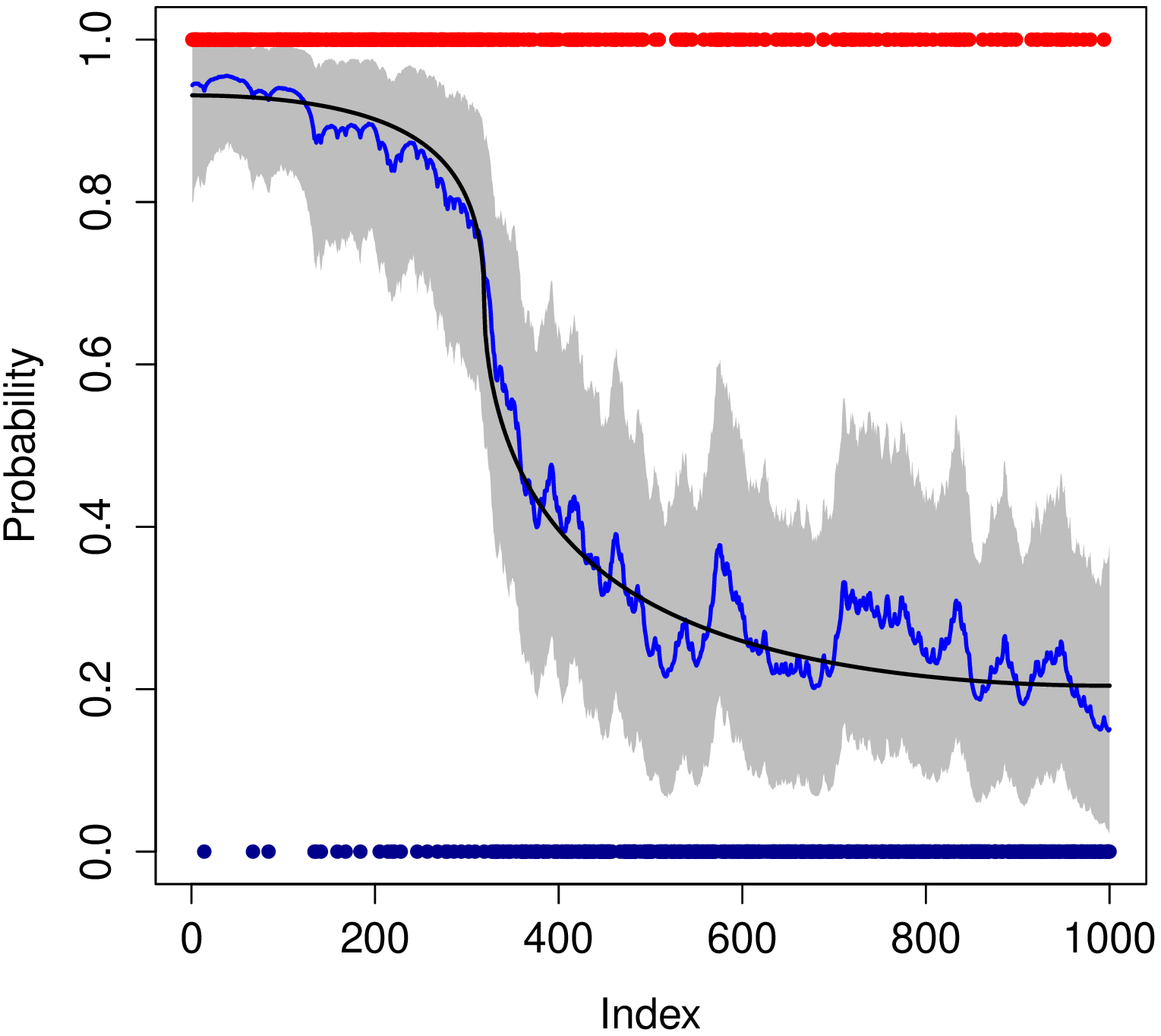}
\includegraphics[width = \textwidth]{Path1000q1_5par1-0-0.eps}
\includegraphics[width = \textwidth]{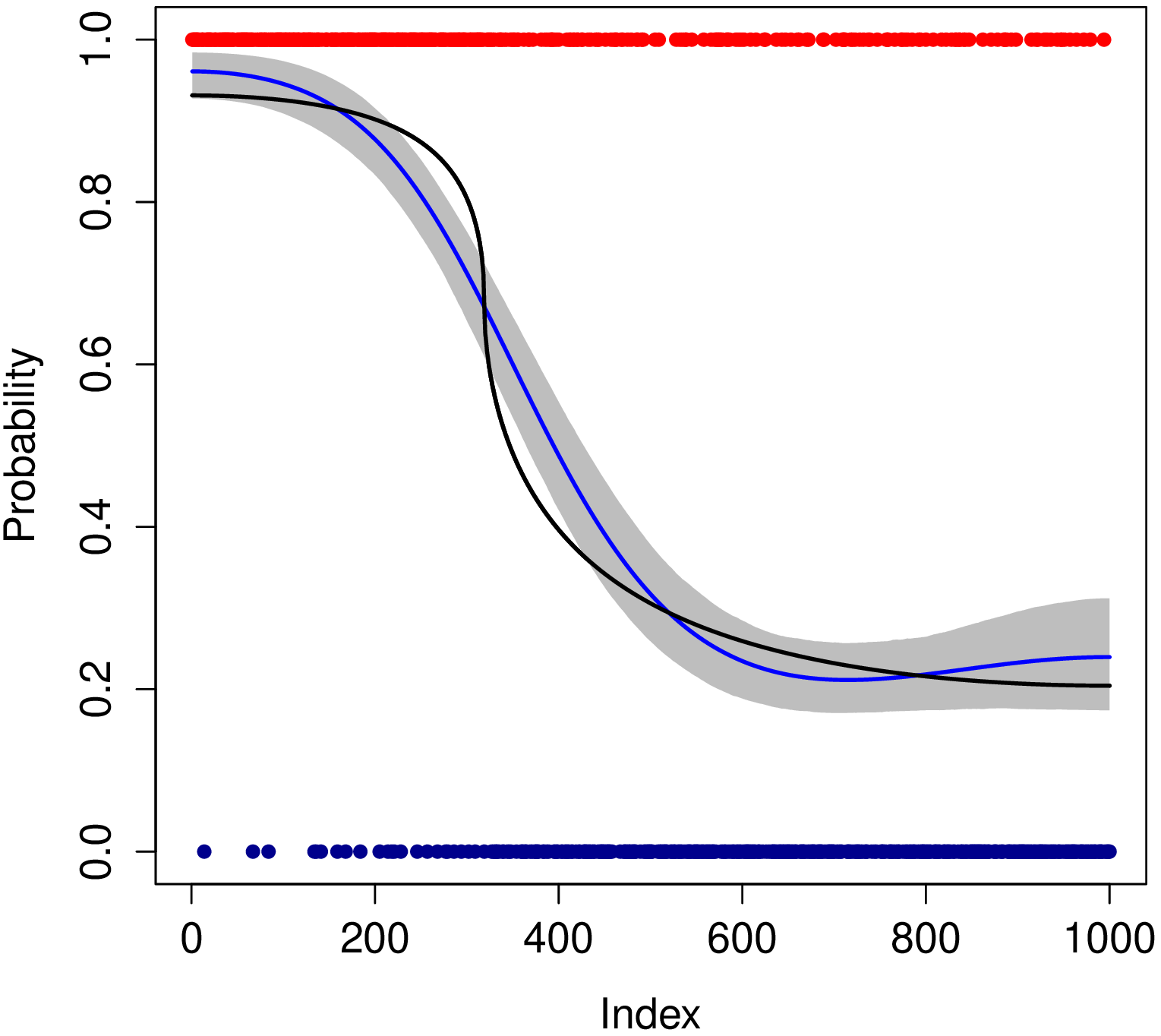}
\end{minipage}
\caption{Posteriors for $\ell$ when using the ordinary gamma prior on $c$
with $a=b=0$. From 
left two right we have $n = 100, 500, 1000$. From top to bottom we have
$q = \alpha + 1/2$, with $\alpha = 0.2, 1$ and $5$ respectively.}\label{fig: q2}
\end{figure}

Figure \ref{fig: q2} gives the same plots when using the ordinary gamma prior on $c$, 
with $a=b=0$. We see essentially the same effects, but the effect of choosing $q$
too small is a bit more pronounced. 
In terms of posterior coverage the ordinary gamma prior does a bit better, although it
gives  too conservative credible sets when $q$ is set too low.

\subsection{Small-world graph}

\begin{figure}[H]
\centering
\begin{minipage}{0.49\textwidth}
\includegraphics[width = \textwidth]{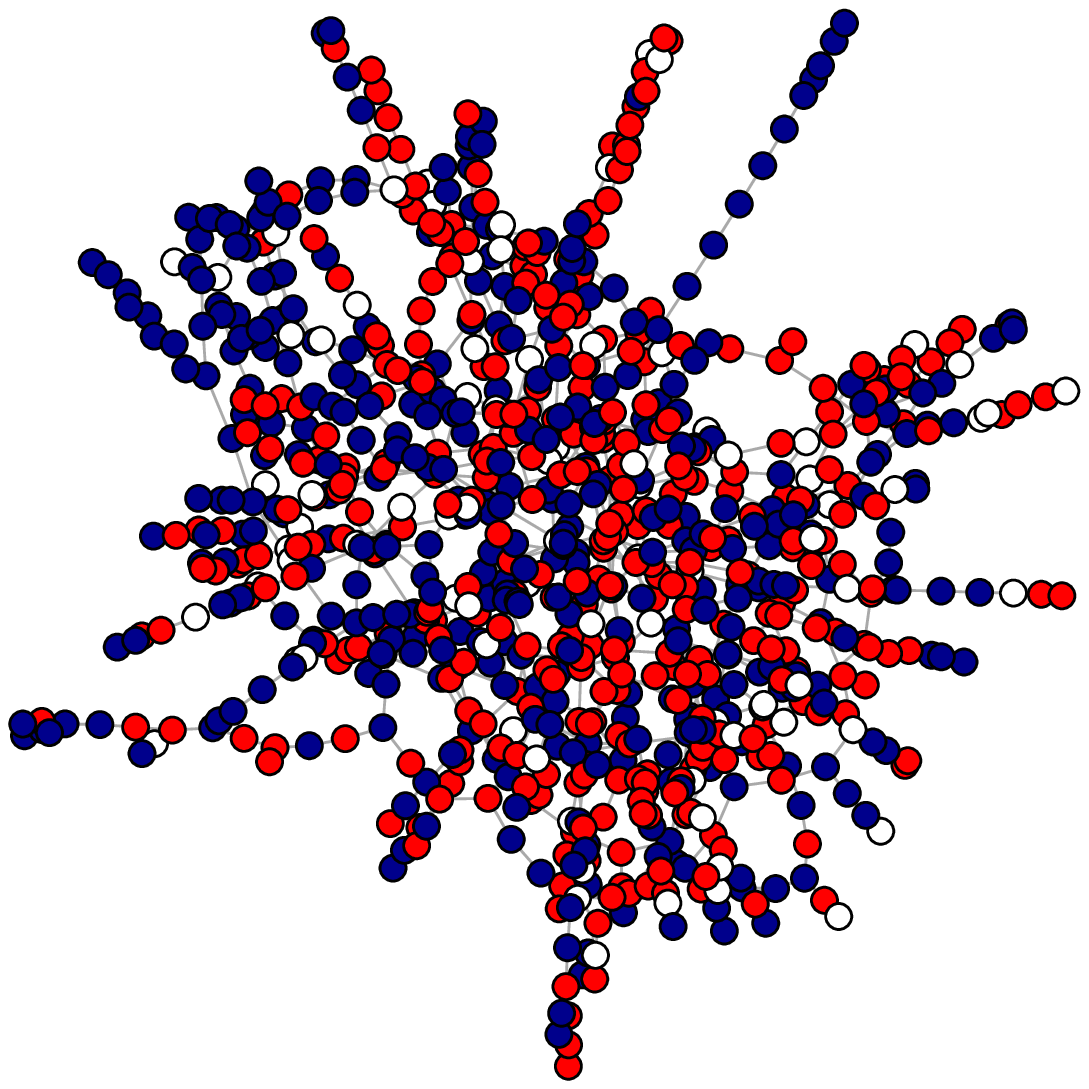}
\end{minipage}
\begin{minipage}{0.49\textwidth}
\includegraphics[width = \textwidth]{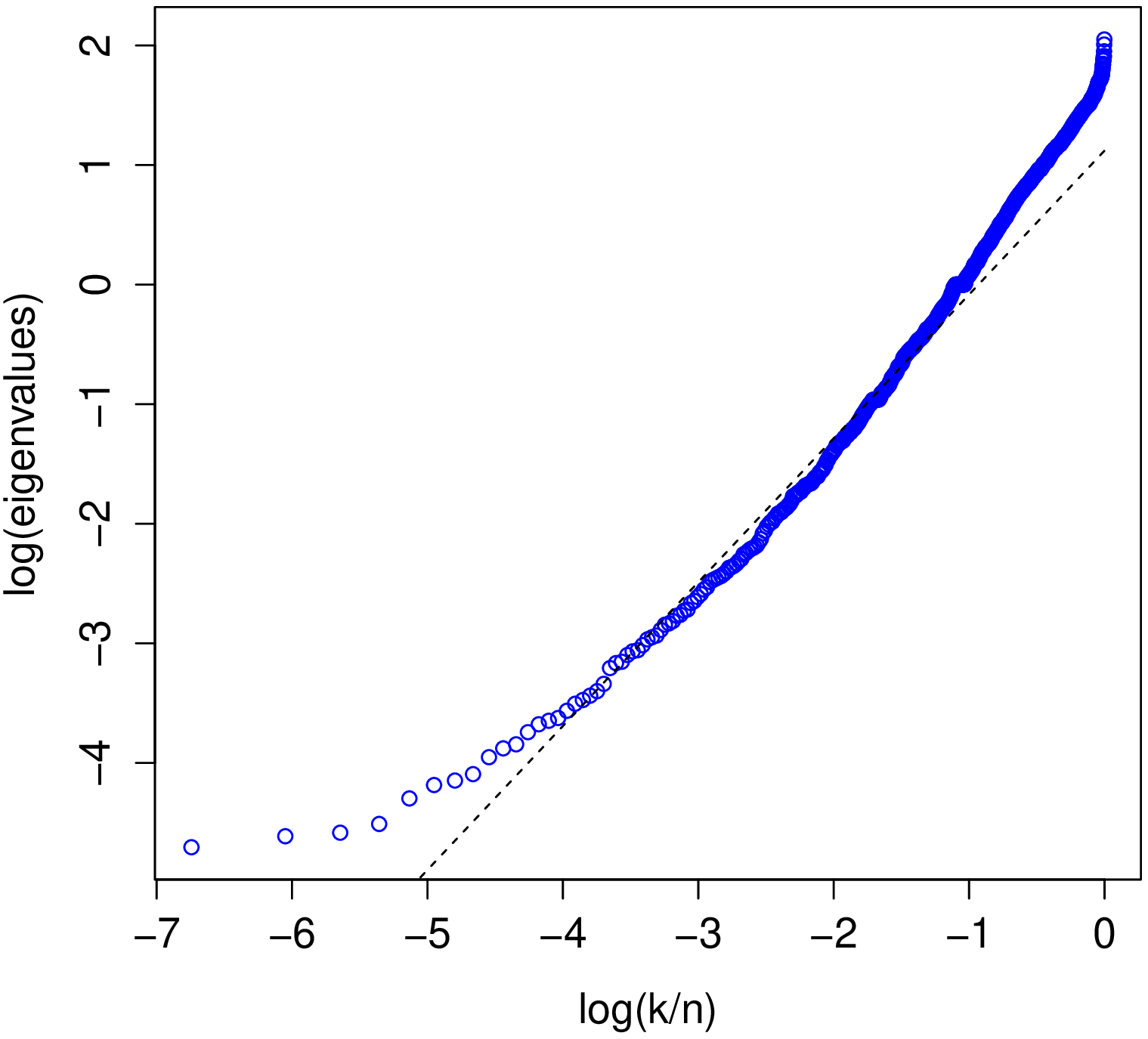}
\end{minipage}
\caption{Left: small world graph with two types of labels. White labels are missing.
Right: eigenvalue plot for the small-world graph. The linear fit has slope $1.20$ corresponding to geometry condition with $r = 1.7$.}
\label{fig:watts}
\end{figure}

In this section we consider simulated data on a small-world graph obtained as a realization of the Watts-Strogatz model (\cite{watts1998}). The graph is obtained by first considering  a ring graph of $1000$ nodes. Then we loop through the nodes and uniformly rewire each edge with probability $0.25$. We keep the largest connected component and delete multiple edges and loops resulting in the graph in Figure \ref{fig:watts} with $848$ nodes. 

We numerically determine the eigenvalues $\lambda_k$ and eigenfunctions $u^{(k)}$
of the graph Laplacian and define a function $f_0$ on the graph by 
\begin{equation*}
f_0 = \sum_{k = 1}^{n - 1} a_k u^{(k)},
\end{equation*}
where we choose $a_k = \sqrt{n}k^{-2/r-1/2}\sin k$ for $k = 1, \ldots, n - 1$ to have 
Sobolev-type smoothness $\beta = 2$ (cf.\ \cite{kirichenko2017}). 
As before we assign labels to the graph according to probabilities $P(Y_i = 1) = \ell_0(i) = \Phi(f_0(i))$, where $\Phi$ is the distribution function of the standard normal distribution. We remove the label of $10\%$ of the nodes. The aim is to predict these using the observed labels.

\begin{figure}[H]
\begin{minipage}{0.32\textwidth}
\includegraphics[width=\textwidth]{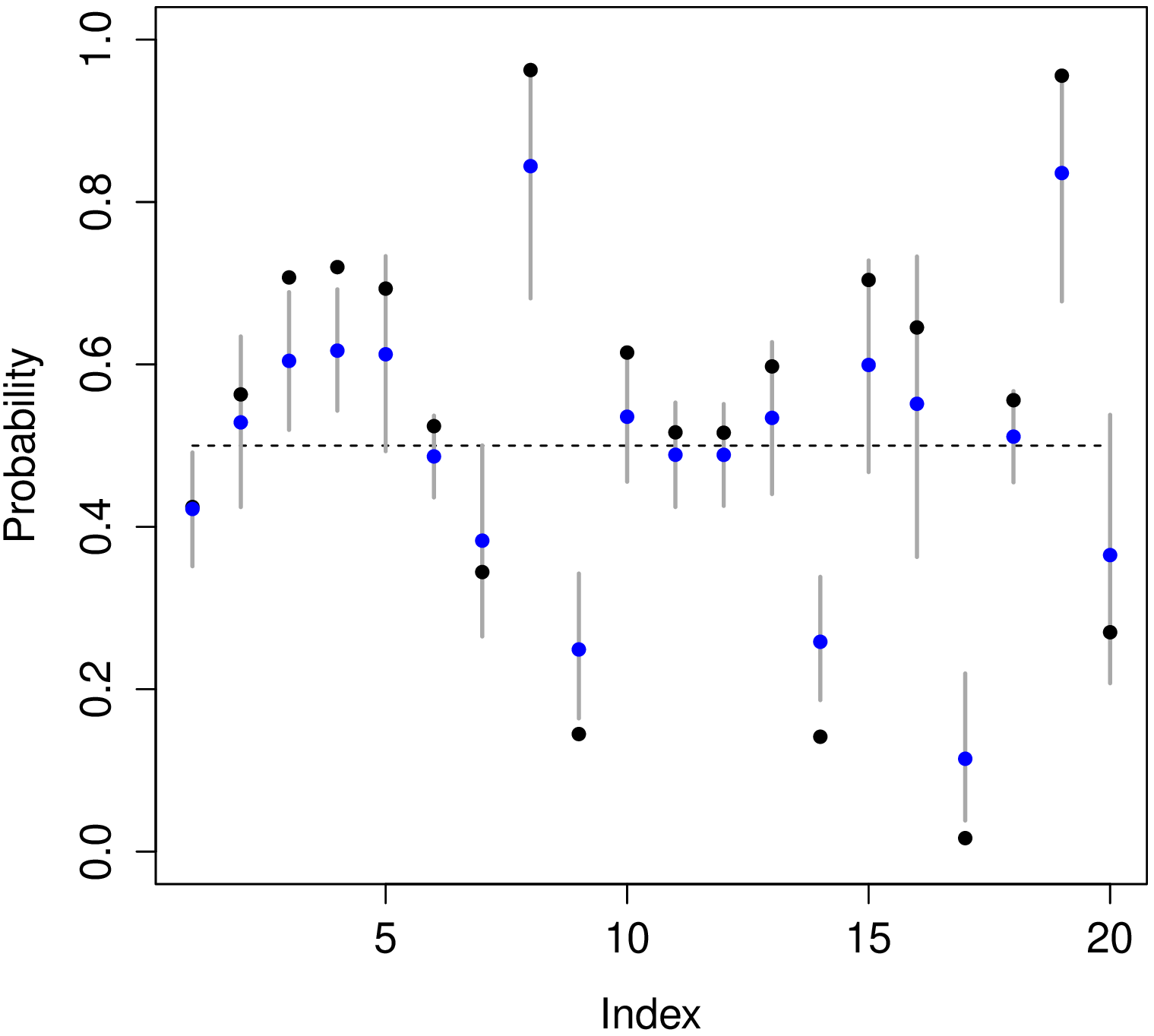}
\includegraphics[width=\textwidth]{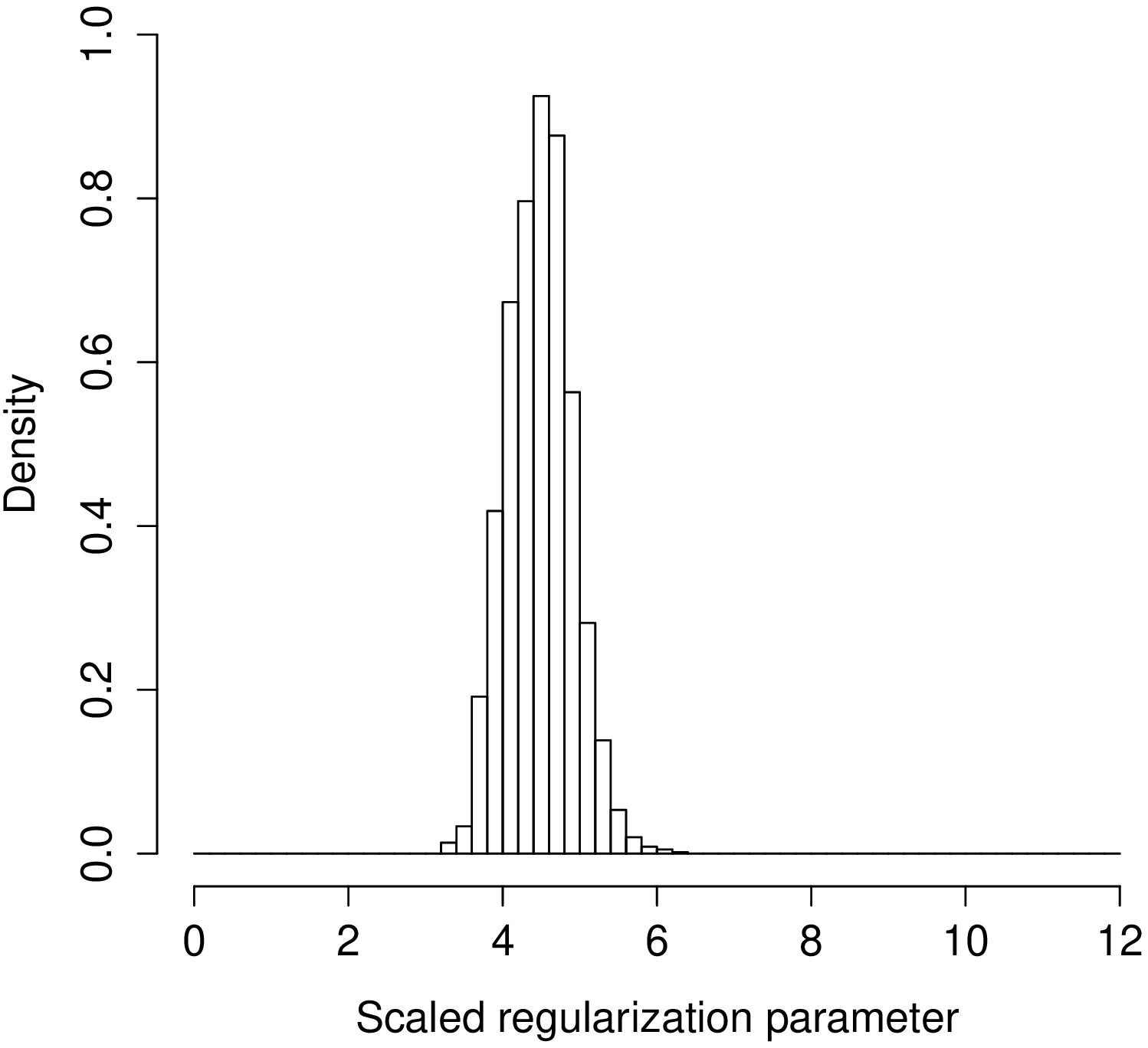}
\end{minipage}
\begin{minipage}{0.32\textwidth}
\includegraphics[width=\textwidth]{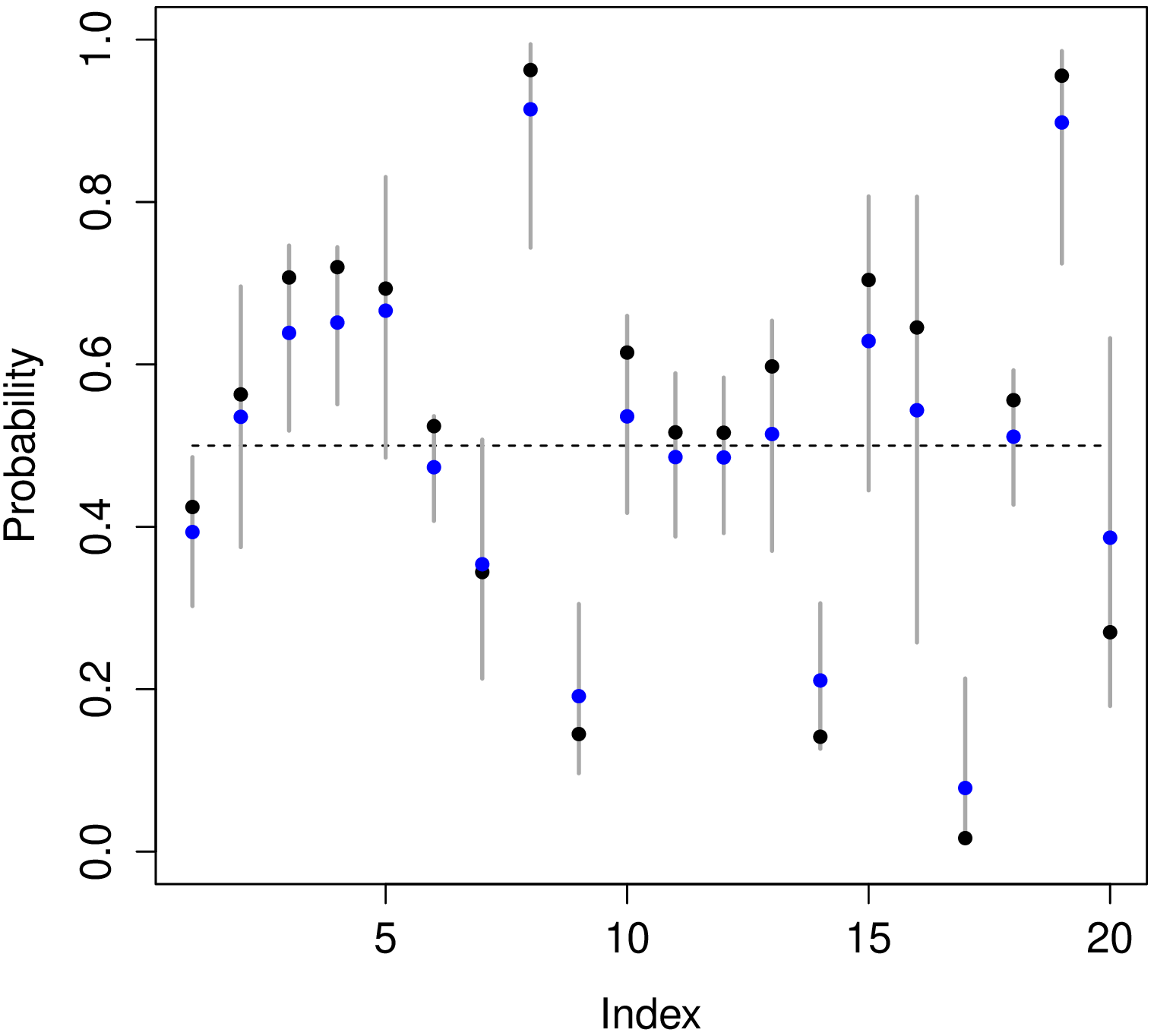}
\includegraphics[width=\textwidth]{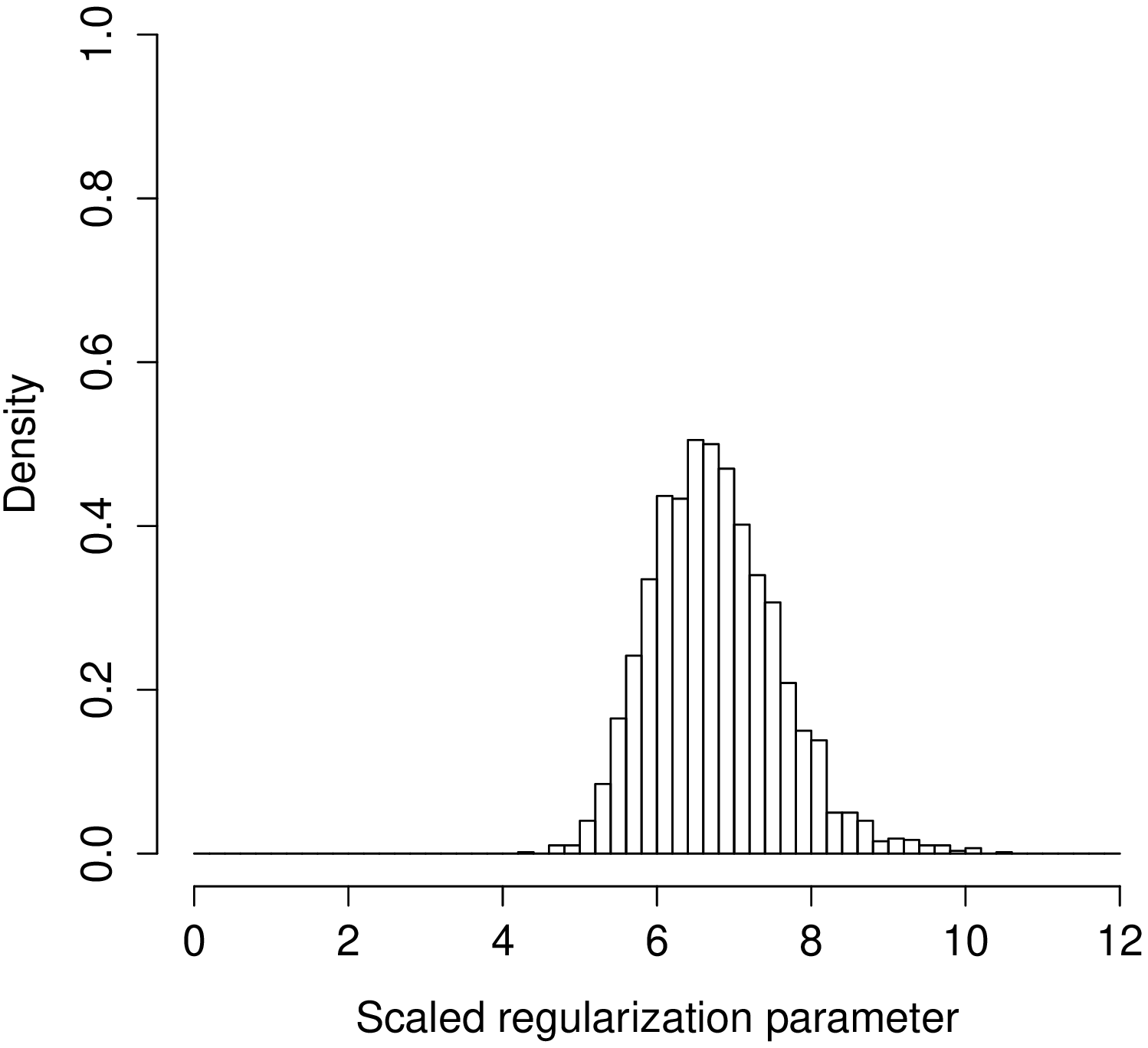}
\end{minipage}
\begin{minipage}{0.32\textwidth}
\includegraphics[width=\textwidth]{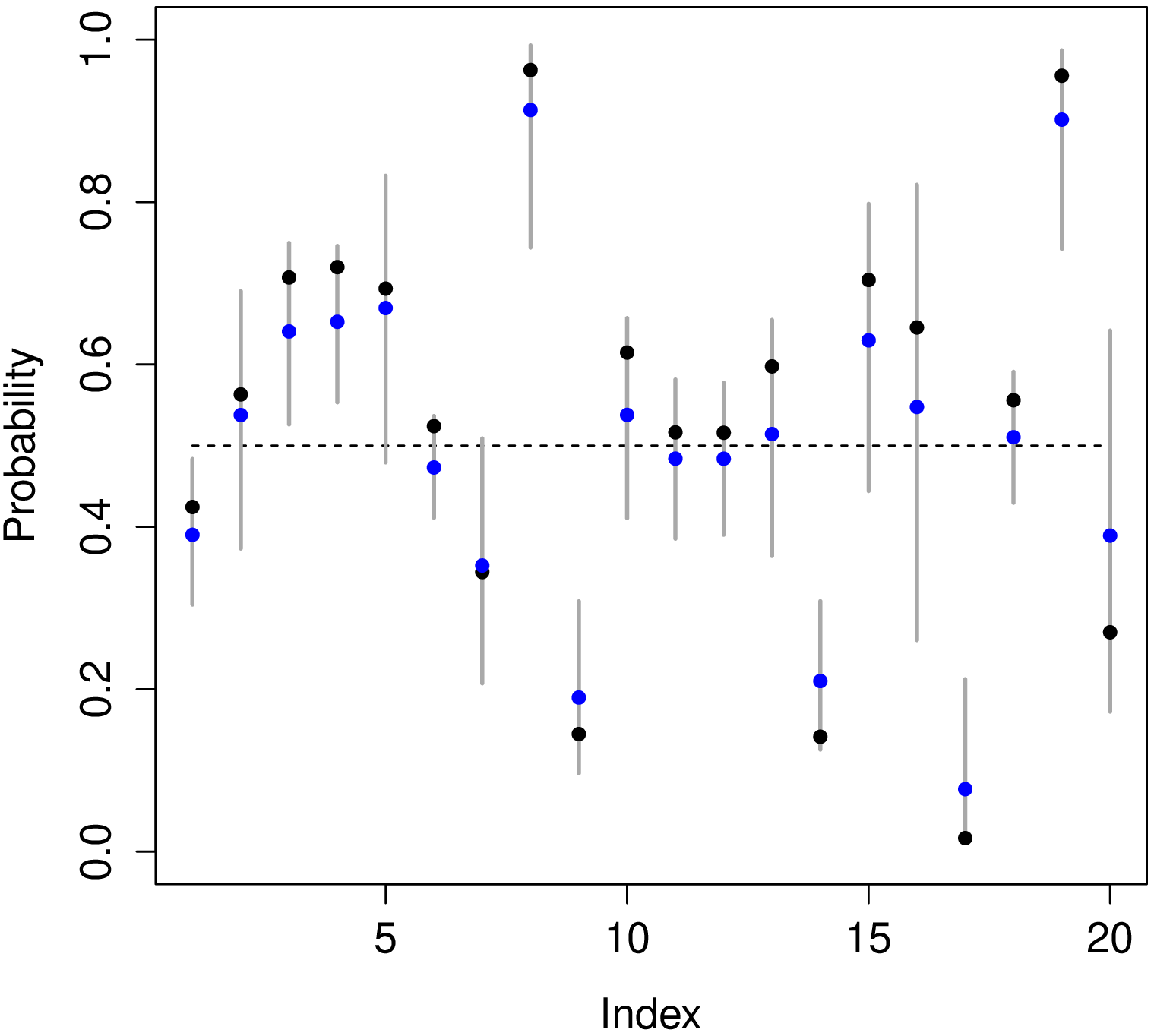}
\includegraphics[width=\textwidth]{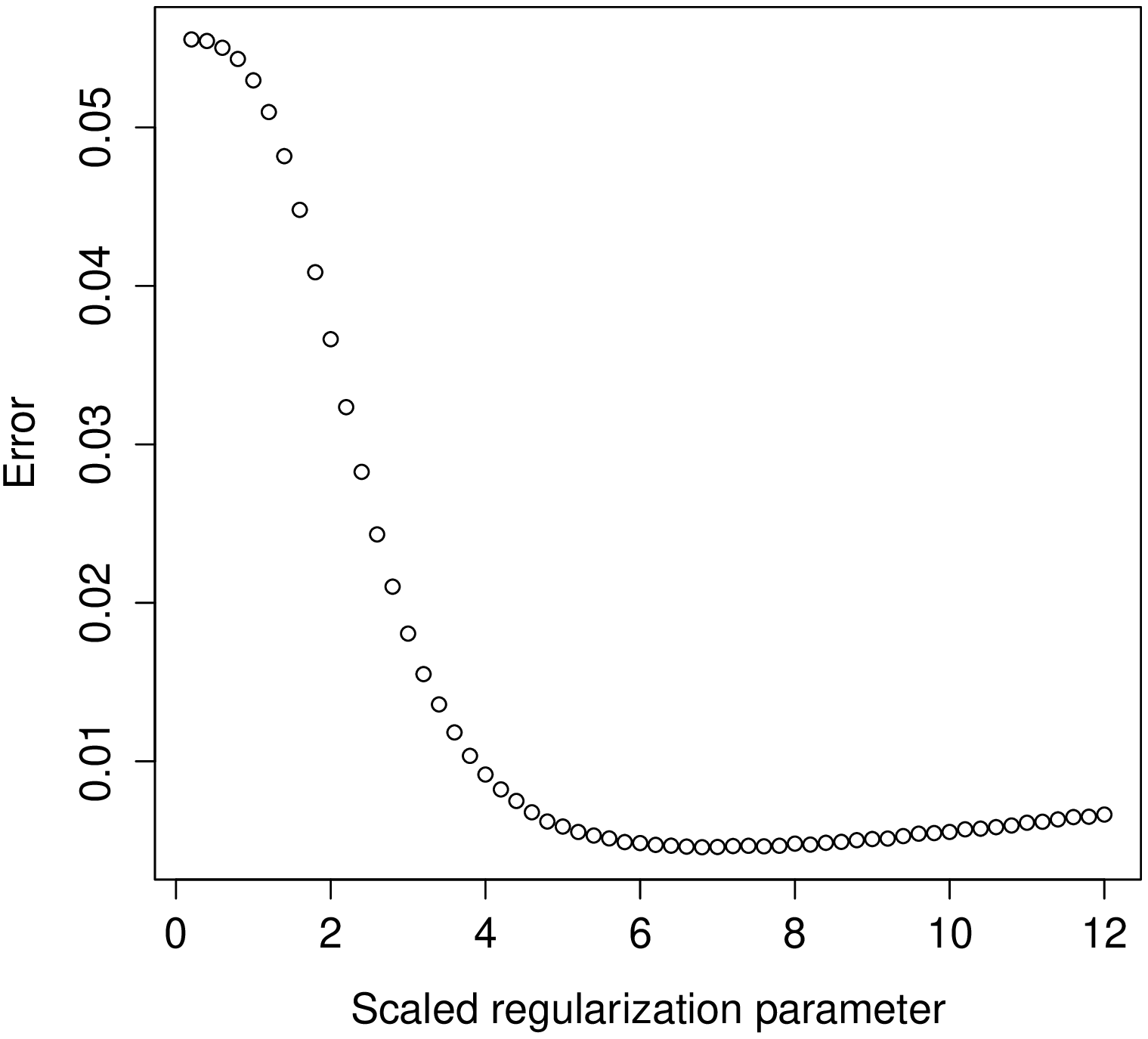}
\end{minipage}
\caption{Top row: credible intervals for soft label function at $20$ vertices with missing labels.
Blue dots are the posterior means, black dots are the true function values.  
From left to right the plots correspond to the procedure with 
the generalised gamma prior on $c$, the ordinary gamma prior with $a=b=0$, 
and with the fixed, oracle choice of $c$, respectively.
Bottom row: histograms of posterior for $c$ (scaled).}\label{fig: sw1}
\end{figure}

In this case it is hard to visualise smooth functions on the graph and hence
to visualise the entire posterior distribution of  the soft label function. Instead 
we analyse the quality of the procedure by plotting $95\%$ credible intervals for the 
soft label function at $20$ randomly selected vertices of which we have not observed
the noisy labels.
Figure \ref{fig: sw1} gives these plots for the procedure with the generalised gamma
prior on $c$ (left), the ordinary gamma prior on $c$ with $a=b=0$ (middle), 
and the fixed oracle choice of $c$ (right). At the bottom  left and middle
the posteriors for $c$ are shown. The bottom right is a plot of the MSE of the posterior mean
corresponding to a fixed $c$ as a function of that $c$. The point where it is minimal is 
the oracle choice of $c$.

Also in this example we observe that with the theoretically optimal  generalised 
gamma prior on $c$ we are shrinking a bit too much,  resulting in particular in credible intervals 
not containing the true soft label. When using the ordinary gamma prior 
the performance is closer to the oracle procedure.
The bottom row of Figure \ref{fig: sw1} confirms that with the ordinary gamma prior 
the posterior for $c$ is closer to the oracle choice.

\subsubsection{Impact of hyper parameters}

We have determined numerically that the graph under consideration  
satisfies the geometry condition \eqref{eq: geom} with $r = 1.7$, 
see the right panel of Figure \ref{fig:watts}. 
The results of \cite{kirichenko2017} thus suggest to use as prior on $f$ 
the Laplacian to the power $q = \alpha+1.7/2$ for parameter  $\alpha$ that determines the 
prior smoothness. Also for this example we have investigated the impact 
of different choices.

\begin{figure}[H]
\begin{minipage}{0.32\textwidth}
\includegraphics[width=\textwidth]{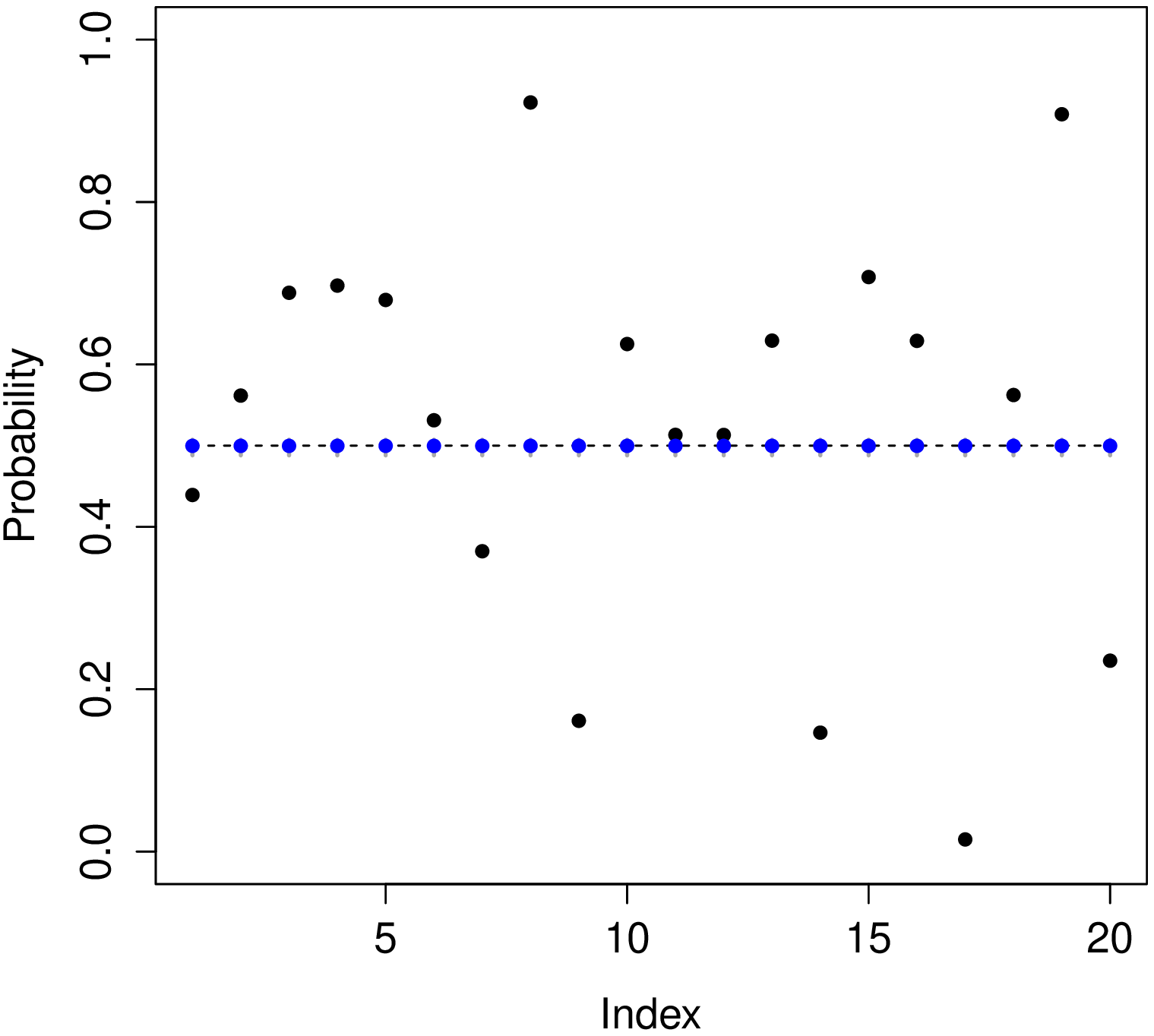}
\includegraphics[width=\textwidth]{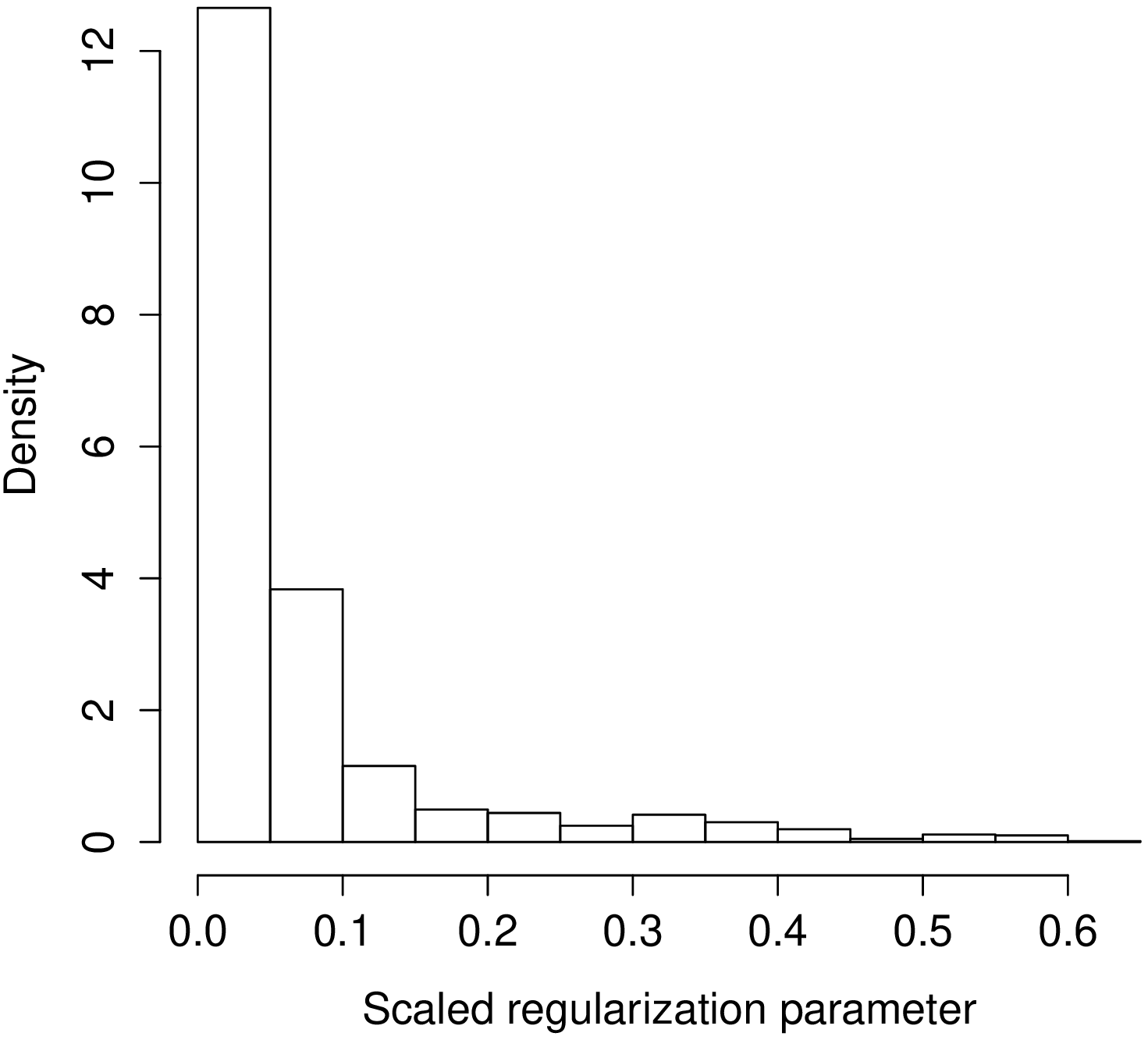}
\end{minipage}
\begin{minipage}{0.32\textwidth}
\includegraphics[width=\textwidth]{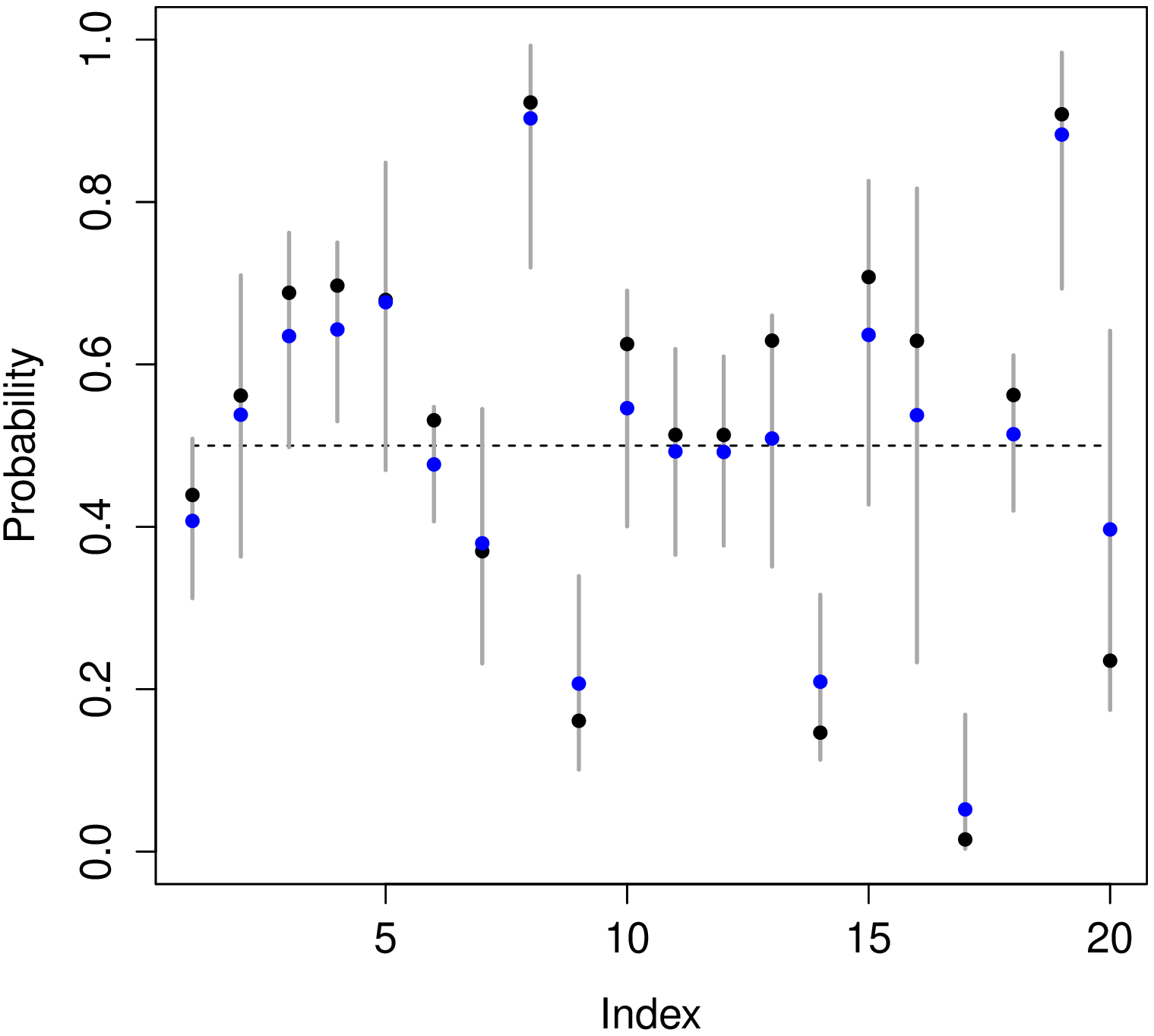}
\includegraphics[width=\textwidth]{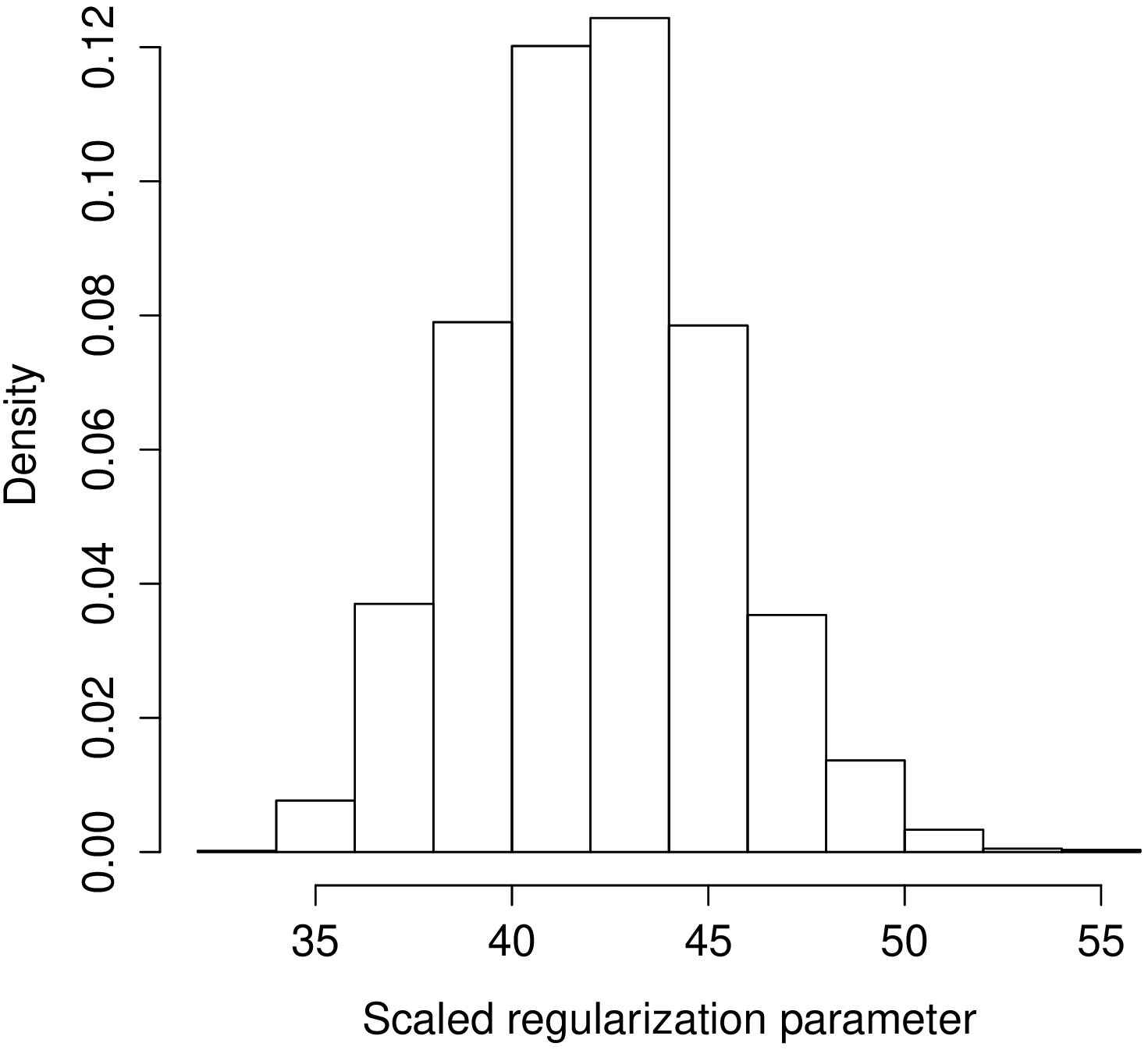}
\end{minipage}
\begin{minipage}{0.32\textwidth}
\includegraphics[width=\textwidth]{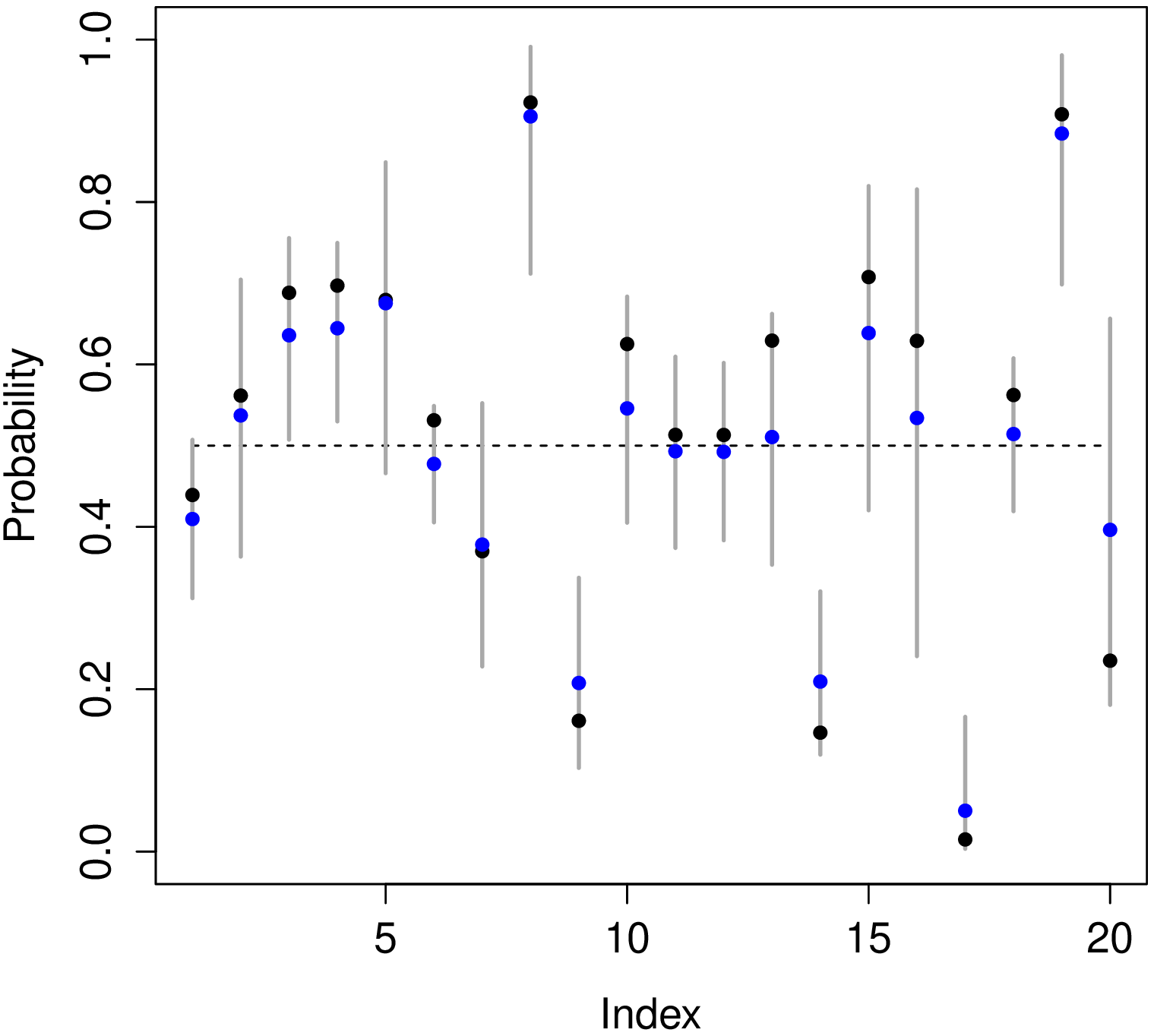}
\includegraphics[width=\textwidth]{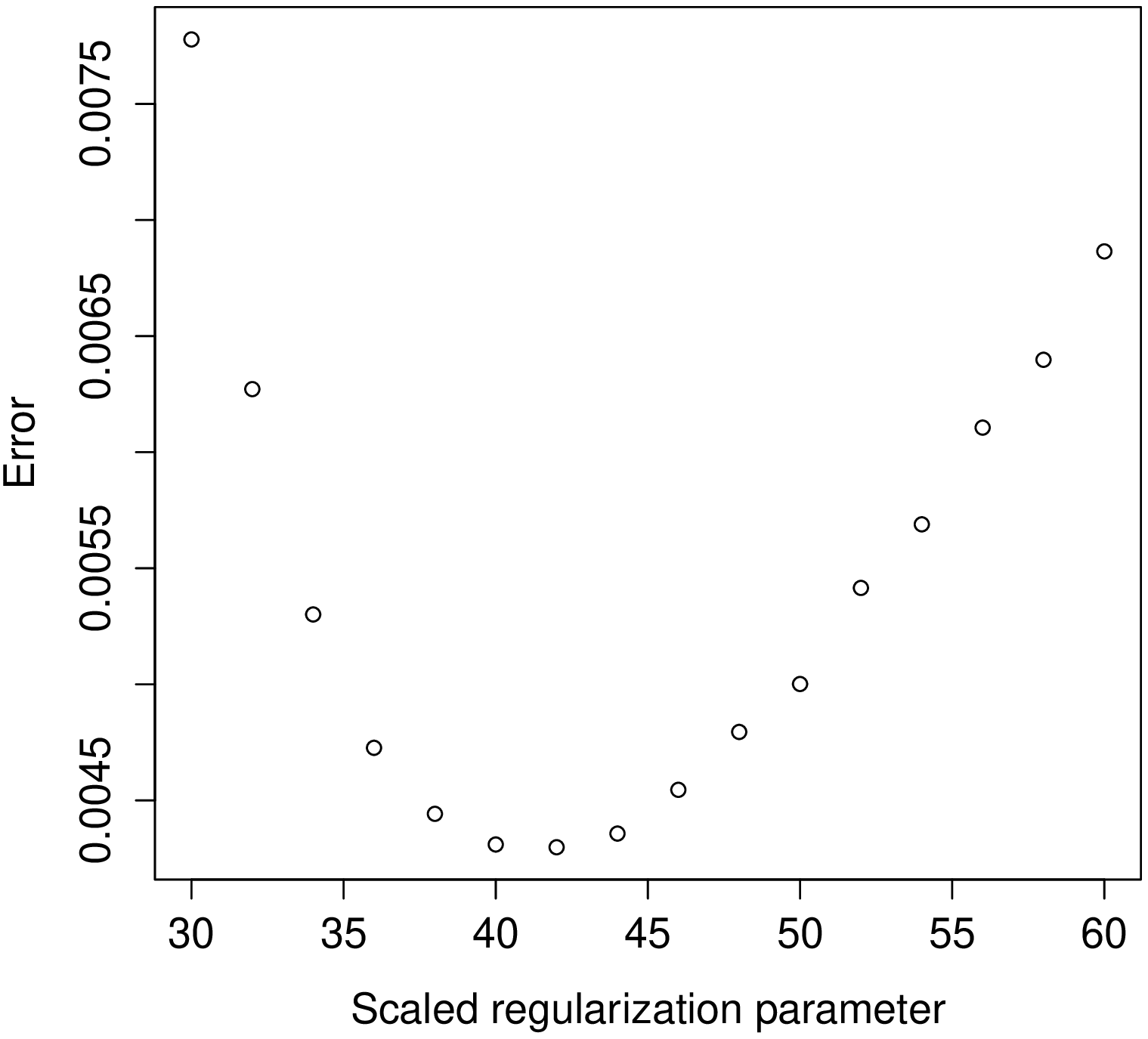}
\end{minipage}
\caption{Same as Figure \ref{fig: sw1}, but now with $r=1$.}
\label{fig: ws1}
\end{figure}

\begin{figure}[H]
\begin{minipage}{0.32\textwidth}
\includegraphics[width=\textwidth]{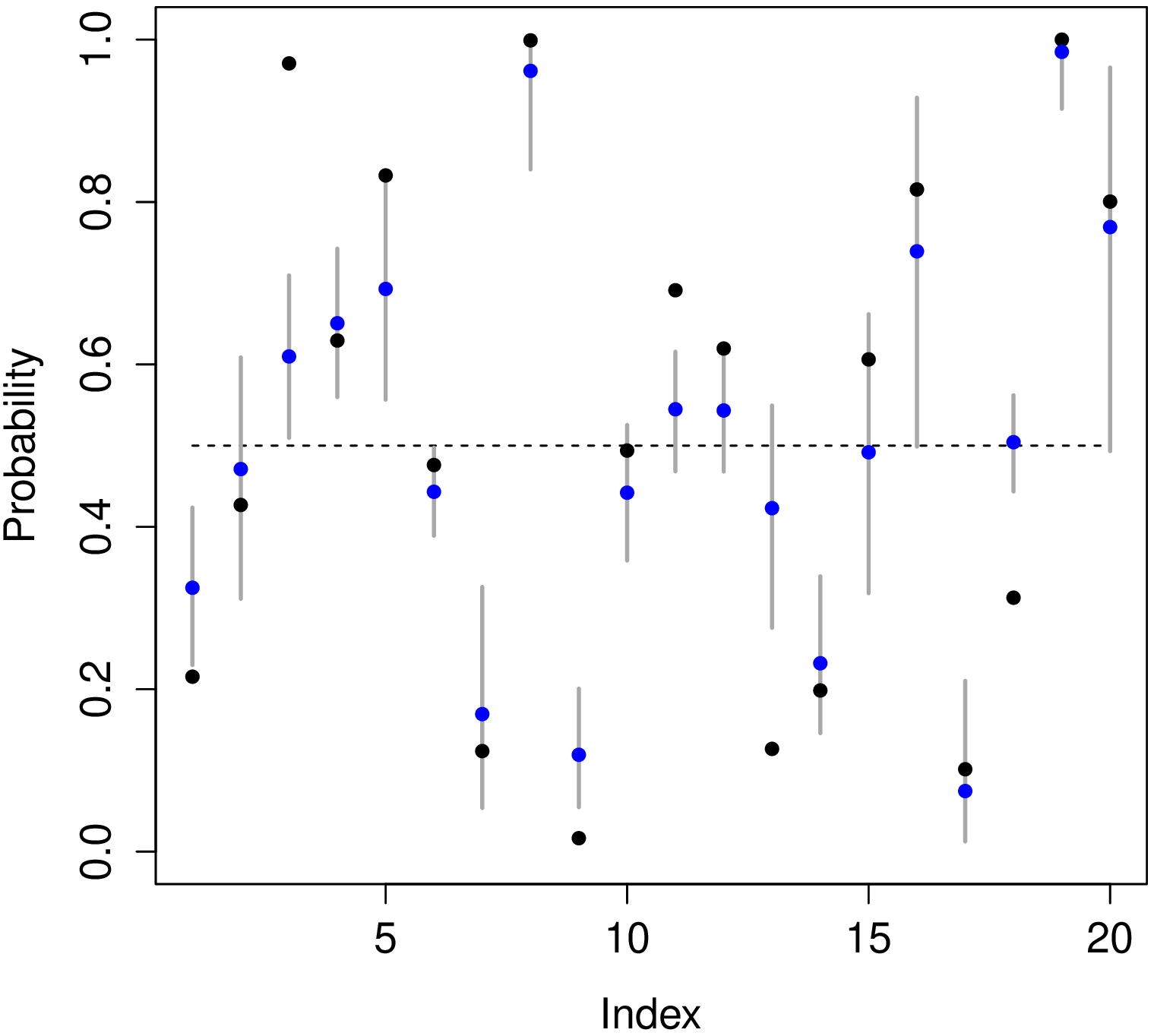}
\includegraphics[width=\textwidth]{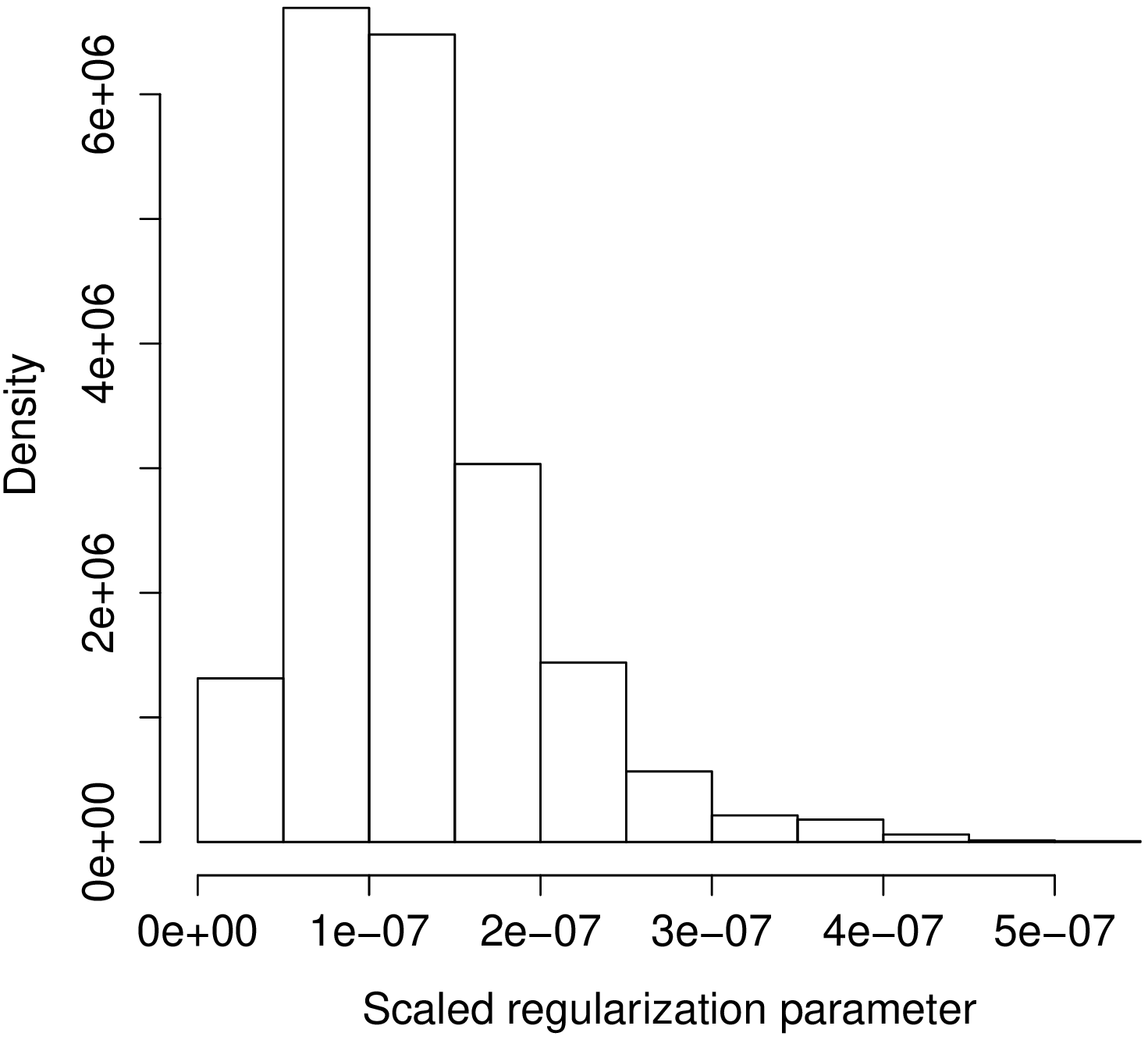}
\end{minipage}
\begin{minipage}{0.32\textwidth}
\includegraphics[width=\textwidth]{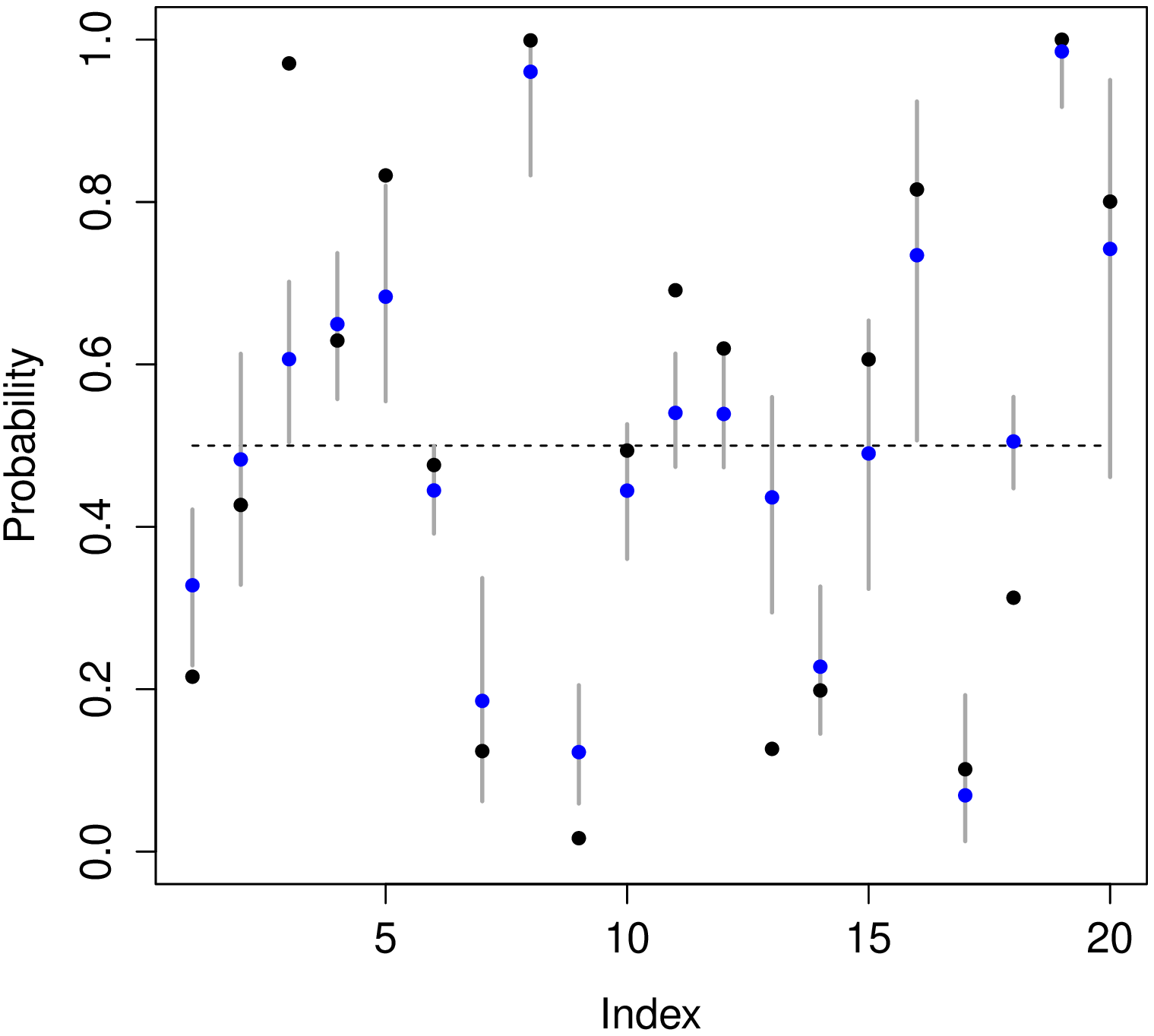}
\includegraphics[width=\textwidth]{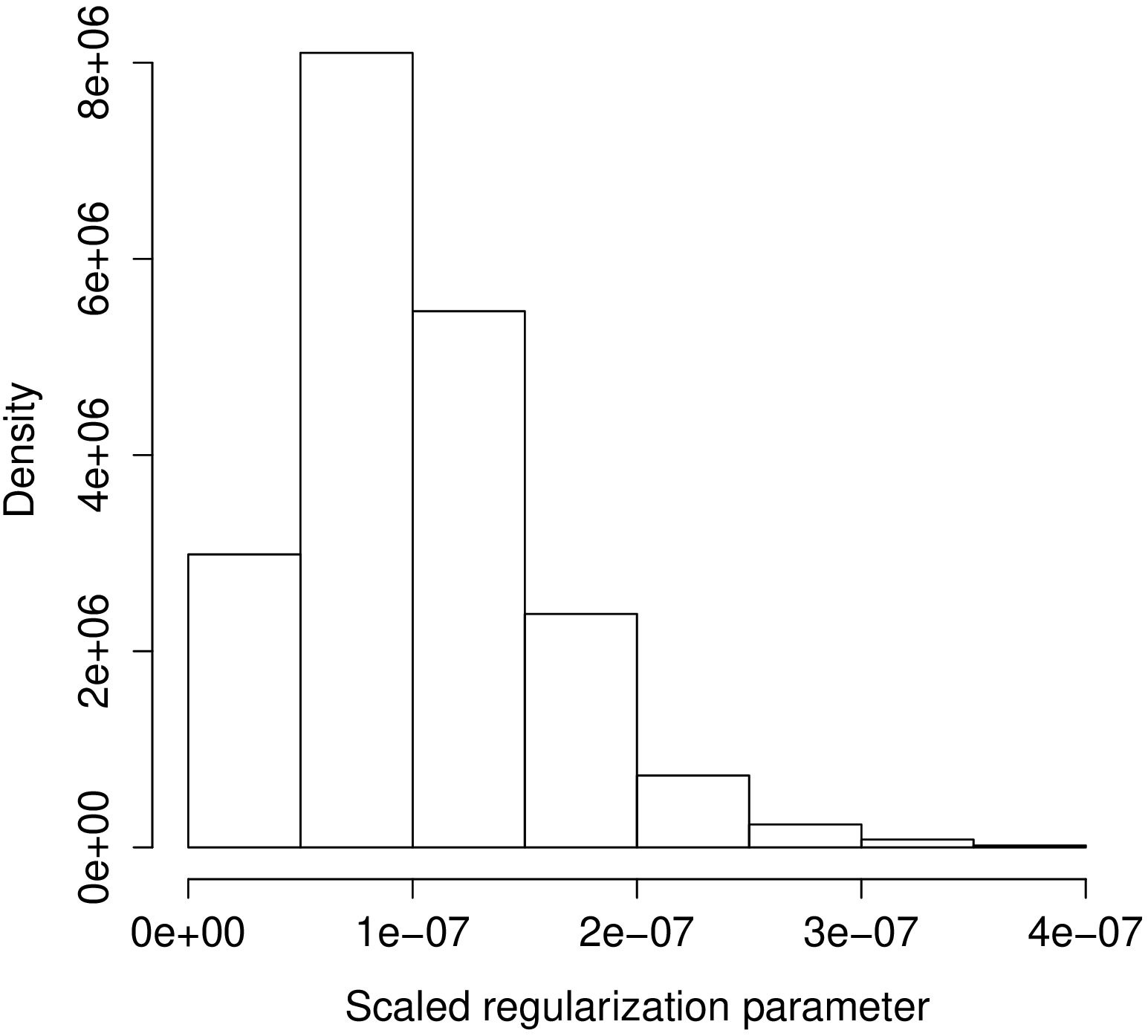}
\end{minipage}
\begin{minipage}{0.32\textwidth}
\includegraphics[width=\textwidth]{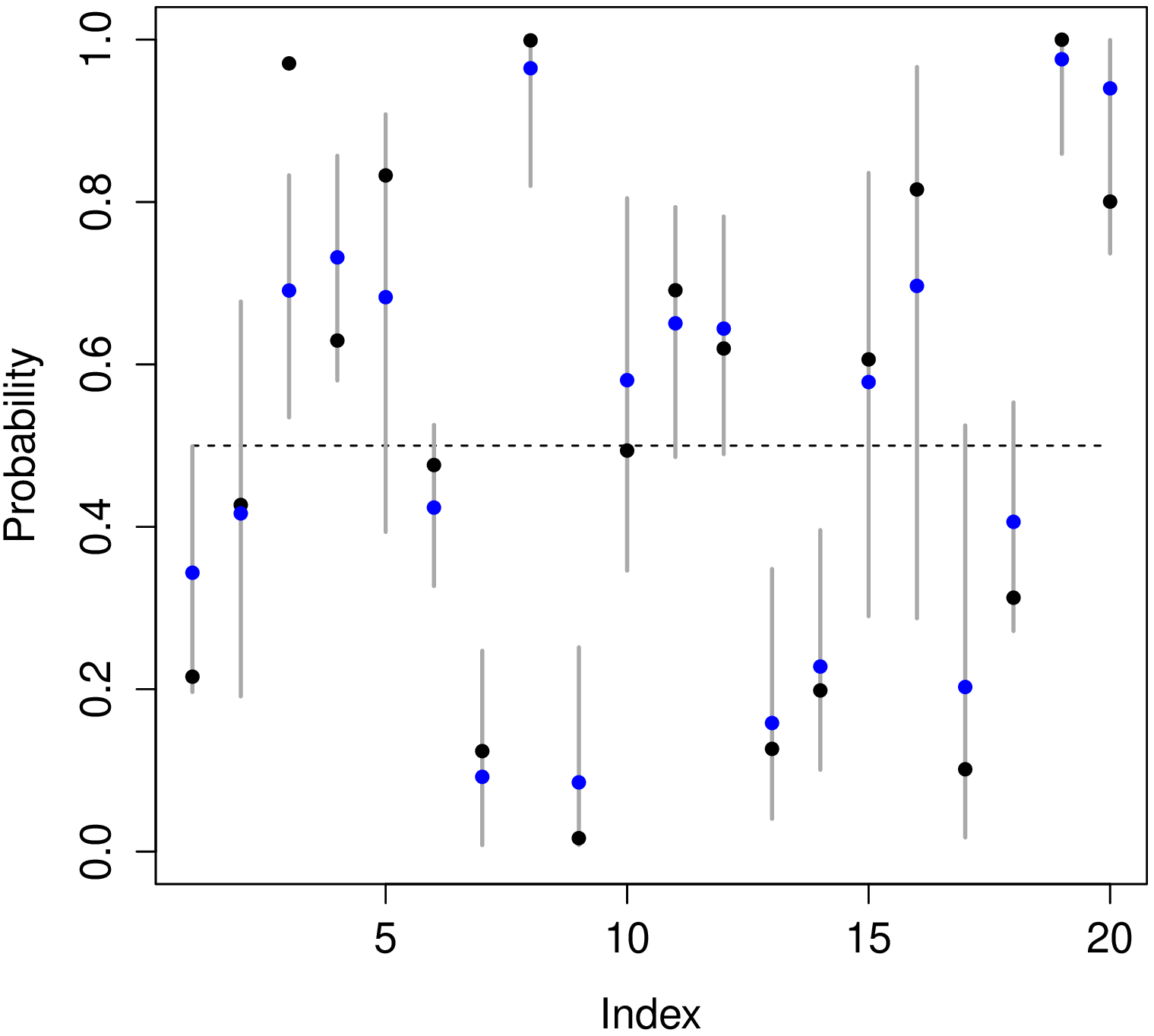}
\includegraphics[width=\textwidth]{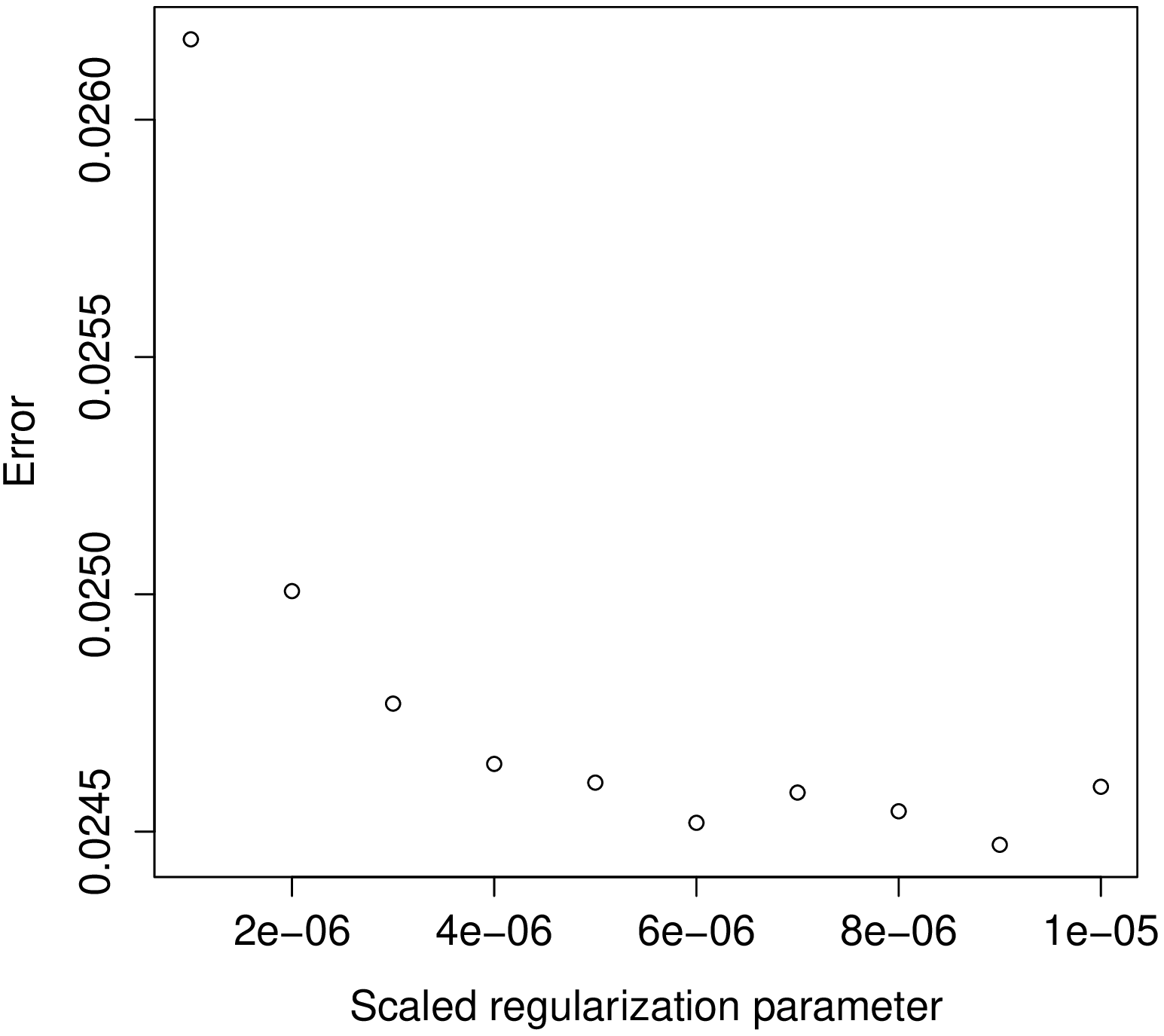}
\end{minipage}
\caption{Same as Figure \ref{fig: sw1}, but now with $r=10$.}
\label{fig: ws2}
\end{figure}

In Figure \ref{fig: ws1} illustrate what happens if $r$ is chosen too low. 
On the left we see that the procedure with the generalised gamma 
prior on $c$ oversmooths quite dramatically. The bottom row of the figure
shows that the posterior for $c$ puts 
too little mass around the oracle $c$ in that case.
The plots in the middle 
corresponds to ordinary gamma prior on $c$ with $a=b=0$. This 
performs much better, close to procedure with oracle choice of $c$ shown on the right.
In Figure \ref{fig: ws2} the parameter $r$ is chosen too high. Here all three 
procedures have comparable performance. All oversmoothing a {bit} too much due 
to the fact that the power of the Laplacian becomes too large. This is in line 
with what we saw for the path graph.

\begin{figure}[H]
\begin{minipage}{0.32\textwidth}
\includegraphics[width=\textwidth]{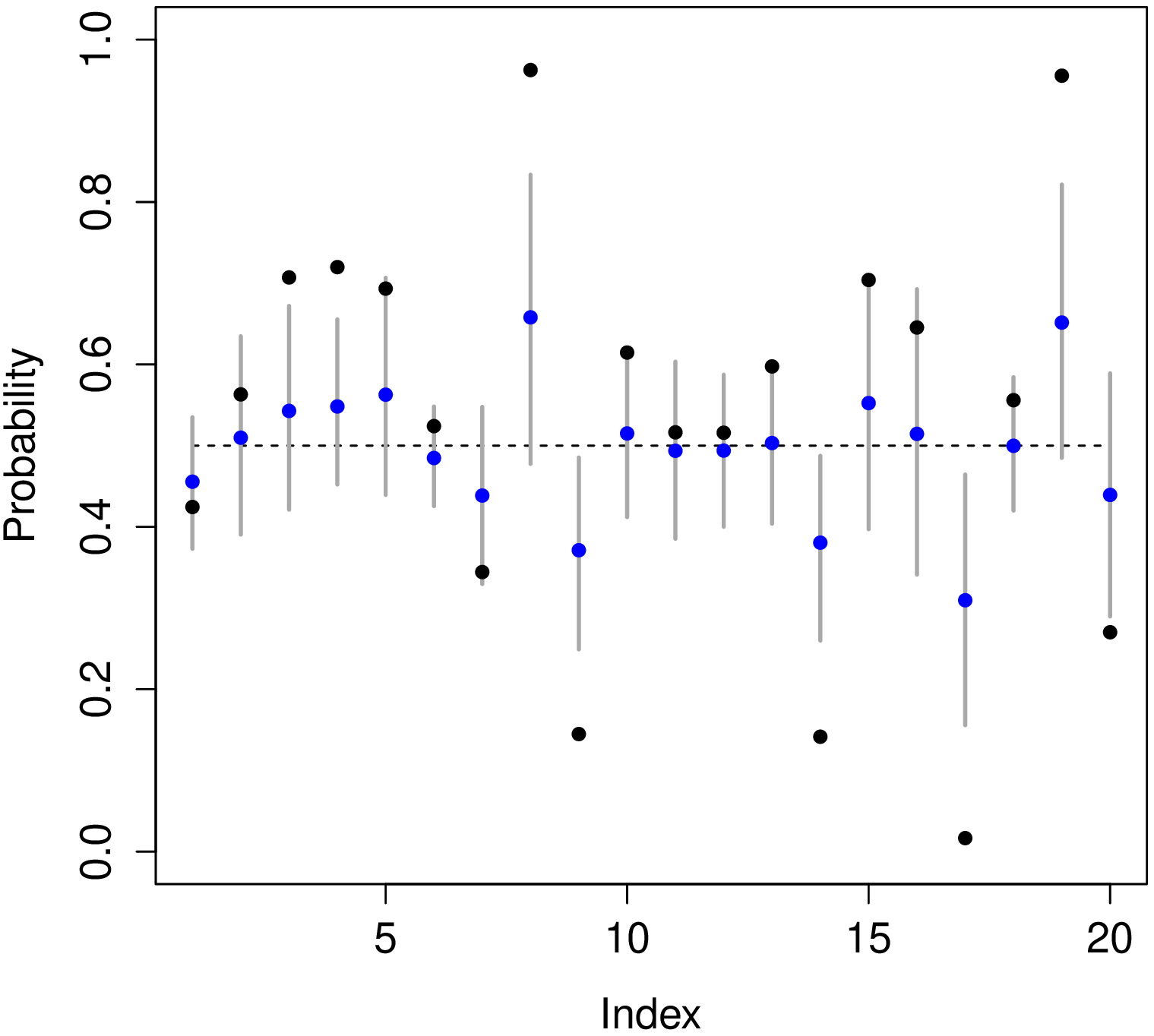}
\includegraphics[width=\textwidth]{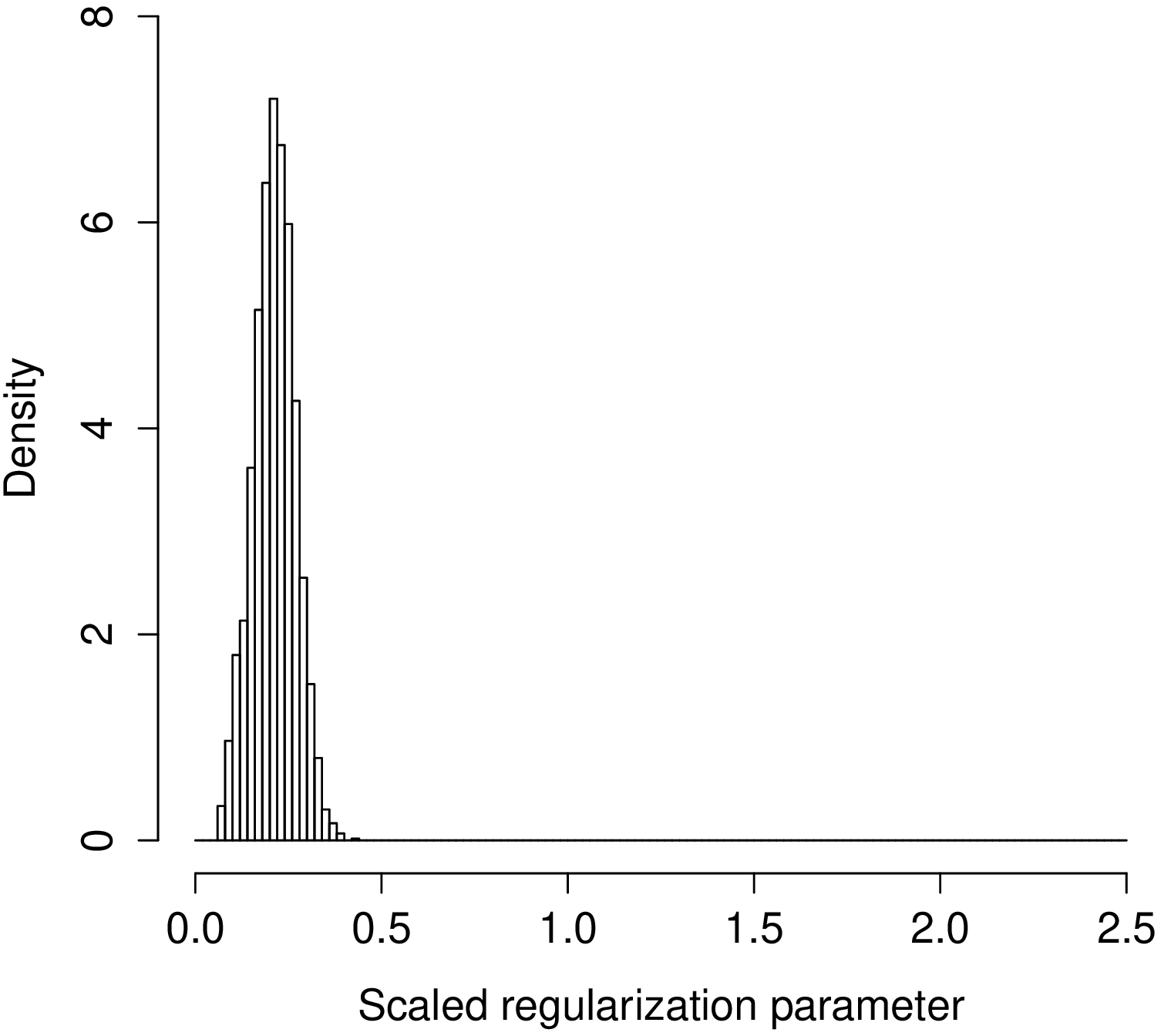}
\end{minipage}
\begin{minipage}{0.32\textwidth}
\includegraphics[width=\textwidth]{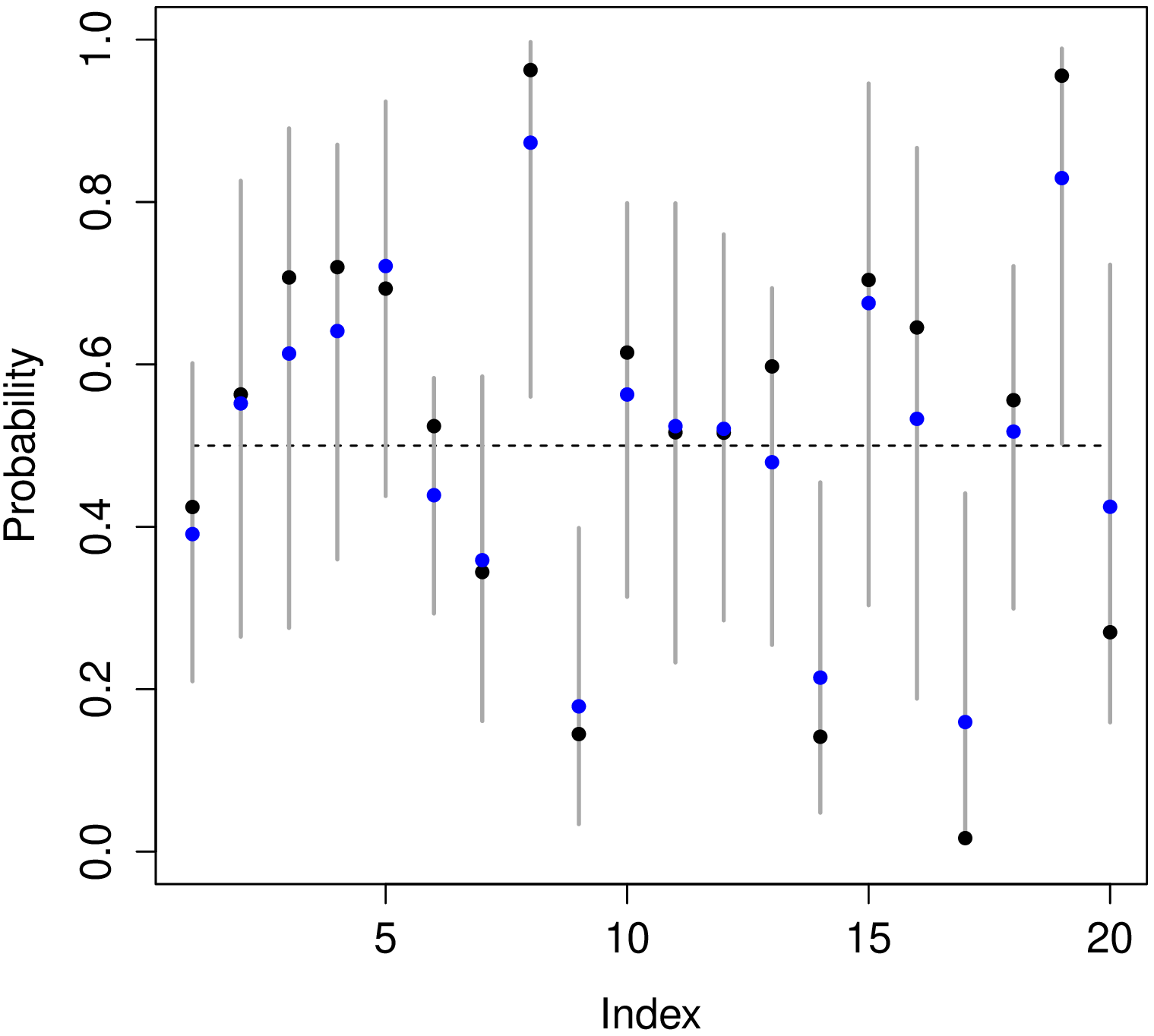}
\includegraphics[width=\textwidth]{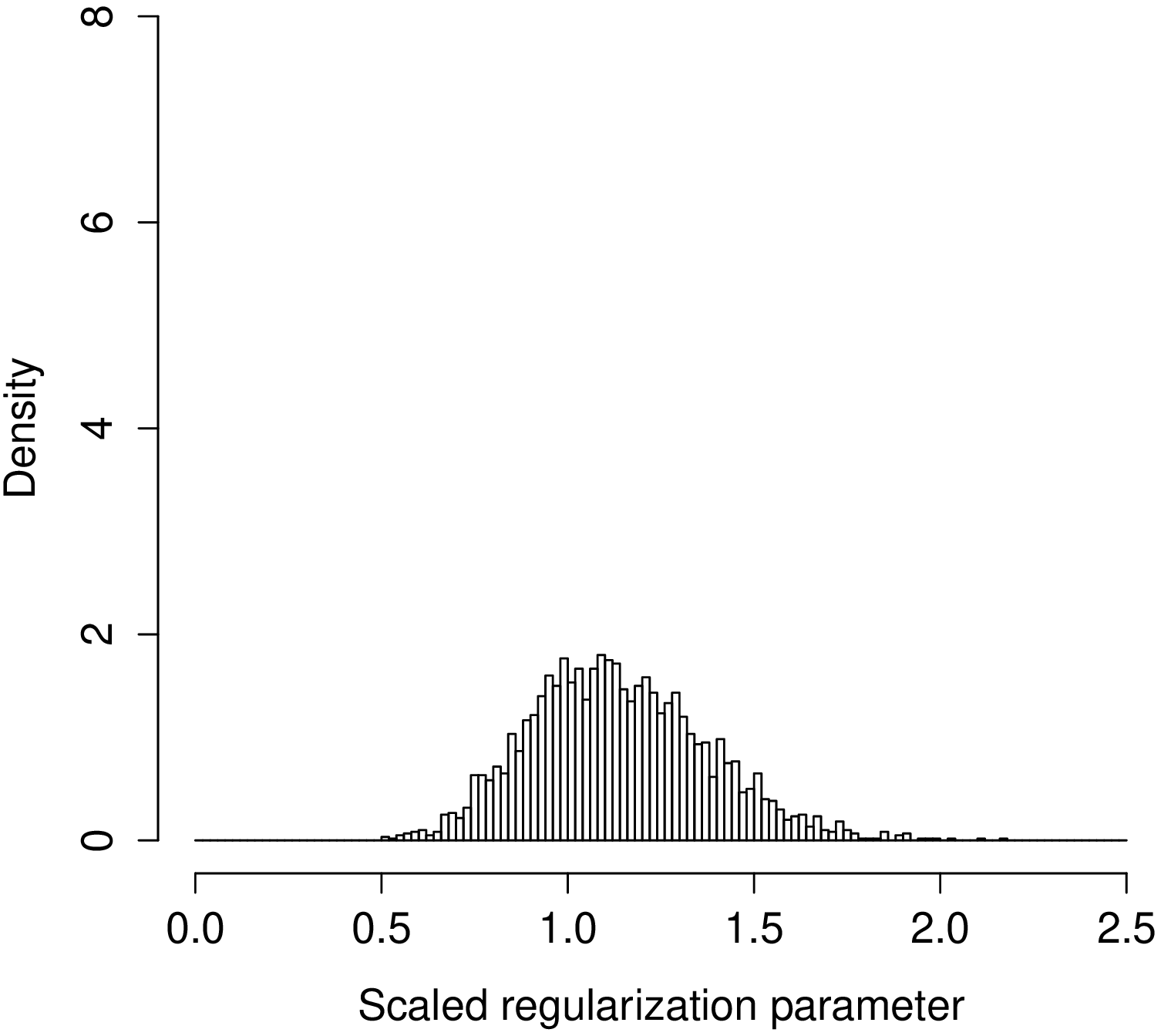}
\end{minipage}
\begin{minipage}{0.32\textwidth}
\includegraphics[width=\textwidth]{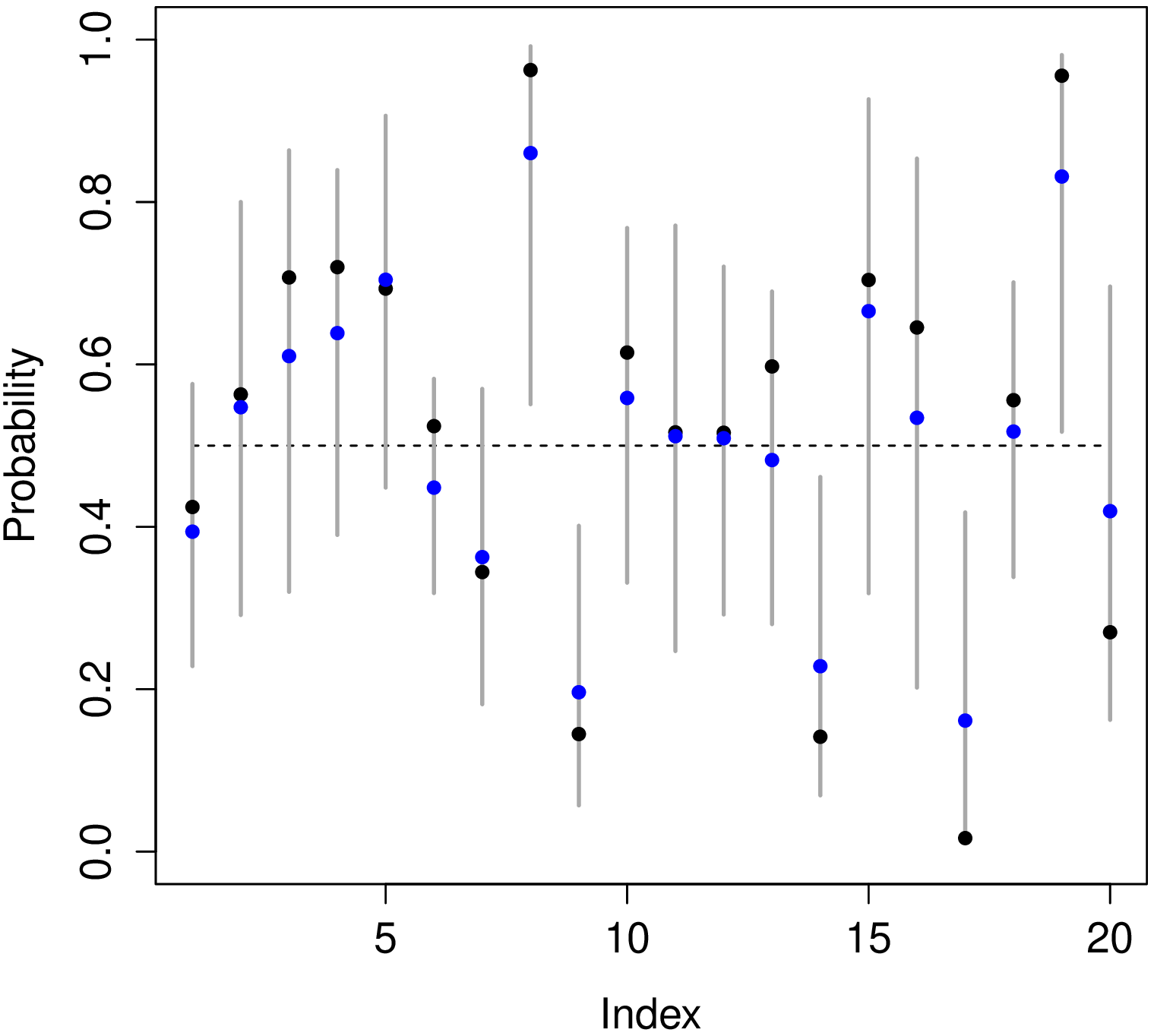}
\includegraphics[width=\textwidth]{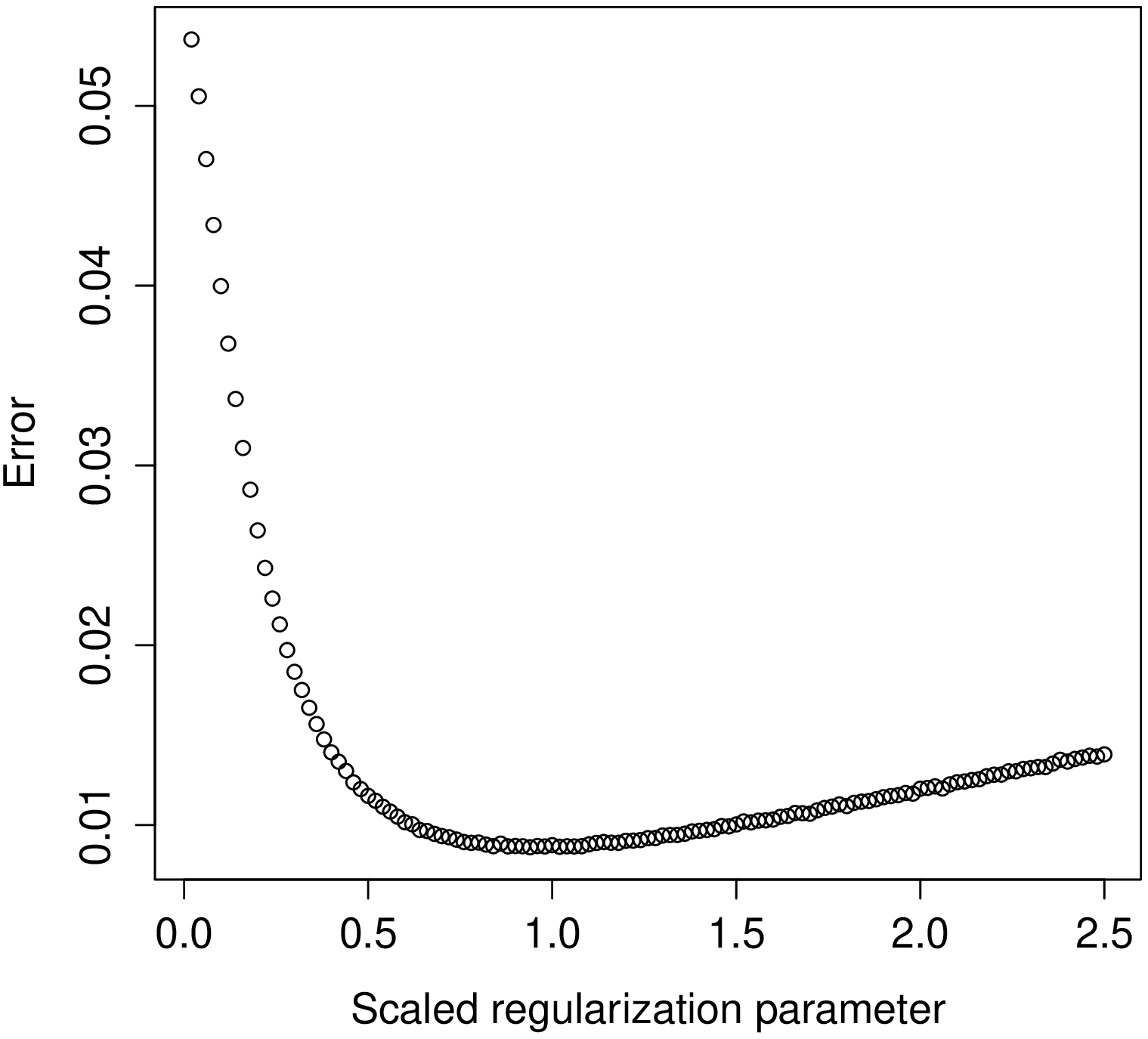}
\end{minipage}
\caption{Same as Figure \ref{fig: sw1}, but now with $\alpha=0.4$.}
\label{fig: a1}
\end{figure}

\begin{figure}[H]
\begin{minipage}{0.32\textwidth}
\includegraphics[width=\textwidth]{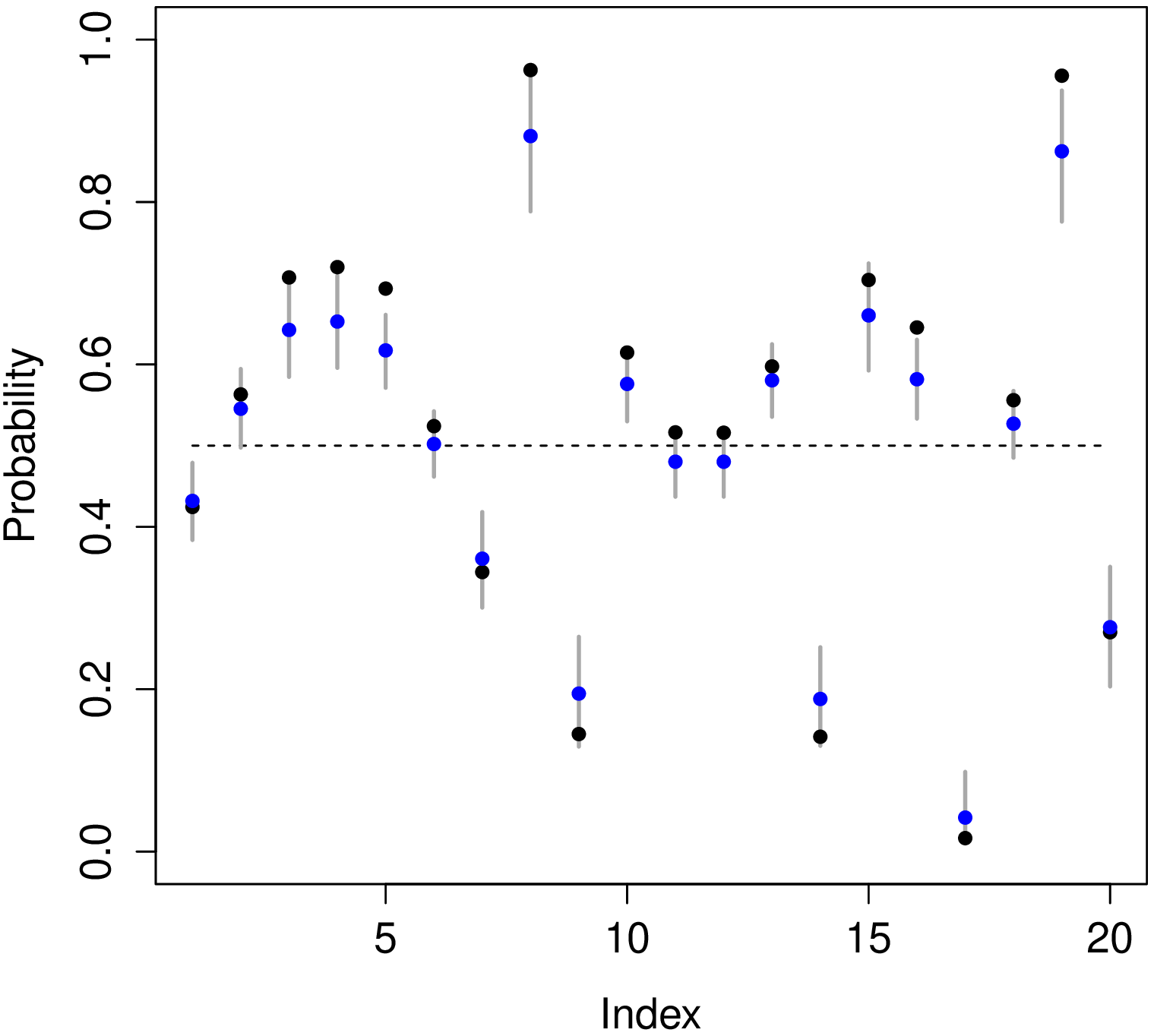}
\includegraphics[width=\textwidth]{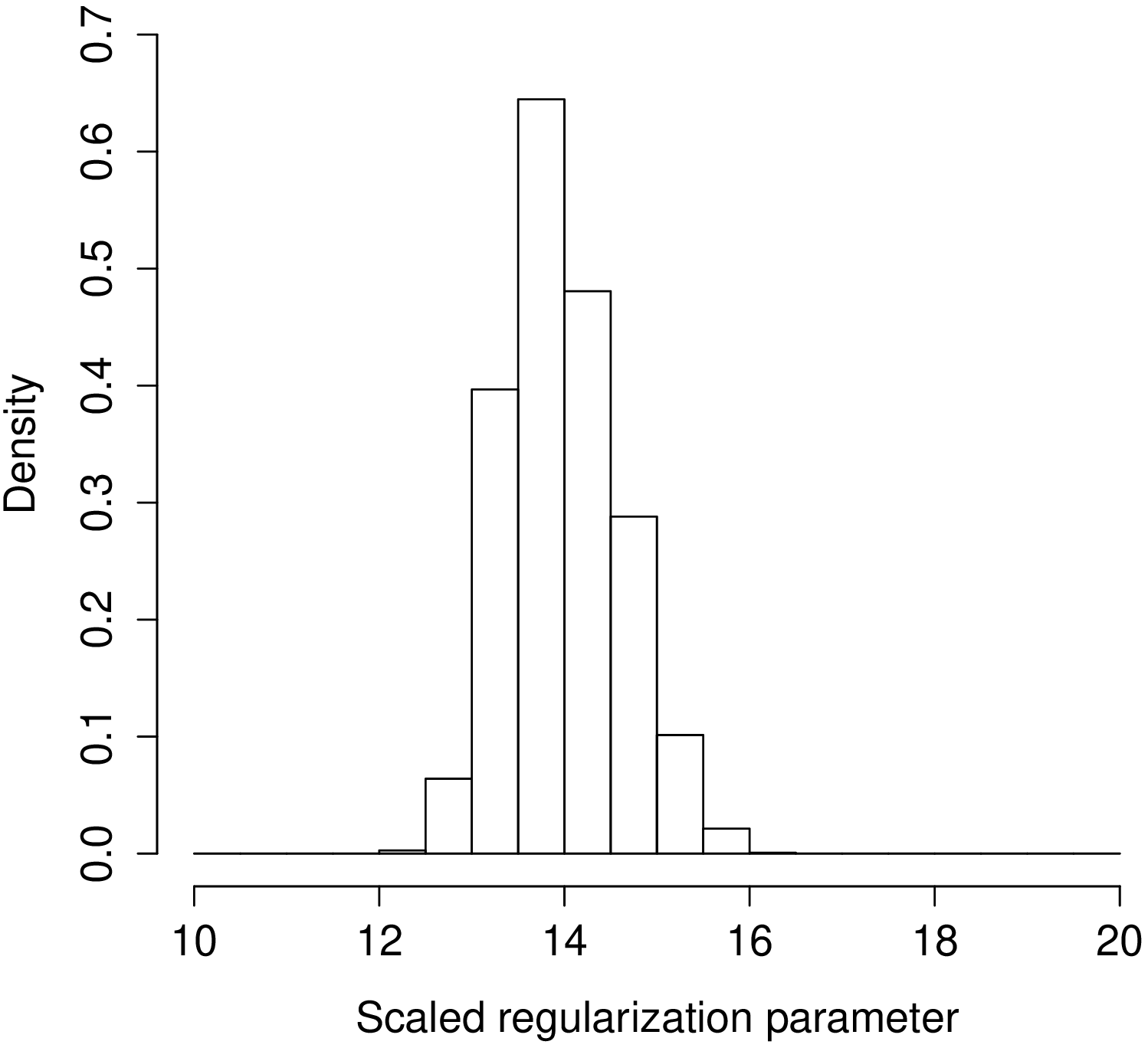}
\end{minipage}
\begin{minipage}{0.32\textwidth}
\includegraphics[width=\textwidth]{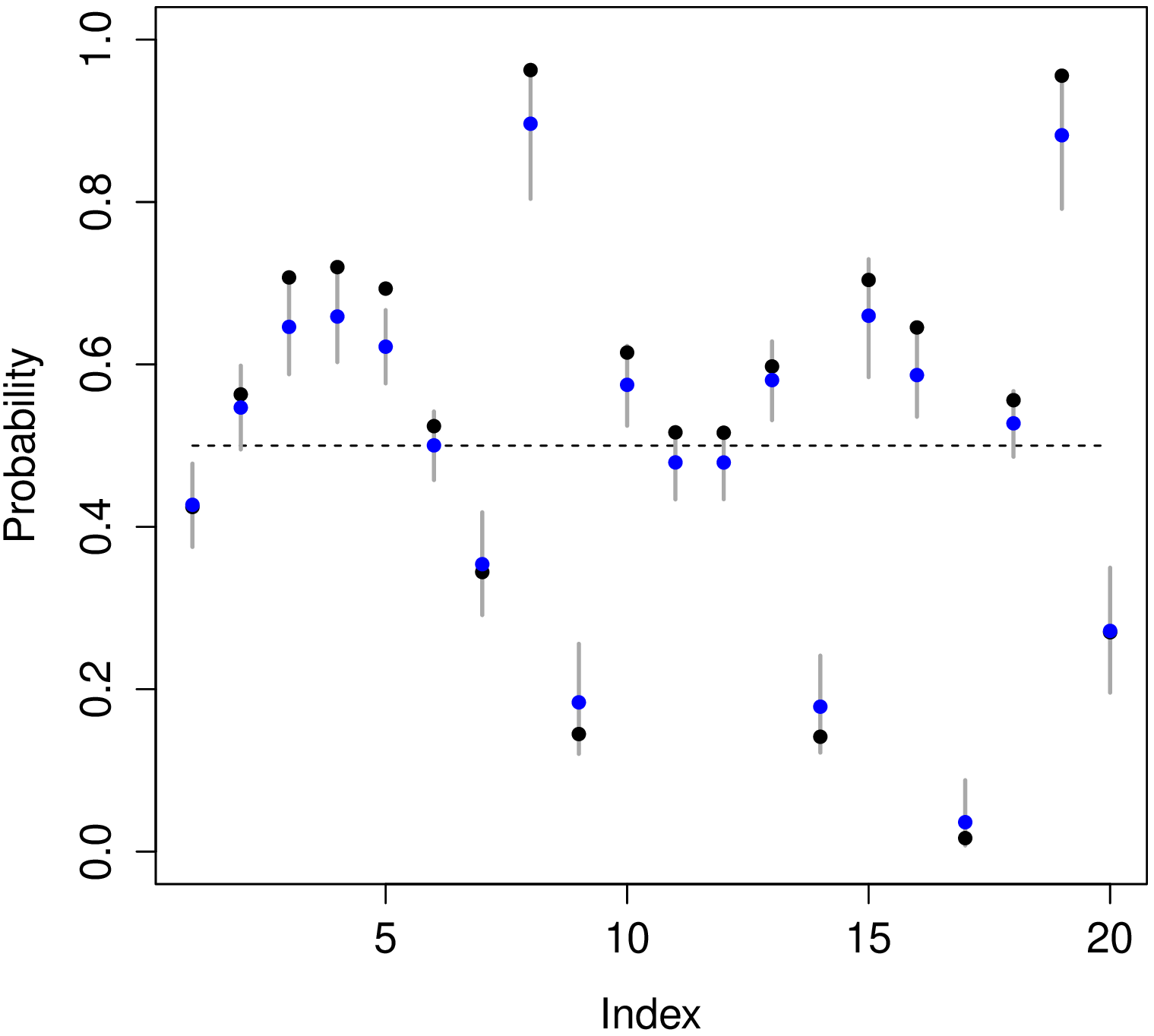}
\includegraphics[width=\textwidth]{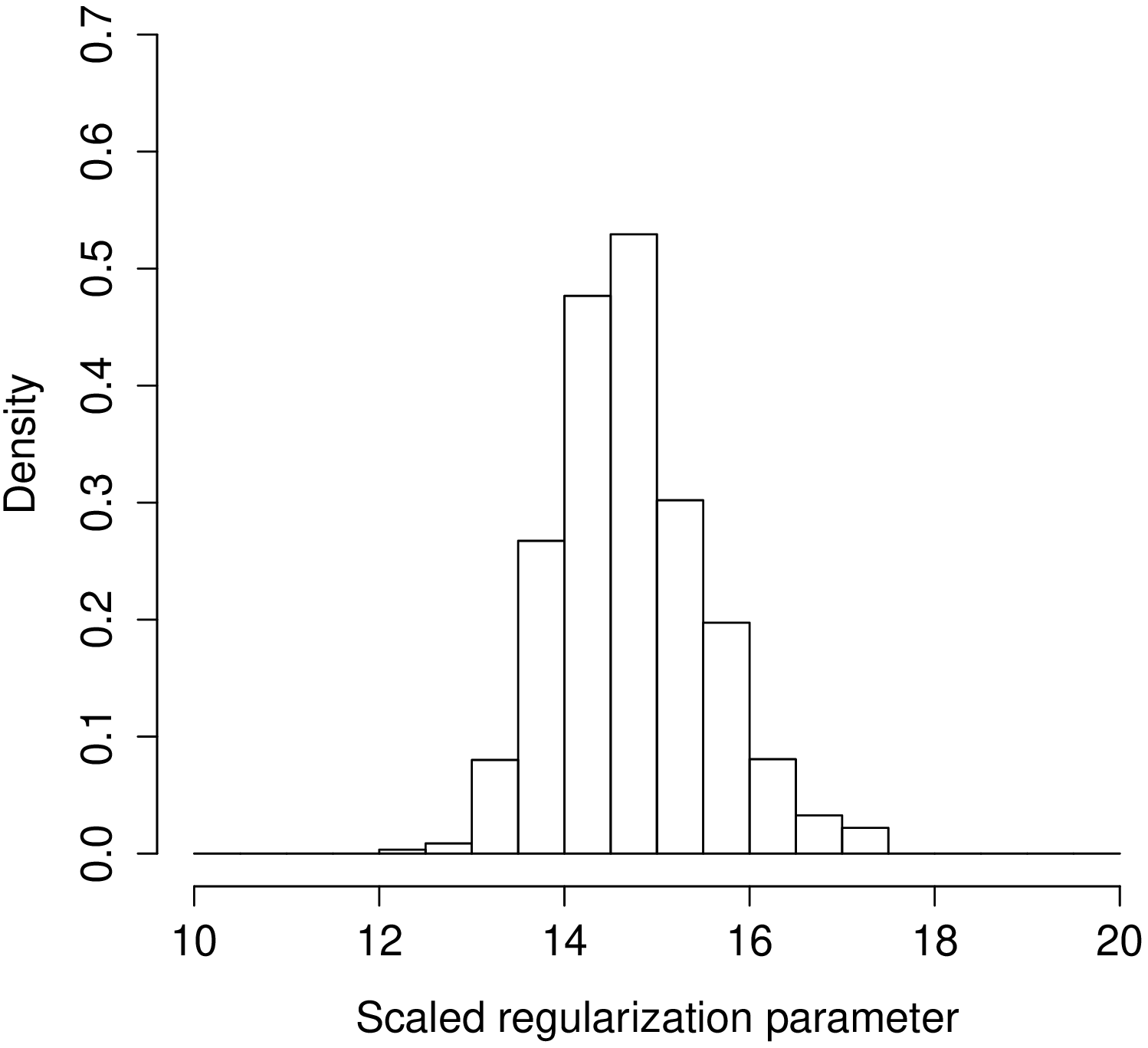}
\end{minipage}
\begin{minipage}{0.32\textwidth}
\includegraphics[width=\textwidth]{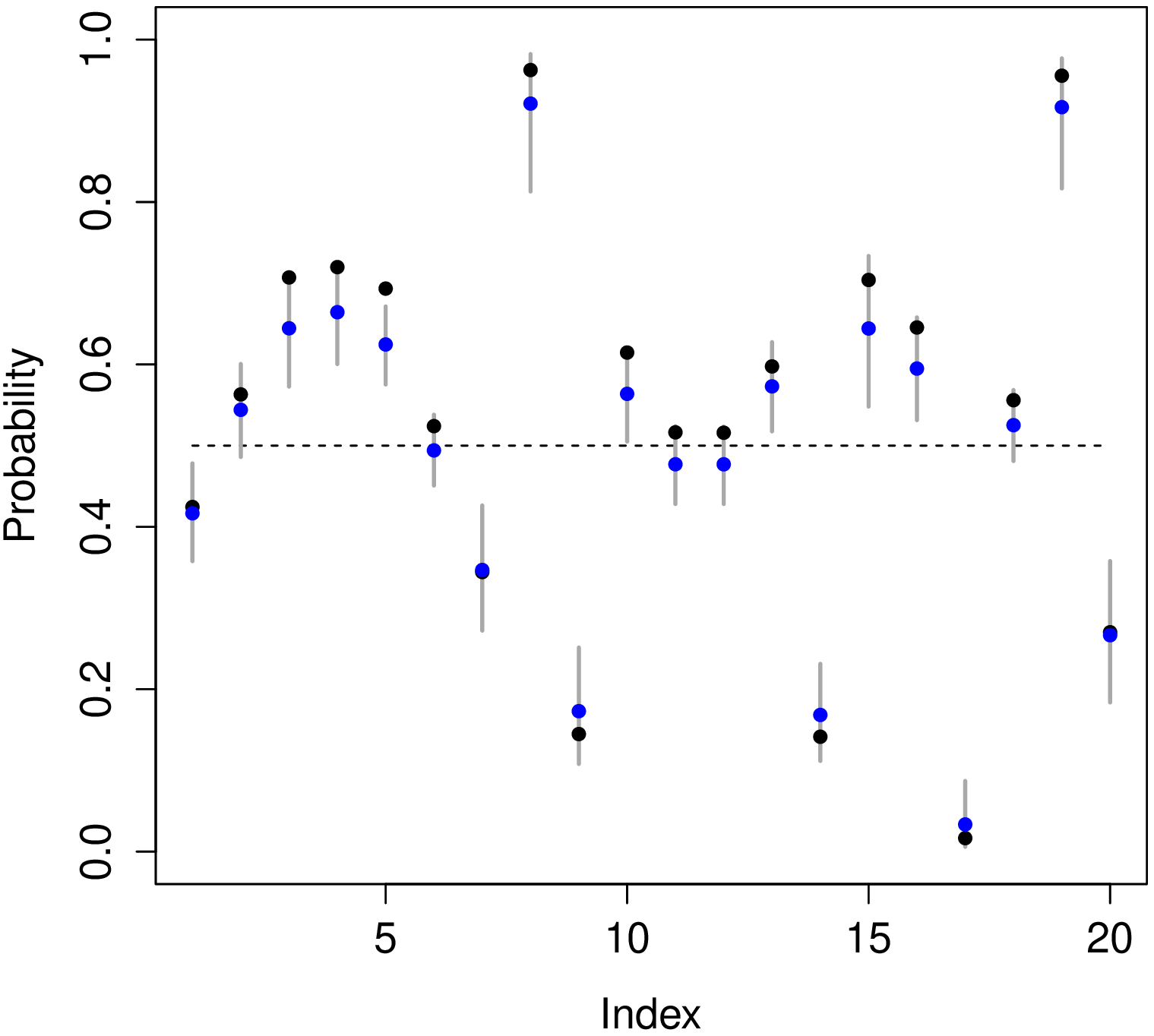}
\includegraphics[width=\textwidth]{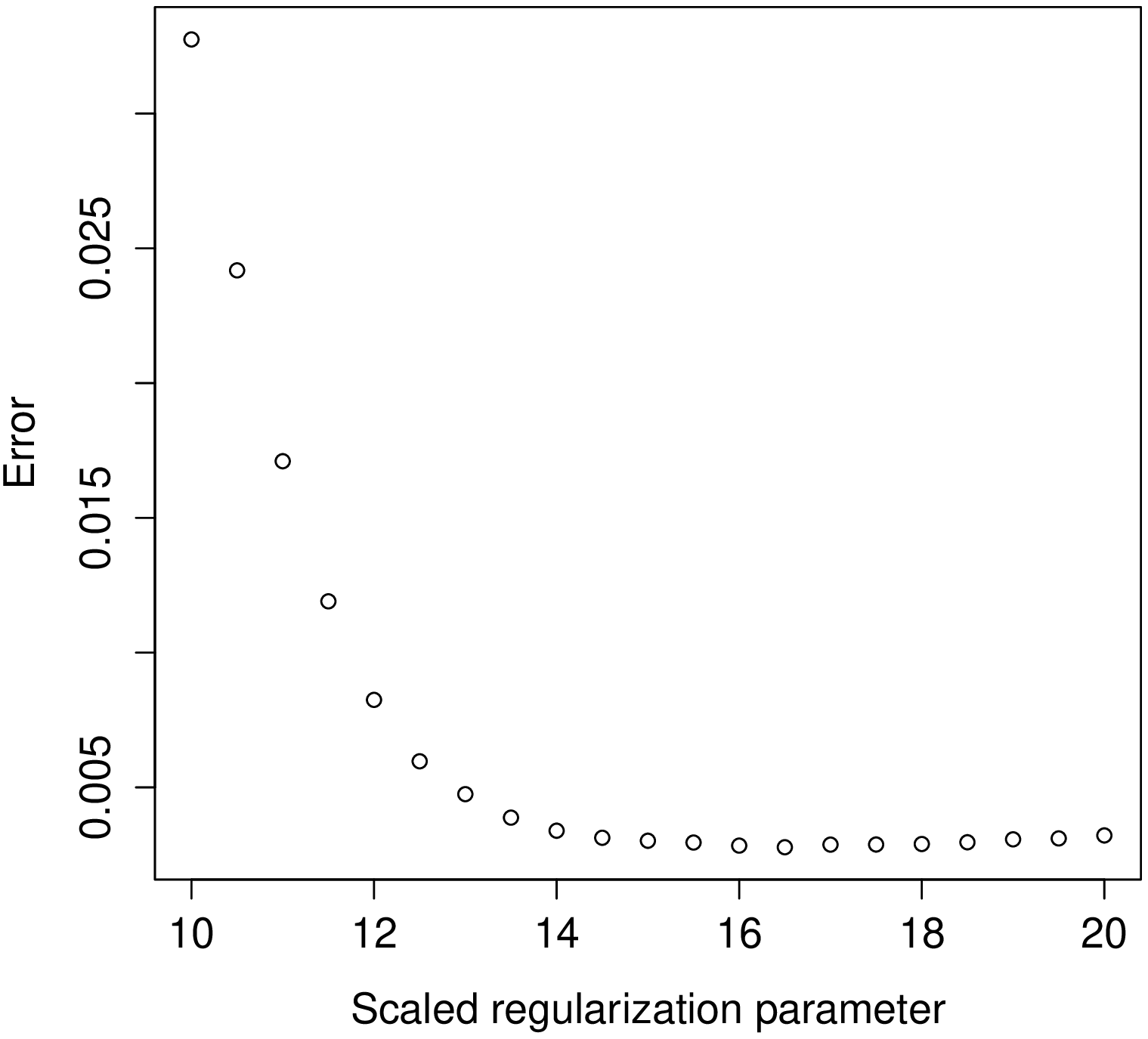}
\end{minipage}
\caption{Same as Figure \ref{fig: sw1}, but now with $\alpha=10$.}
\label{fig: a2}
\end{figure}

If we choose the parameter $r$ correctly, i.e.\ $r = 1.7$ is this case, and 
only choose different values for the hyper parameter $\alpha$ we only change 
the prior smoothness of $f$, without changing the prior on $c$. Figures 
\ref{fig: a1} and \ref{fig: a2} illustrate the effect. In the first 
we set $\alpha = 0.4$, which is too low relative to the smoothness
of the true soft label function, in the second one $\alpha = 10$, 
which is too high. We clearly see the effect on the width of the credible 
intervals. The effect on coverage is not very large.

\subsubsection{Impact of missing observations}

\begin{figure}[H]
\begin{minipage}{0.32\textwidth}
\includegraphics[width=\textwidth]{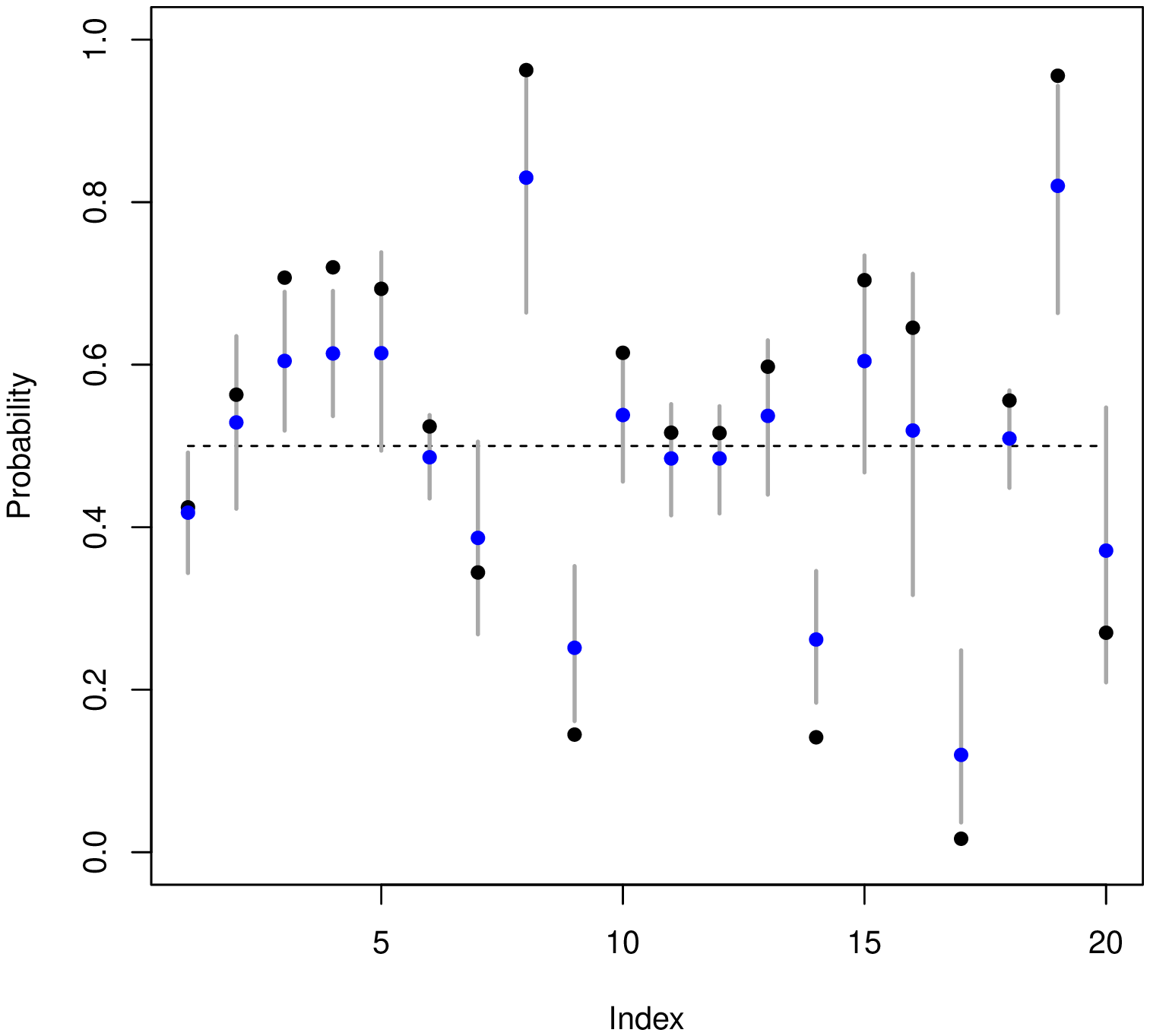}
\includegraphics[width=\textwidth]{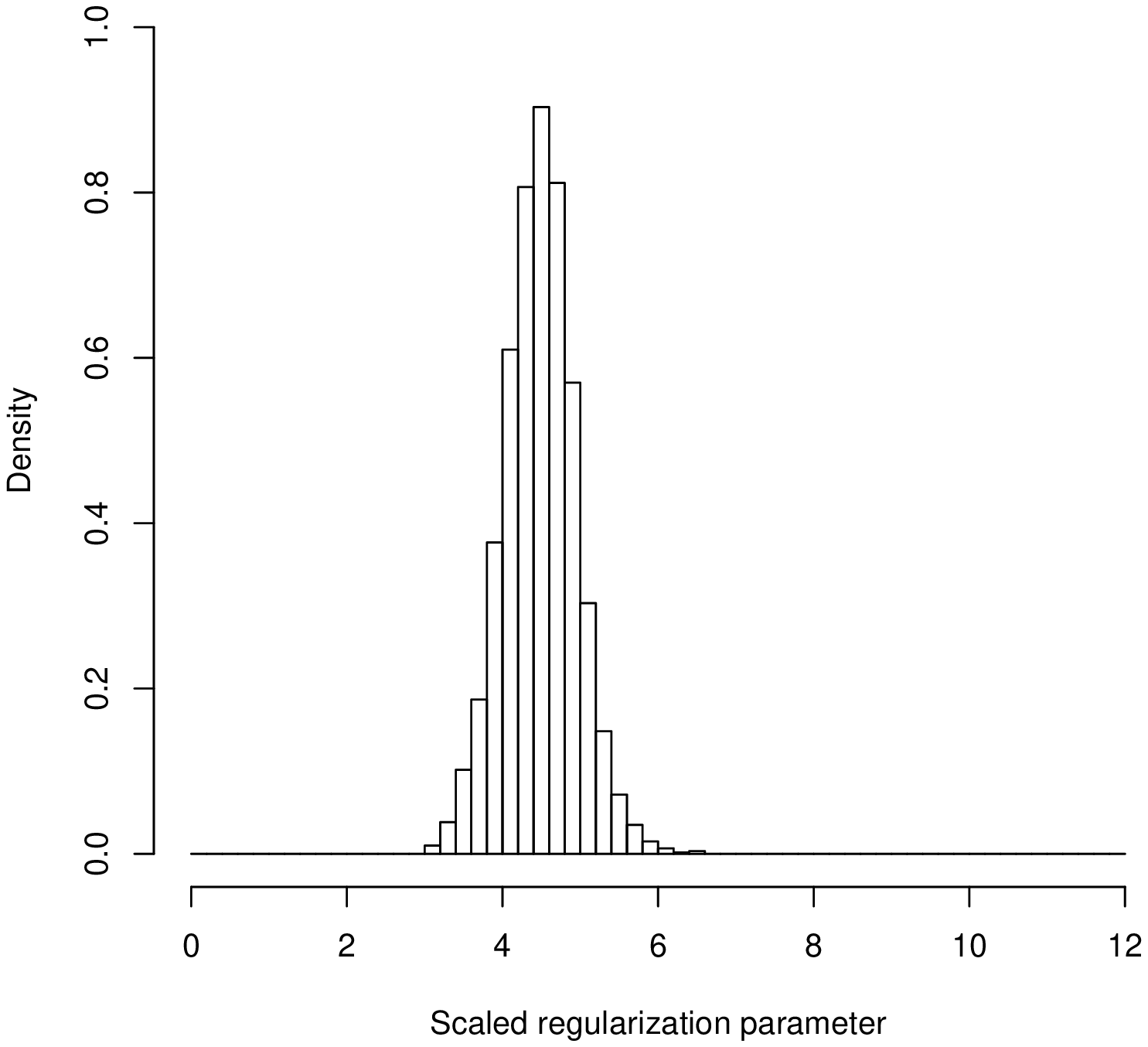}
\end{minipage}
\begin{minipage}{0.32\textwidth}
\includegraphics[width=\textwidth]{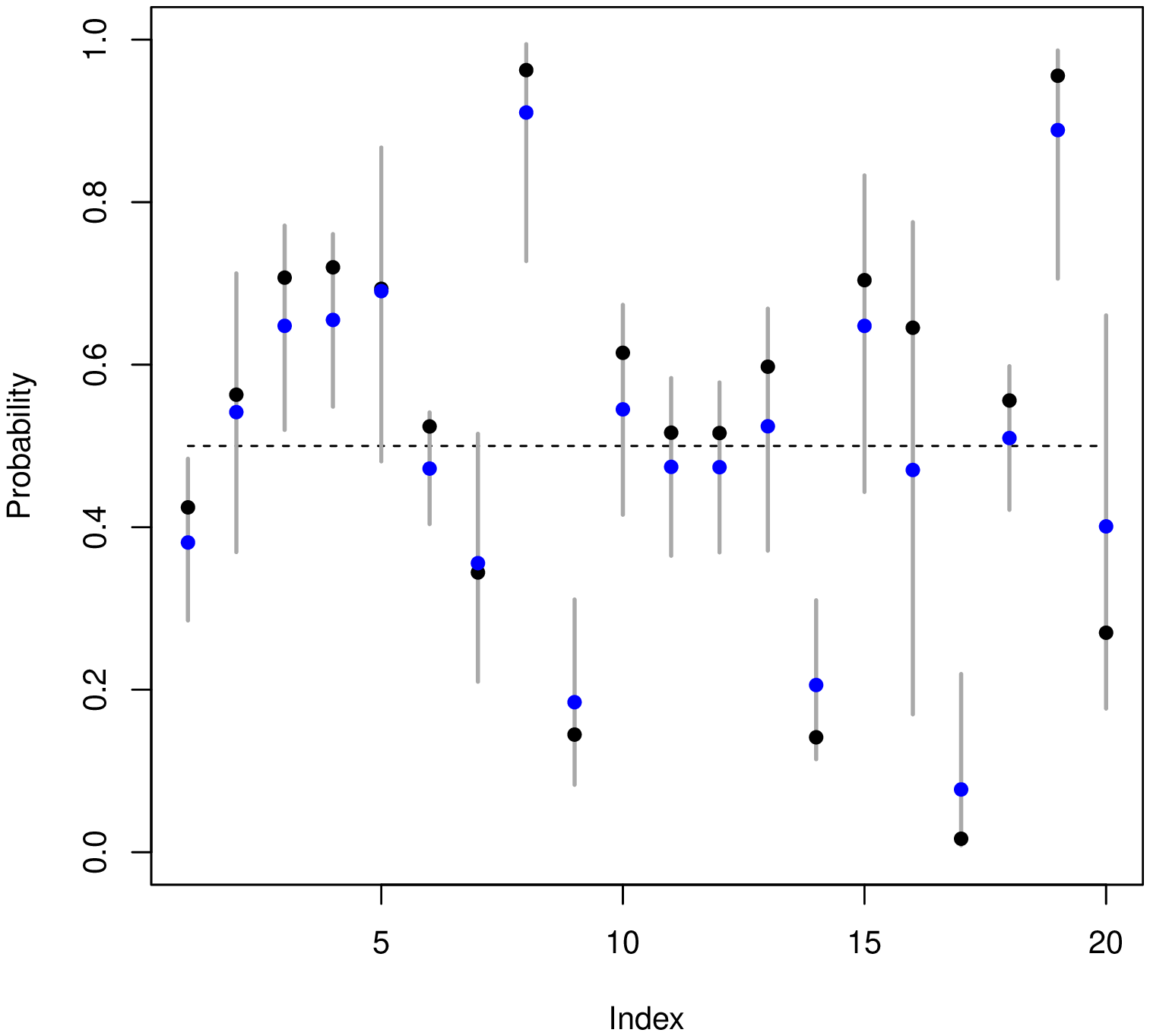}
\includegraphics[width=\textwidth]{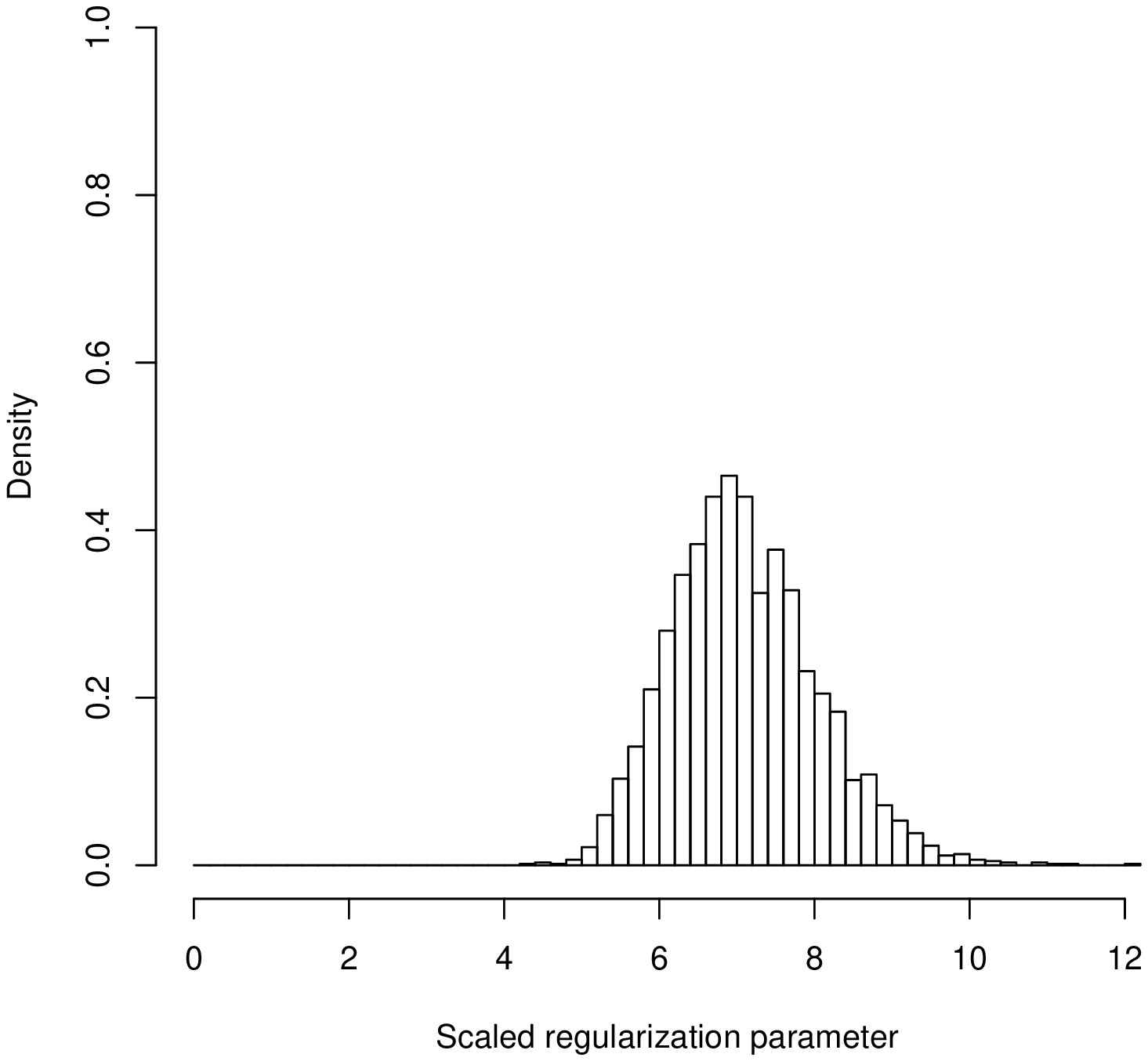}
\end{minipage}
\begin{minipage}{0.32\textwidth}
\includegraphics[width=\textwidth]{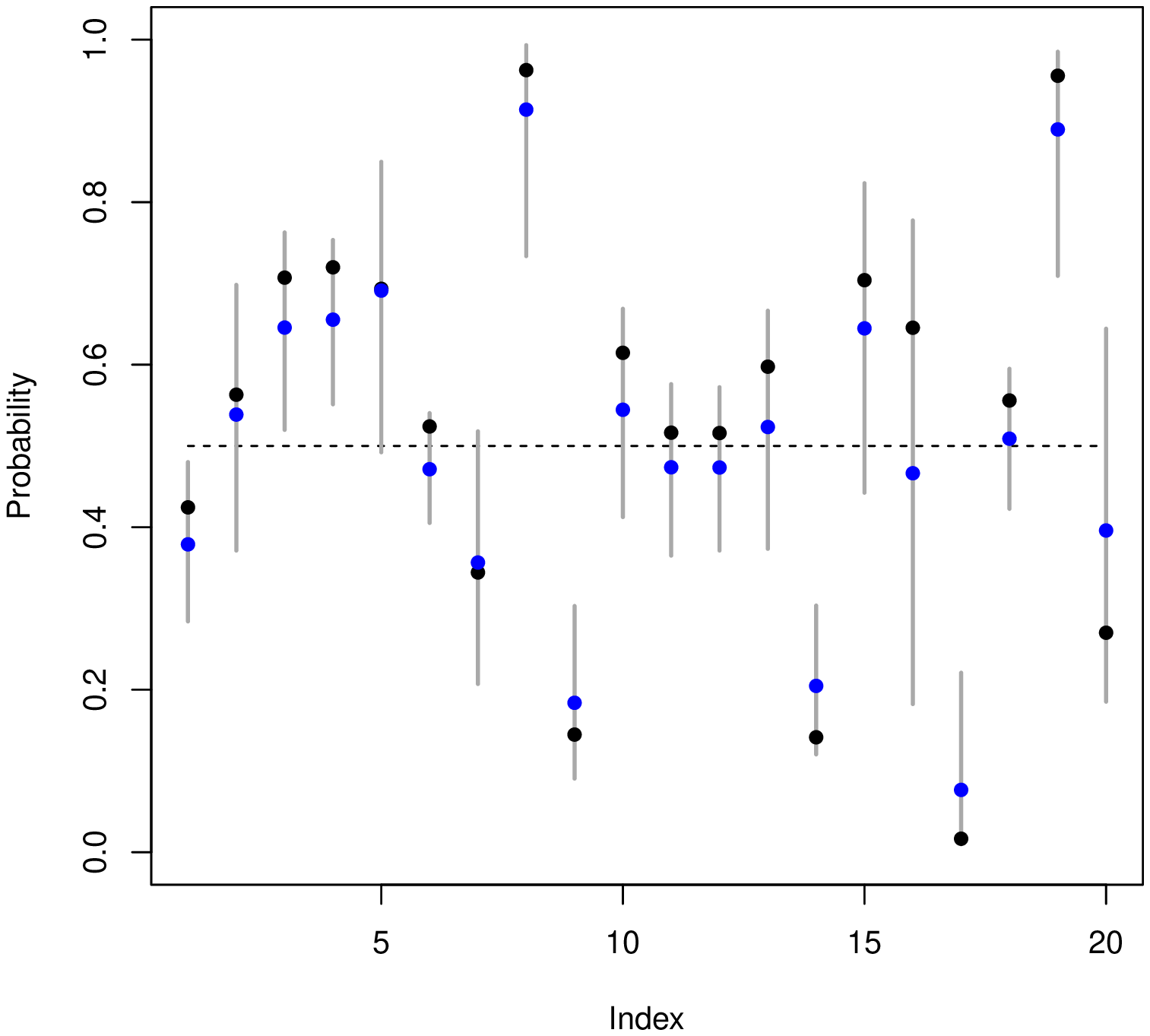}
\includegraphics[width=\textwidth]{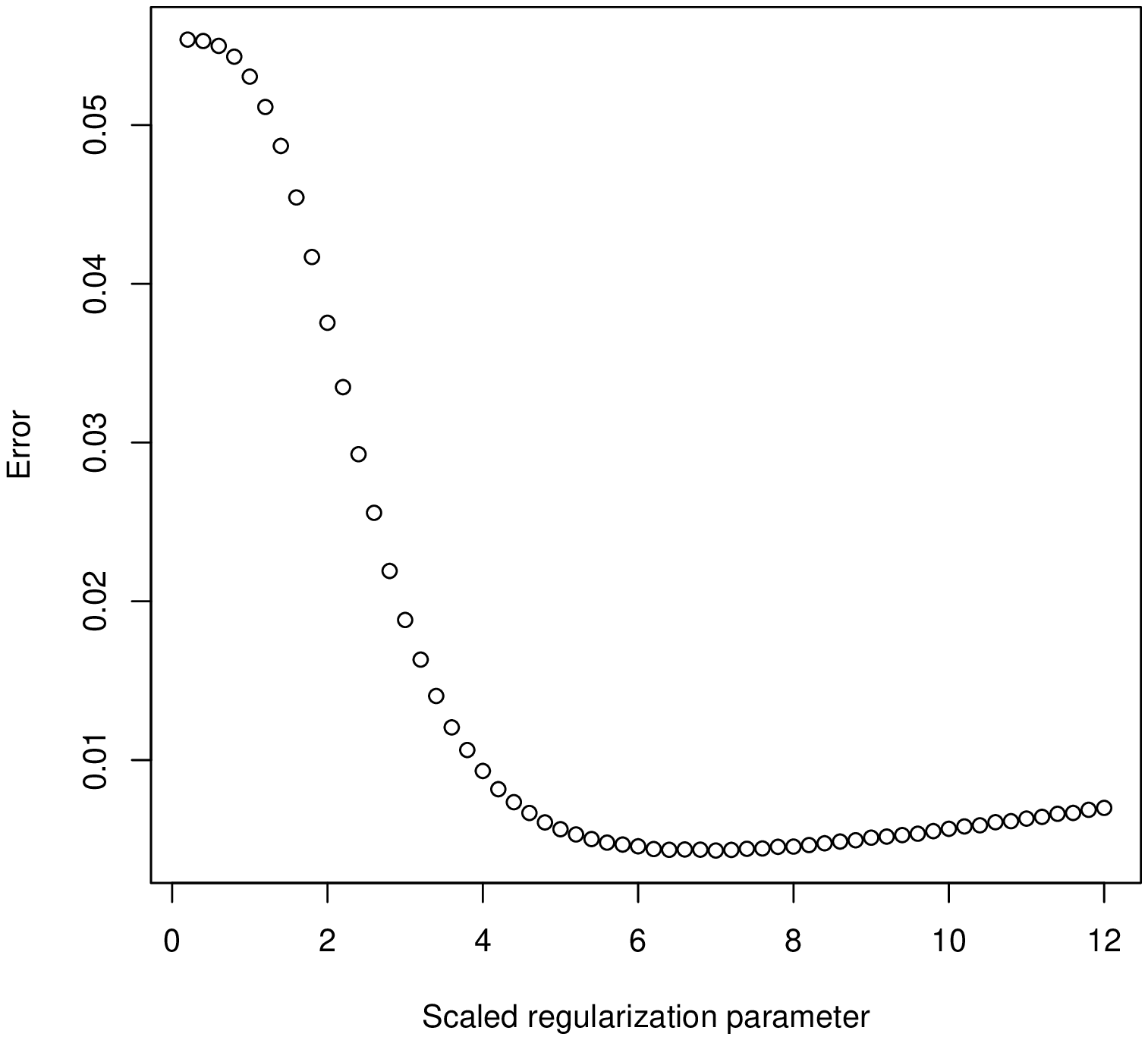}
\end{minipage}
\caption{Same as Figure \ref{fig: sw1}, but now with $20\%$ of the label unobserved.}
\label{fig: miss1}
\end{figure}

\begin{figure}[H]
\begin{minipage}{0.32\textwidth}
\includegraphics[width=\textwidth]{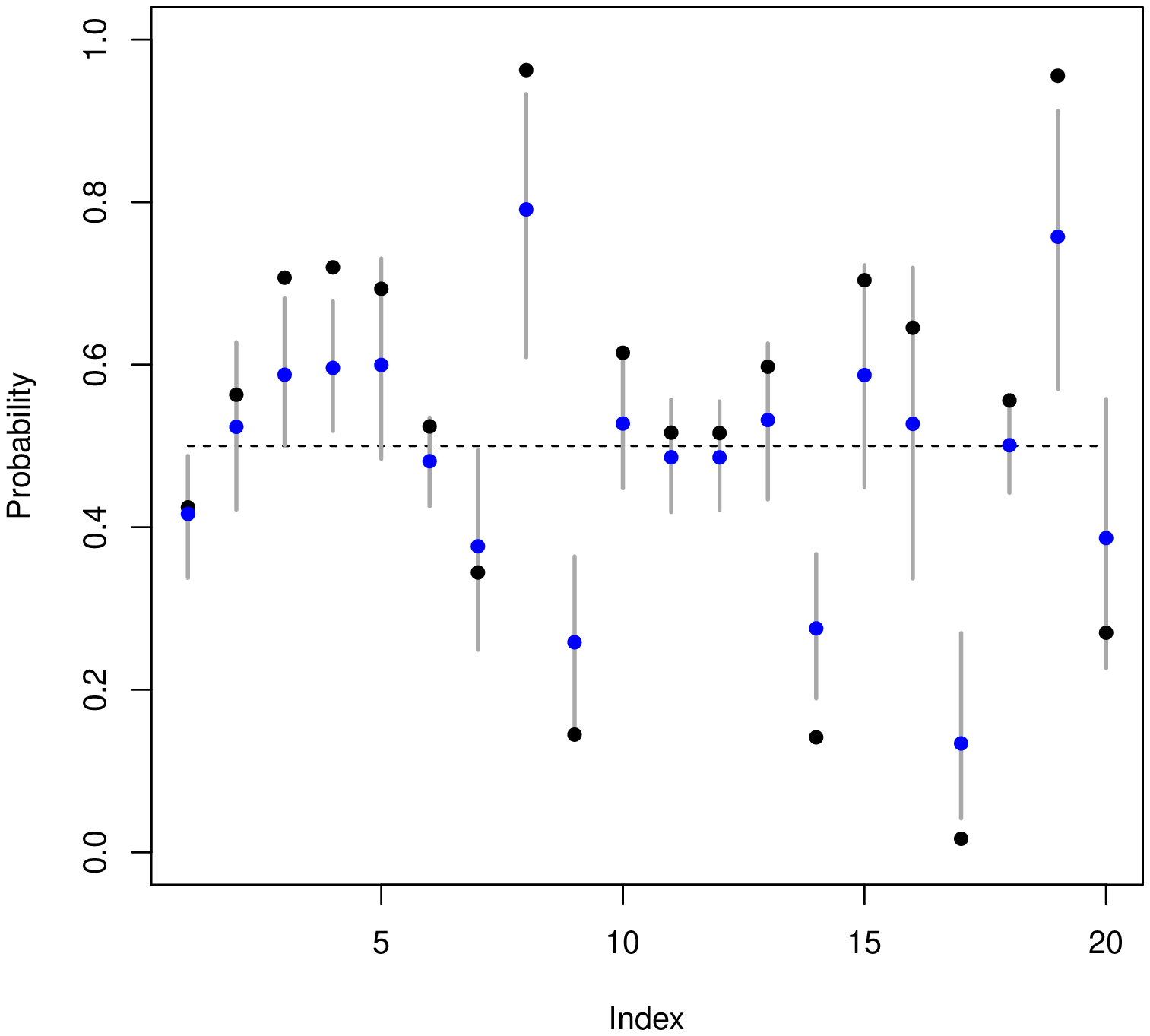}
\includegraphics[width=\textwidth]{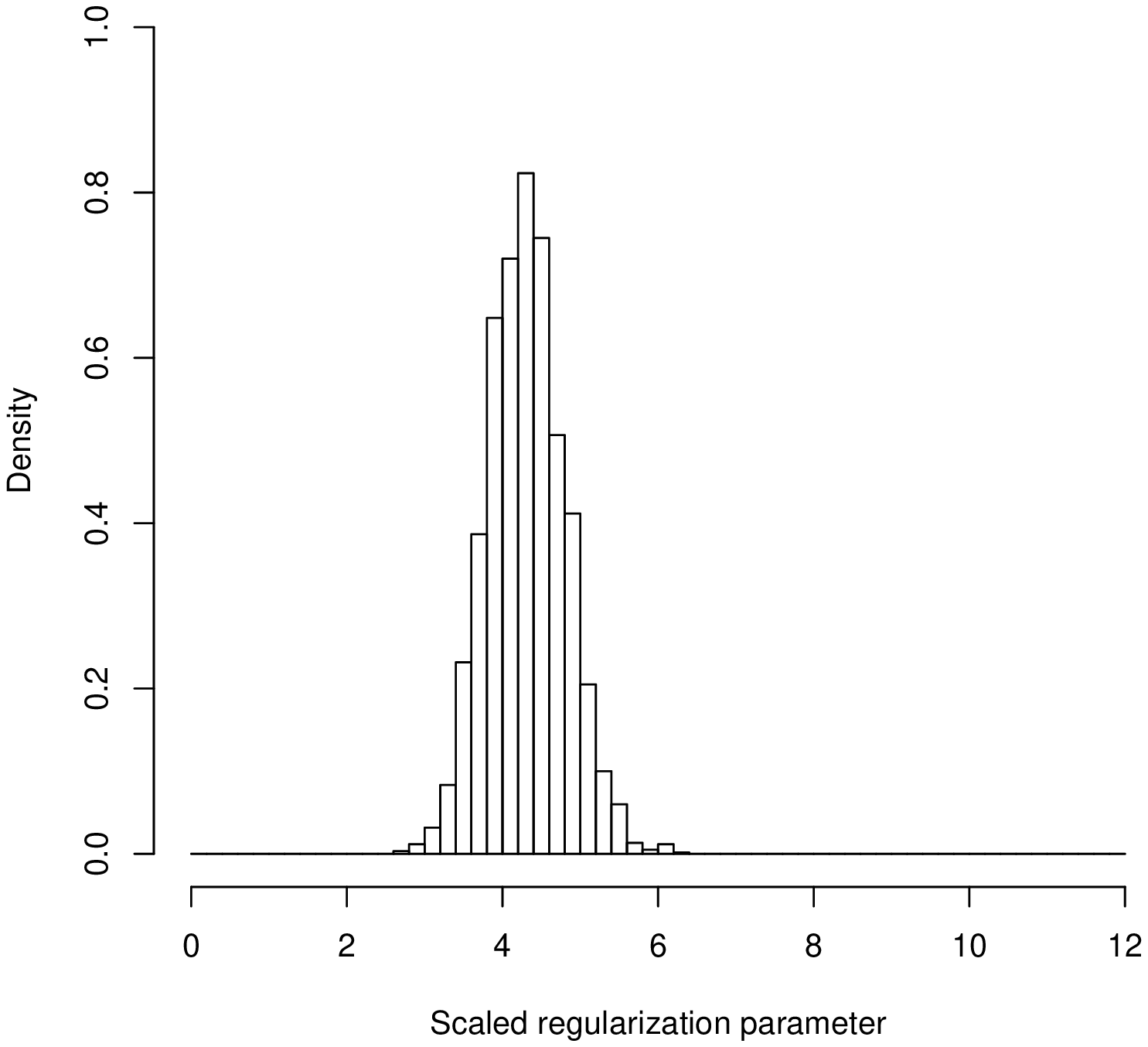}
\end{minipage}
\begin{minipage}{0.32\textwidth}
\includegraphics[width=\textwidth]{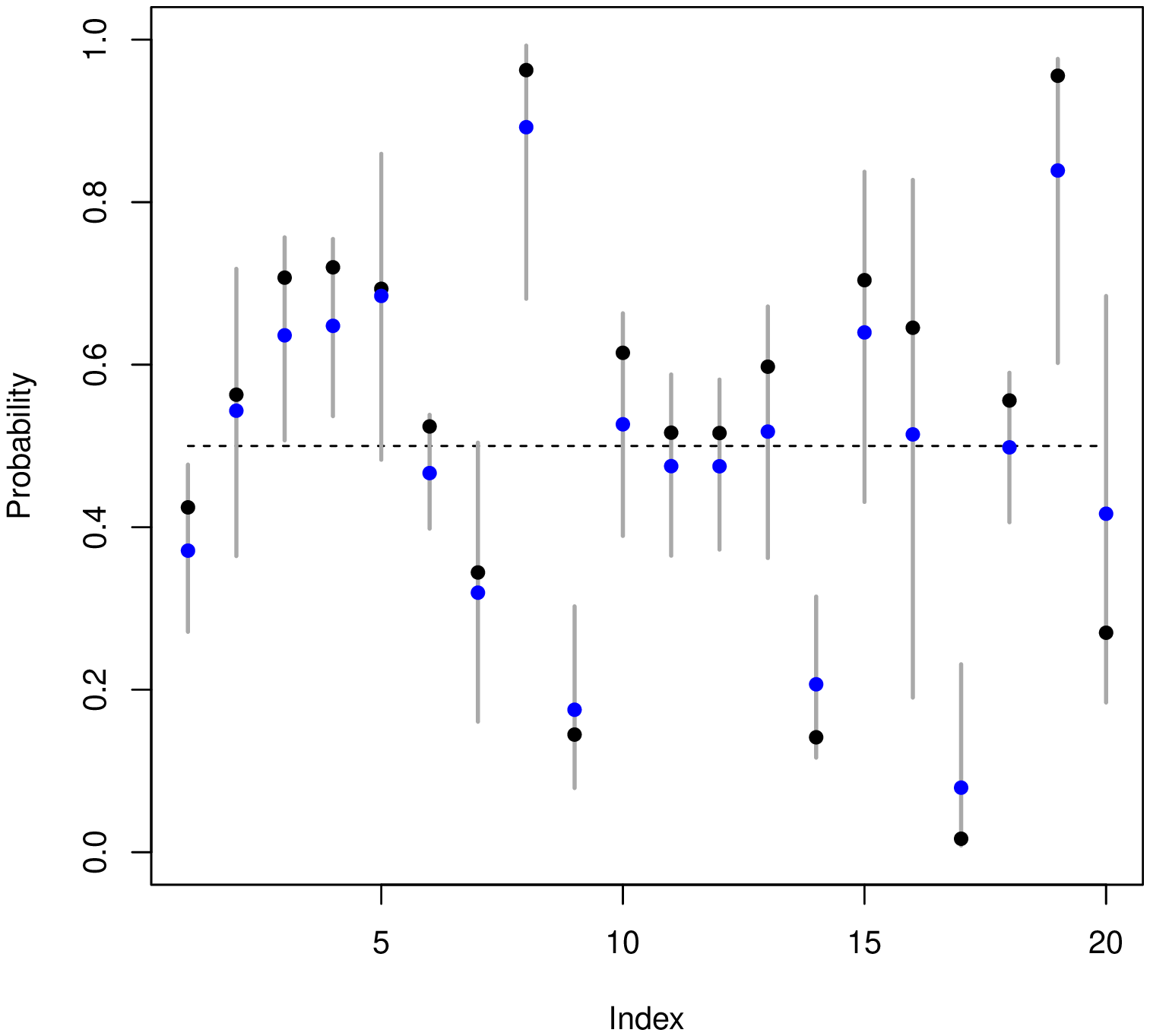}
\includegraphics[width=\textwidth]{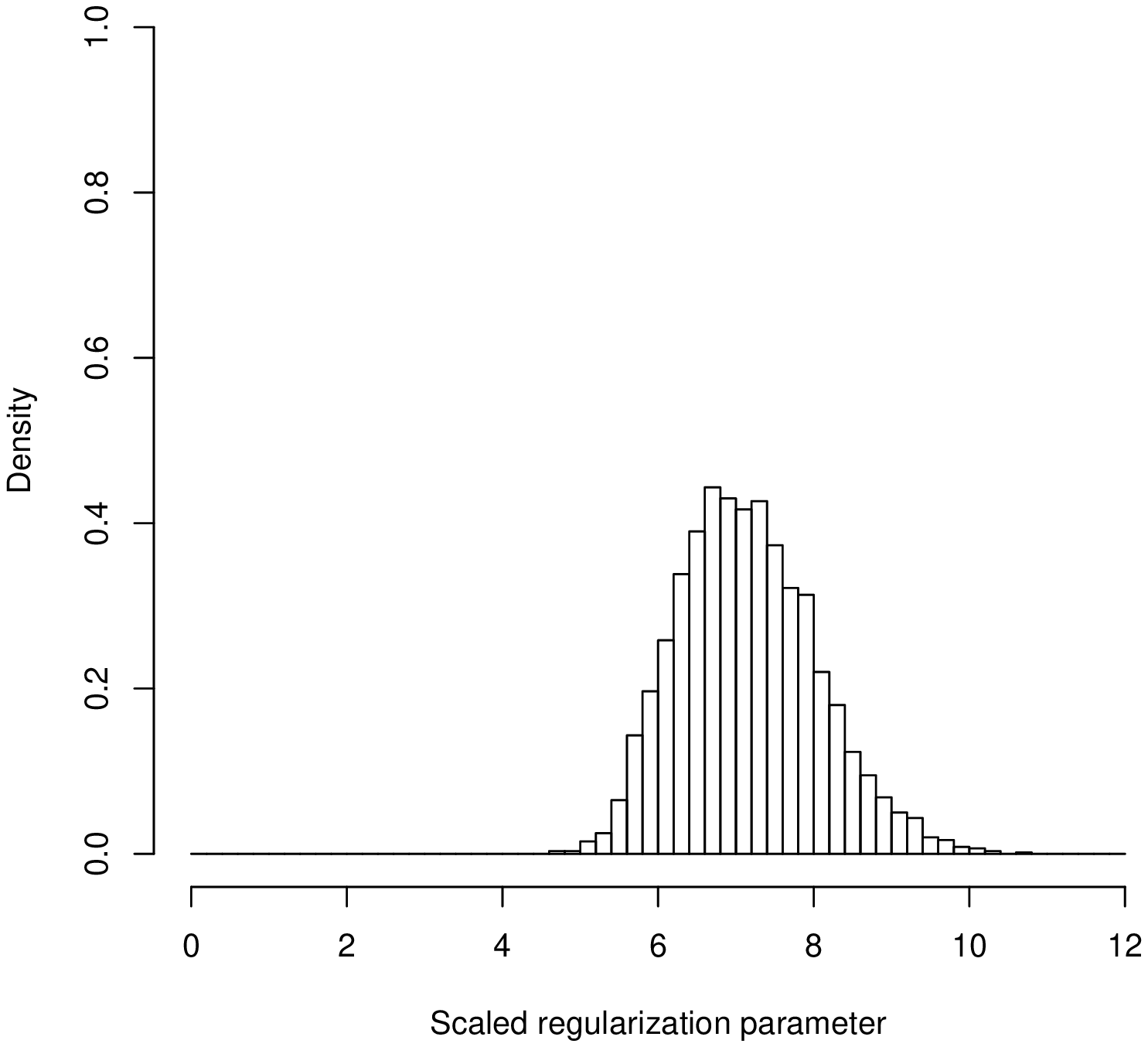}
\end{minipage}
\begin{minipage}{0.32\textwidth}
\includegraphics[width=\textwidth]{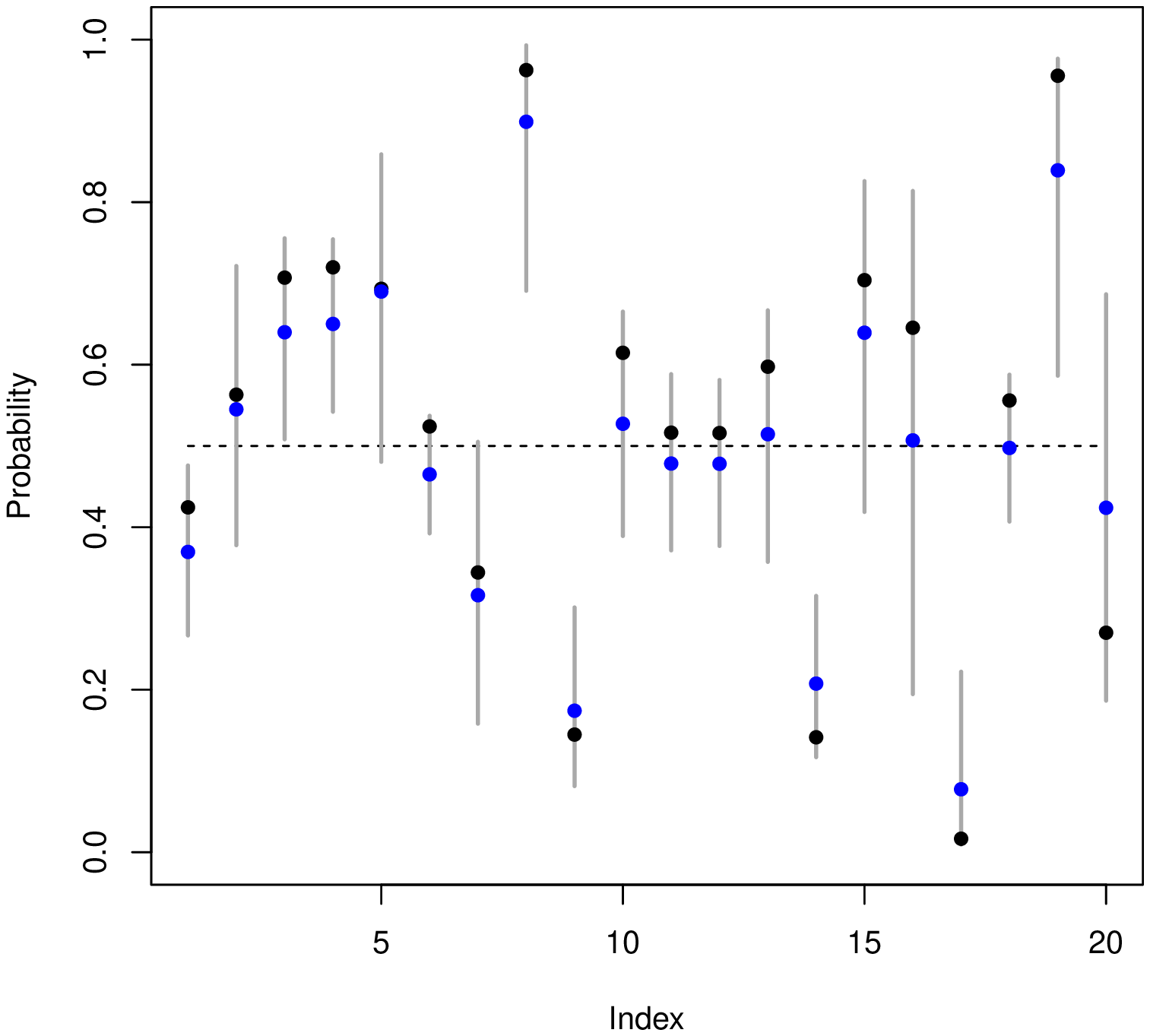}
\includegraphics[width=\textwidth]{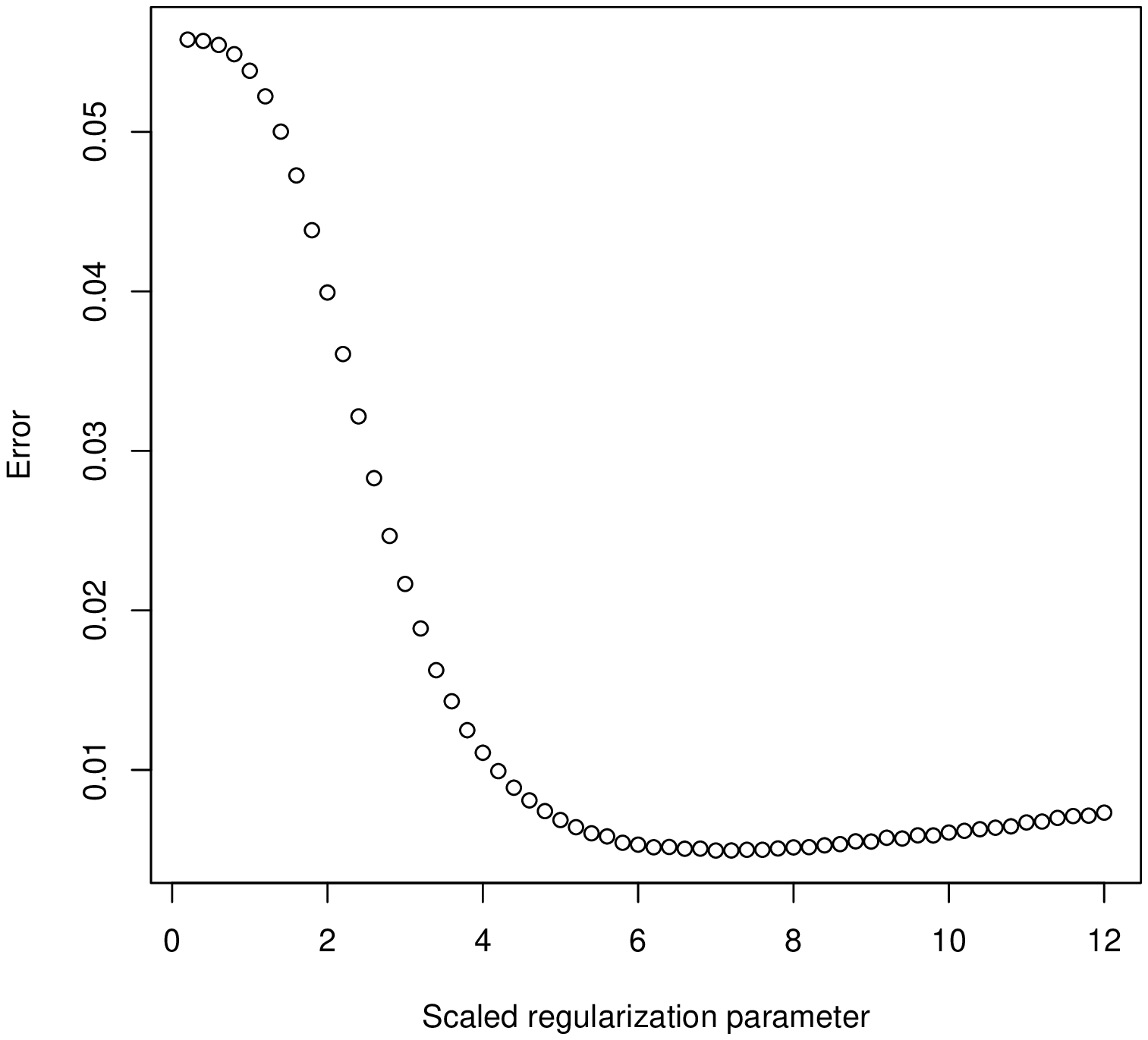}
\end{minipage}
\caption{Same as Figure \ref{fig: sw1}, but now with $30\%$ of the label unobserved.}
\label{fig: miss2}
\end{figure}

\begin{figure}[H]
\begin{minipage}{0.32\textwidth}
\includegraphics[width=\textwidth]{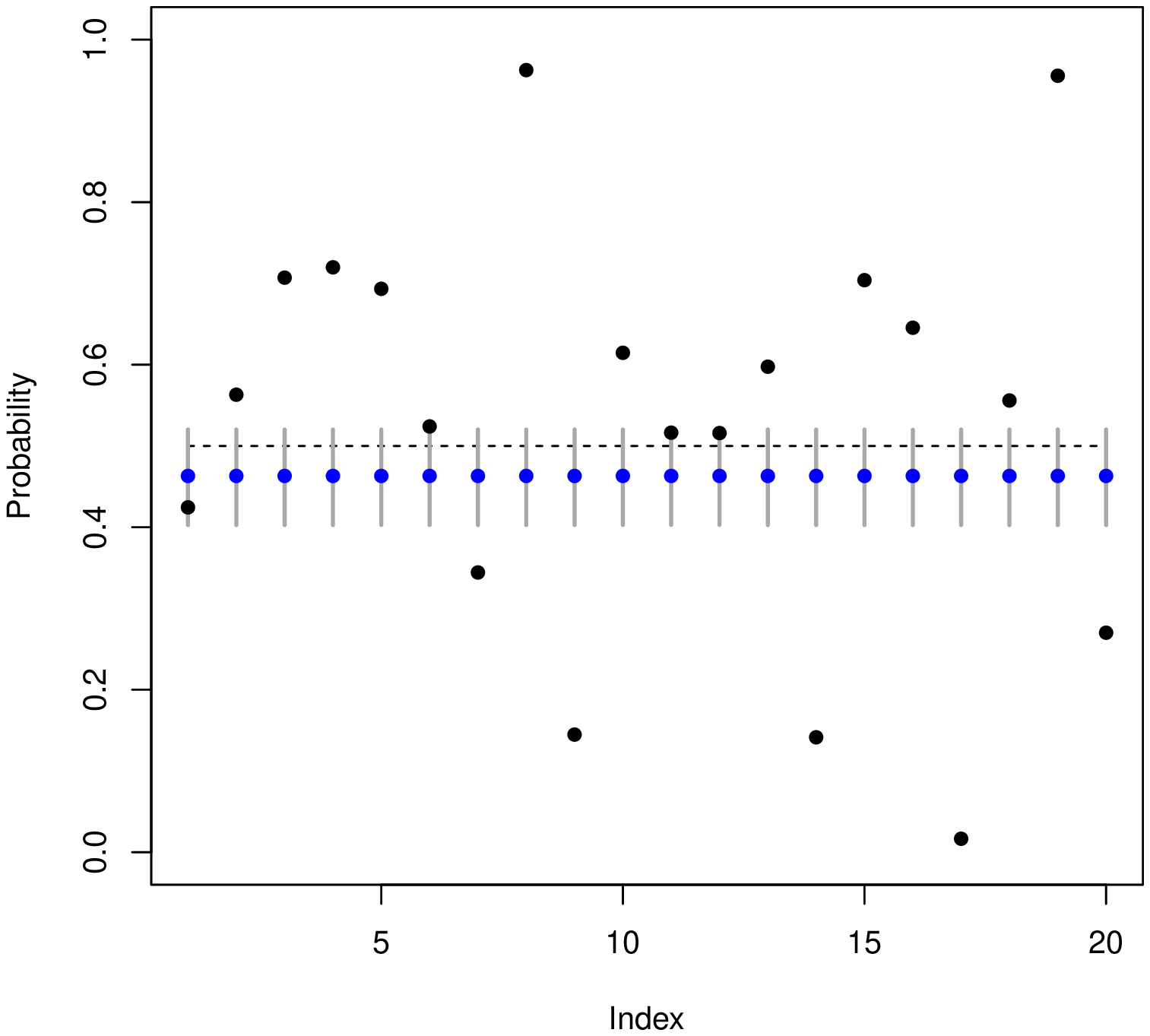}
\includegraphics[width=\textwidth]{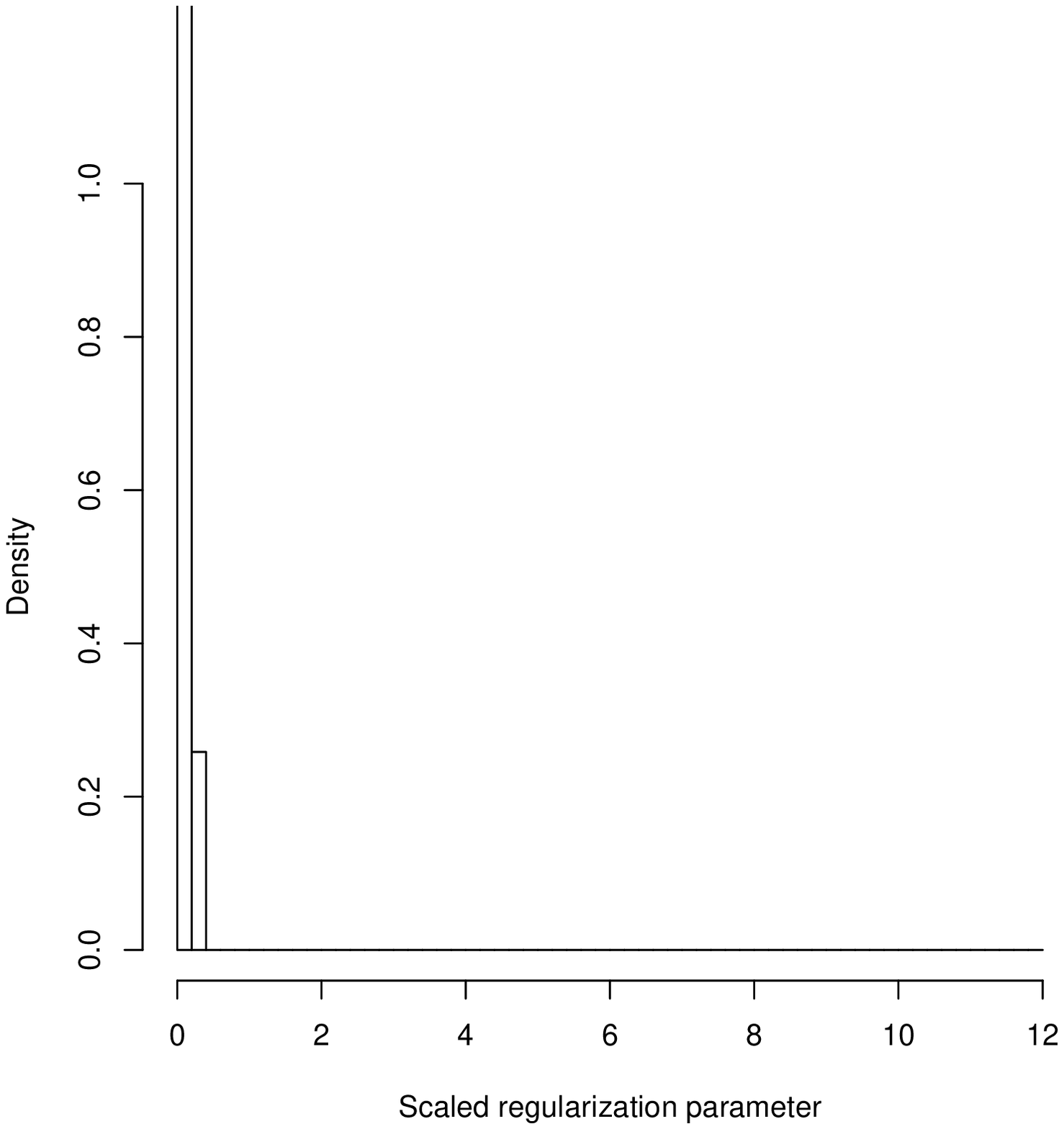}
\end{minipage}
\begin{minipage}{0.32\textwidth}
\includegraphics[width=\textwidth]{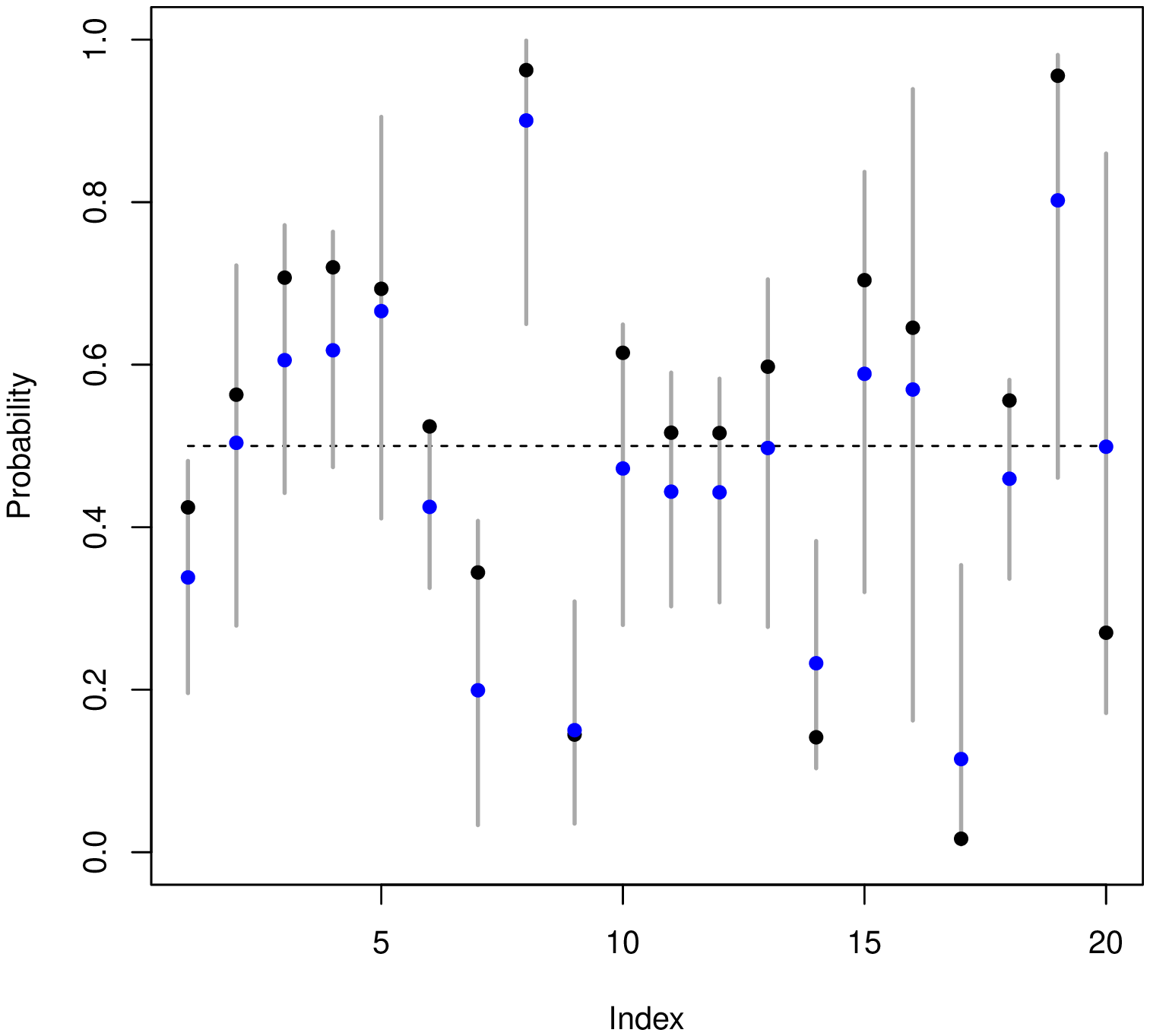}
\includegraphics[width=\textwidth]{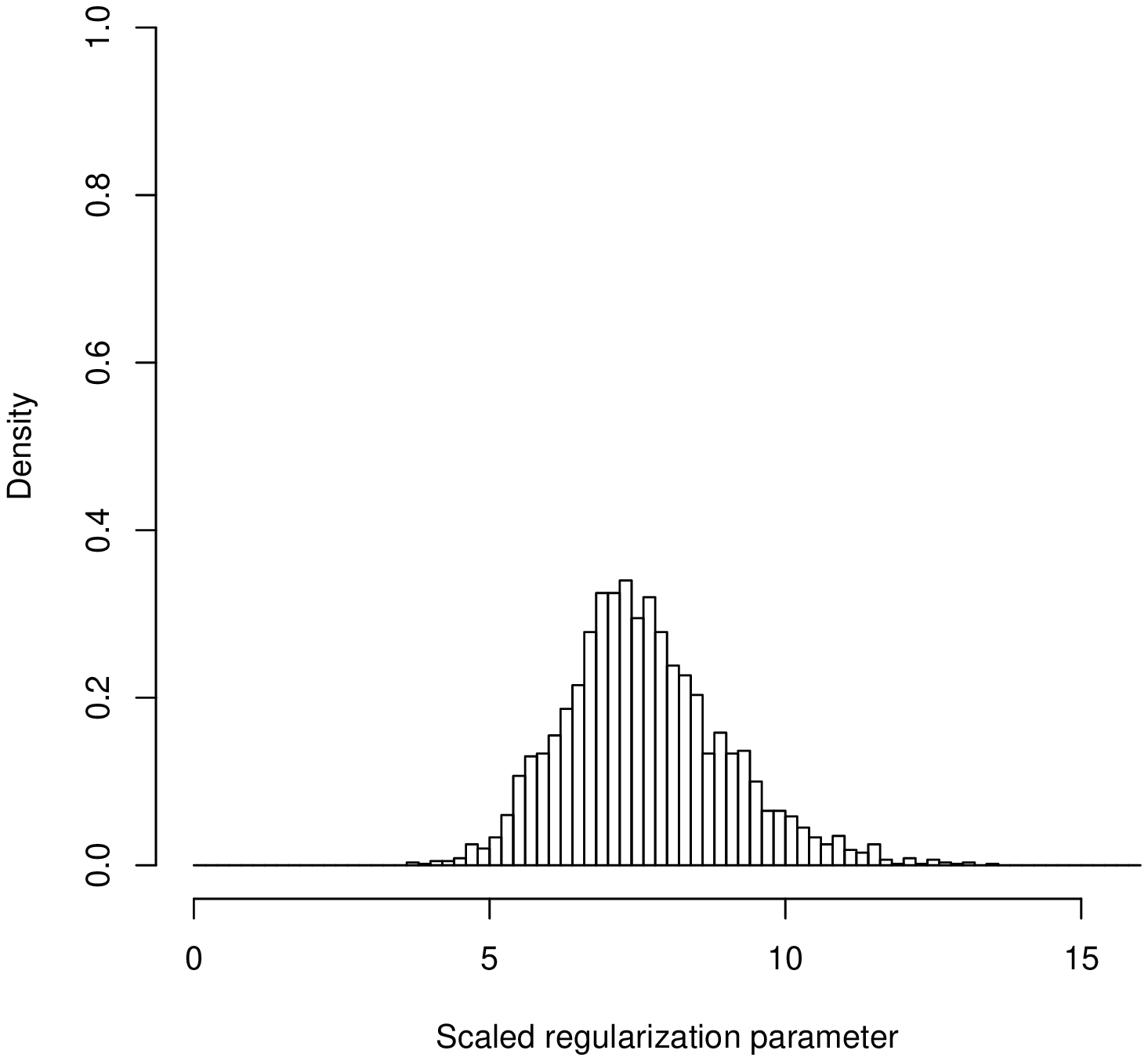}
\end{minipage}
\begin{minipage}{0.32\textwidth}
\includegraphics[width=\textwidth]{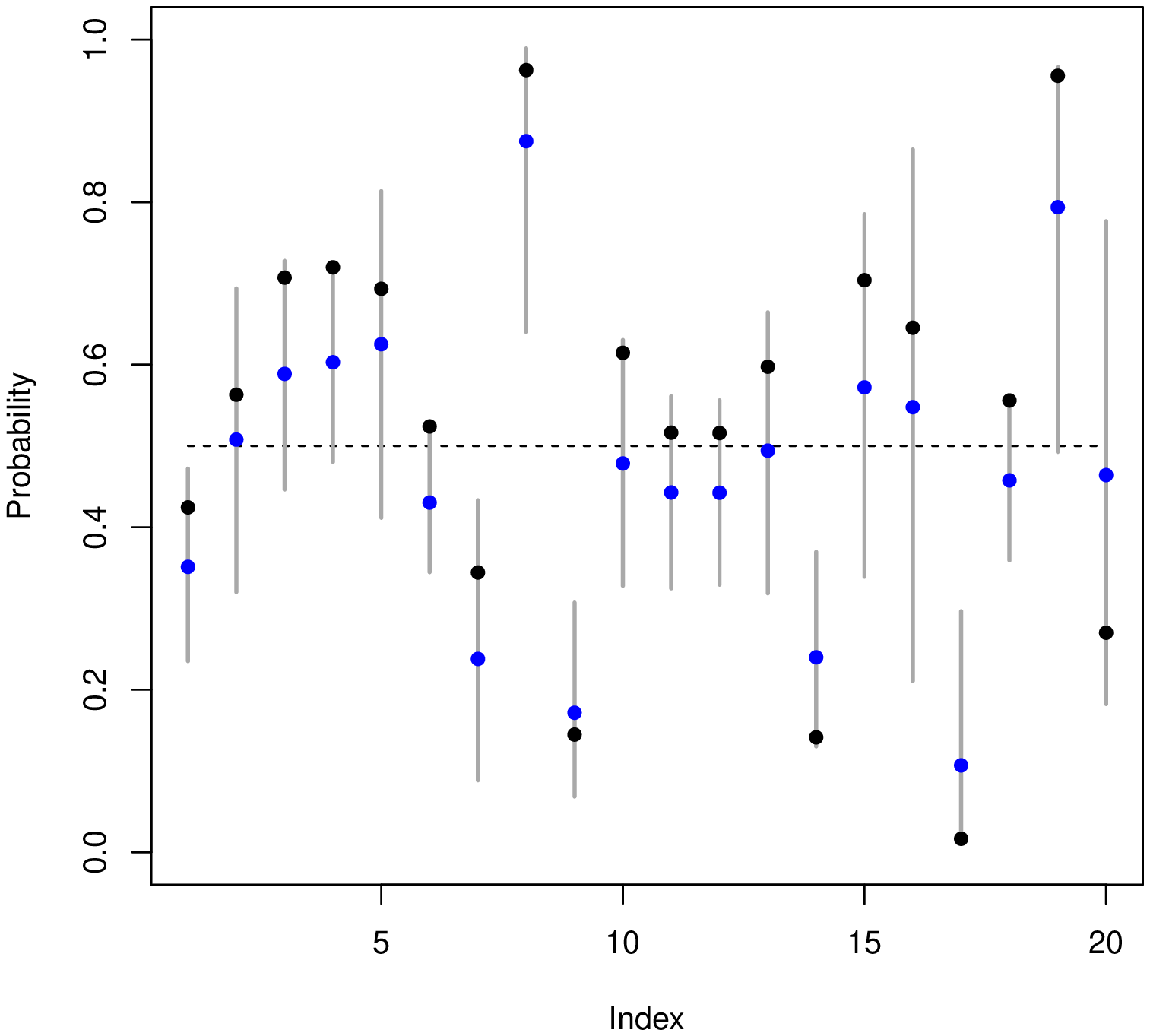}
\includegraphics[width=\textwidth]{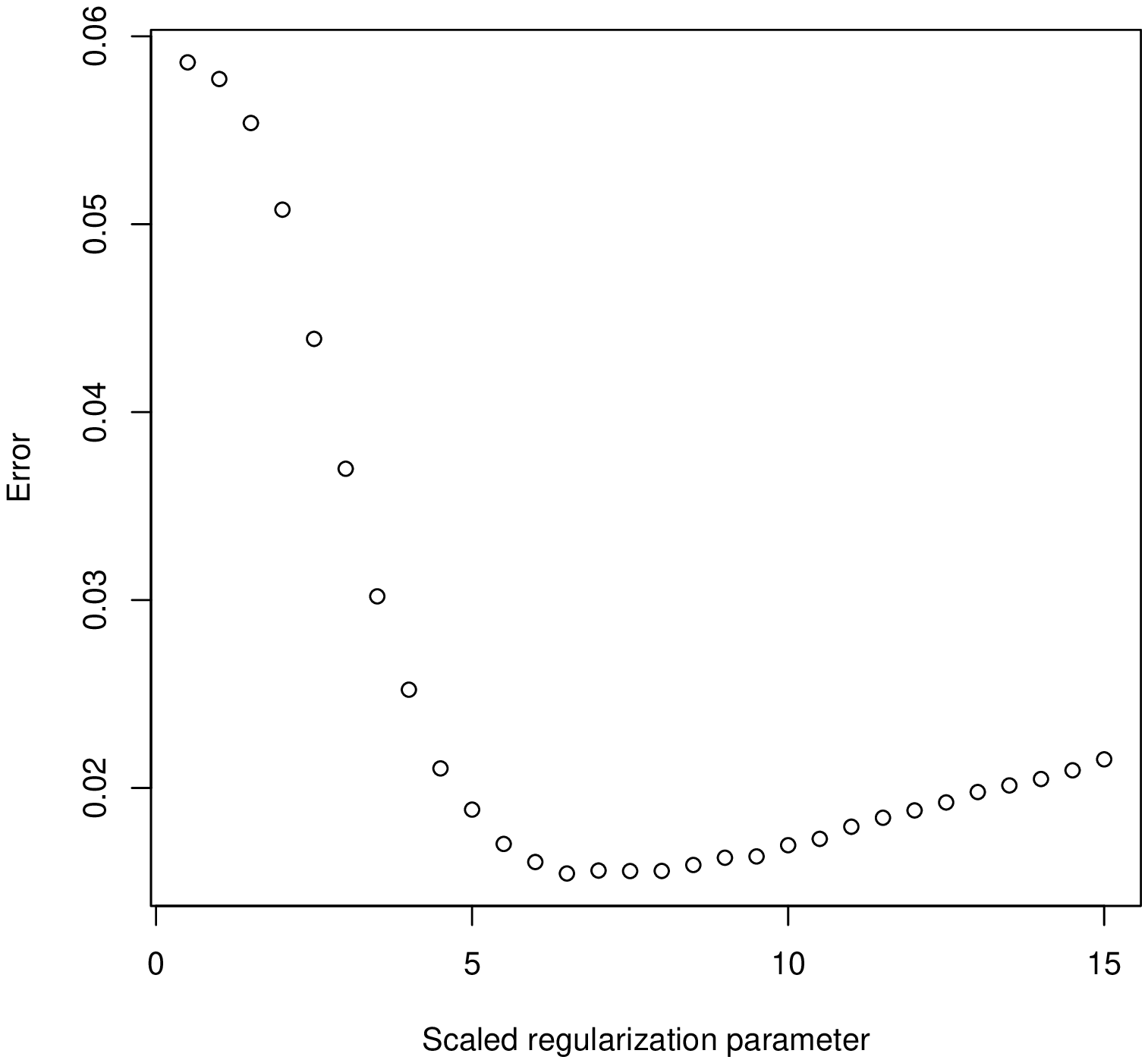}
\end{minipage}
\caption{Same as Figure \ref{fig: sw1}, but now with $70\%$ of the label unobserved.}
\label{fig: miss3}
\end{figure}

To assess the impact of the percentage of missing observations we increase the level from $10\%$ to $20\%$, $30\%$ and $70\%$. We observe in Figures \ref{fig: miss1}, \ref{fig: miss2} and \ref{fig: miss3} that as the percentage of missing observations increases, the posterior of $c$ is more spread out and there is more uncertainty in the function estimates, as is to be expected. 
In the most extreme case of $70\%$ missing labels we observe that the generalised gamma prior on $c$ results in quite severe oversmoothing. In the histogram we see that the posterior for $c$ puts too little mass around the oracle choice of $c$ in that case. The inverse gamma prior has a much better performance, and remains comparable to the oracle choice.

\subsection{Function prediction in a protein-protein interaction graph}

To test the nonparametric Bayes procedure on real data 
we adapt the case study from \cite{kolaczyk2009}, Section 8.5. The example is about the prediction of protein function from a network of interactions among proteins that are responsible for cell communication in yeast. For more information about the background of the experiment set-up, see \cite{kolaczyk2009}. 

\begin{figure}[H]
\begin{minipage}{0.49\textwidth}
\includegraphics[width = \textwidth]{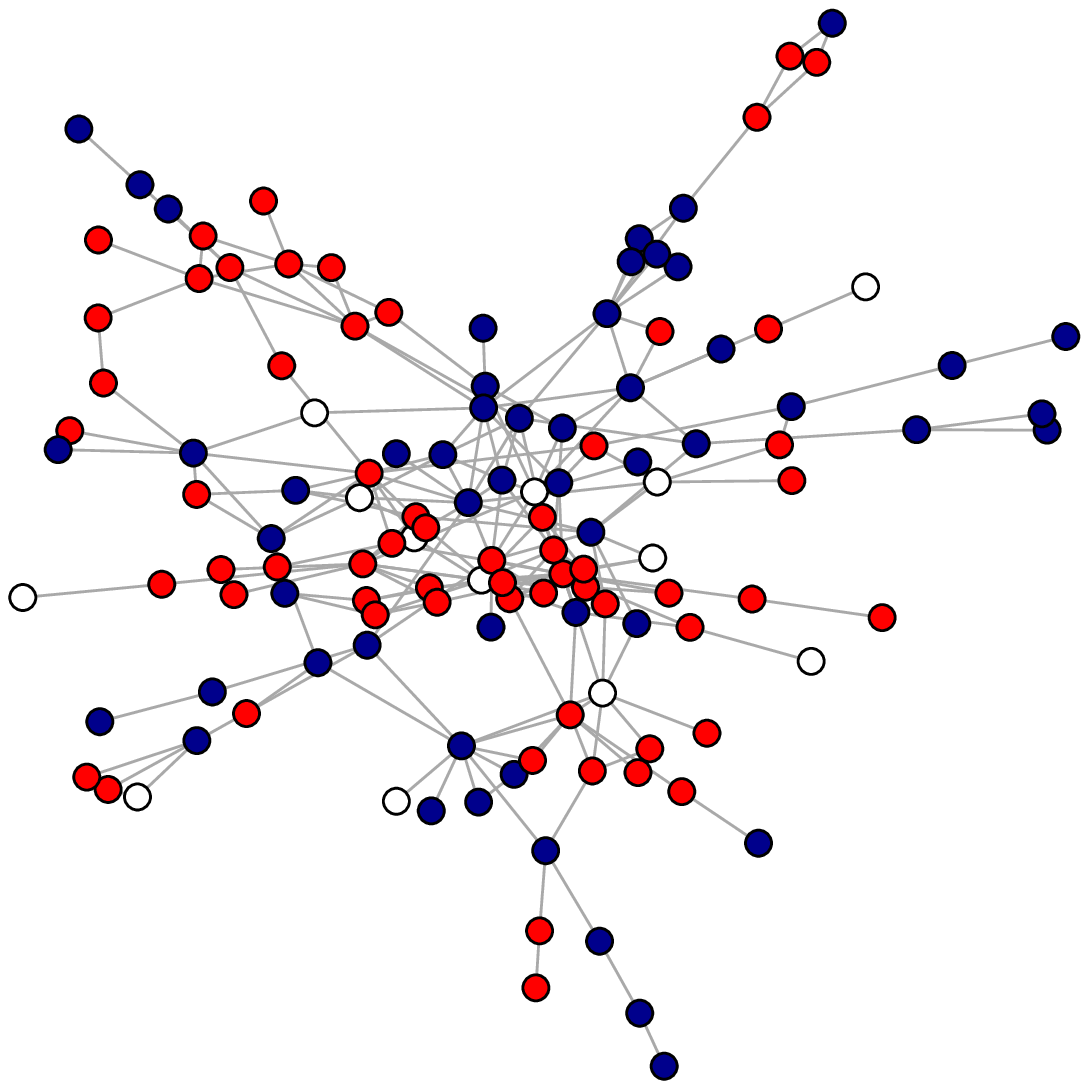}
\end{minipage}
\begin{minipage}{0.49\textwidth}
\includegraphics[width = \textwidth]{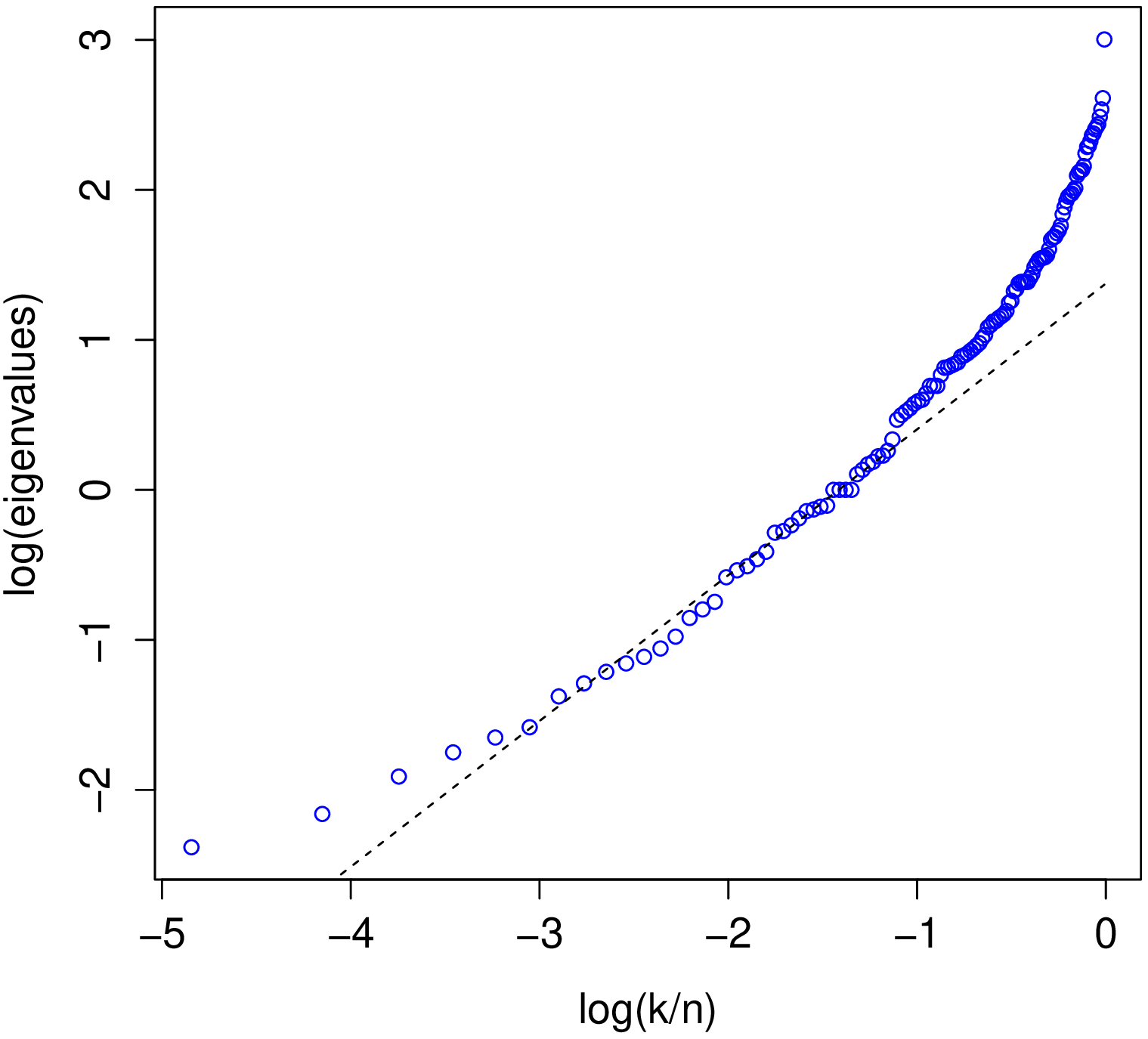}
\end{minipage}
\caption{Left: protein-protein interaction graph. The red nodes in the graph are involved in ICSC, the blue nodes are not involved and the white nodes are unknown.
Right: Laplacian eigenvalues. 
The linear fit has slope $0.97$ corresponding to geometry condition with $r = 2.1$. }
\label{fig:ppi}
\end{figure}

The protein interaction graph  is shown on the left in Figure
\ref{fig:ppi}. A vertex in the graph is labelled according to whether or 
not the corresponding protein is involved in so-called intracellular signaling cascades (ICSC), which is a specific form of communication. 
We have randomly removed $12$ of the labels and we 
apply our Bayesian prediction procedure to try and recover them from the observed labels.

In view of the findings in the preceding sections we apply the procedure with 
the ordinary gamma prior on $c$ with $a=b=0$. Numerical computation of the 
Laplacian eigenvalues shows that in this case the geometry condition is 
fulfilled with $r=2.1$, see the right panel of Figure \ref{fig:ppi}. We use 
this value in the procedure. The parameter $\alpha$ that determines 
the prior smoothness of $f$ is set to $\alpha=1$. This is a conservative 
choice, in order to avoid oversmoothing. We now run our algorithm to 
produce credible intervals for the soft label function $\ell$ at the vertices of 
which we did not observe the labels.

\begin{figure}[H]
\centering
\begin{minipage}{0.64\textwidth}
\includegraphics[width=\textwidth]{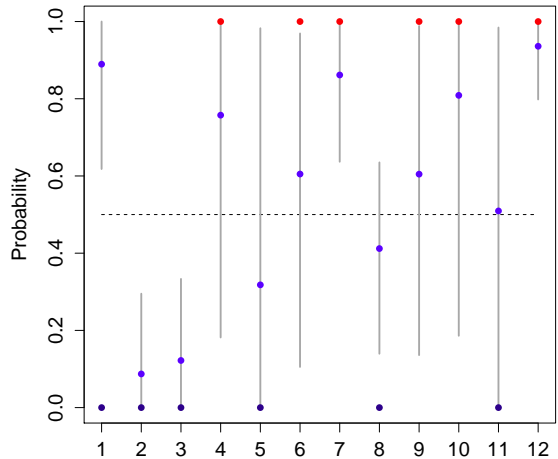}
\end{minipage}
\caption{Credible intervals for the soft threshold function at the nodes with missing labels. 
Blue dots are the posterior means, black and red dots are the true labels.
In this example, we observe a misclassification at threshold $0.5$ in vertices $1$ and $11$.}\label{fig: res}
\end{figure}

%\begin{figure}[H]
%\centering
%\begin{minipage}{0.99\textwidth}
%\includegraphics[width=\textwidth]{PPINN.eps}
%\end{minipage}
%\caption{From left to right, top to bottom, the two step neighborhoods of the missing nodes (yellow).}
%\end{figure}

Figure \ref{fig: res} shows the results, together with the true labels that we removed. 
We see that if we predict the missing labels by thresholding the posterior means at $1/2$,
we have a misclassification rate of $2/12 \approx 16.7\%$. 
If we repeat the procedure $100$ times, every time removing $12$ different labels 
at random, we obtain an average misclassification rate of $27\%$. 
To assess this, we also computed $k$-nearest neighbour ($k$-NN) predictions for various $k$.  
We found average missclassification rates of 
 $32\%$ for one $1$-NN, $28\%$ for $2$-NN and 
 $41\%$ for $3$-NN. Hence in terms of prediction performance our procedure 
 is comparable to $k$-NN with the oracle choice of $k$. This illustrates that 
 in line with the theory, our procedure succeeds in automatically 
tuning the appropriate degree of smoothing. Moreover, the Bayes procedure 
 has the advantage that in addition to predictions, we obtain credible 
 intervals as an indication of the uncertainty in the predictions.

\subsection{MNIST digit prediction}

So far we have considered examples with graphs that satisfy the 
geometry condition \eqref{eq: geom}. For such graph we have
theoretical results that provide some guidelines for the 
construction of the prior and choices of the hyper parameters. 
In principle however, we can also apply our procedure to 
graphs that do not satisfy \eqref{eq: geom} for some $r$. 
It is intuitively clear that a condition like 
\eqref{eq: geom} should not always be necessary for good performance. 
If we use the ordinary gamma prior on $c$ and set $q$ 
at a conservative (not too high) value, we can just apply our procedure 
and should still get reasonable results if the graph geometry
and the distribution of the labels are sufficiently related.
In this section we briefly investigate such as case.

\begin{figure}[H]
\begin{minipage}{0.49\textwidth}
\includegraphics[width = \textwidth]{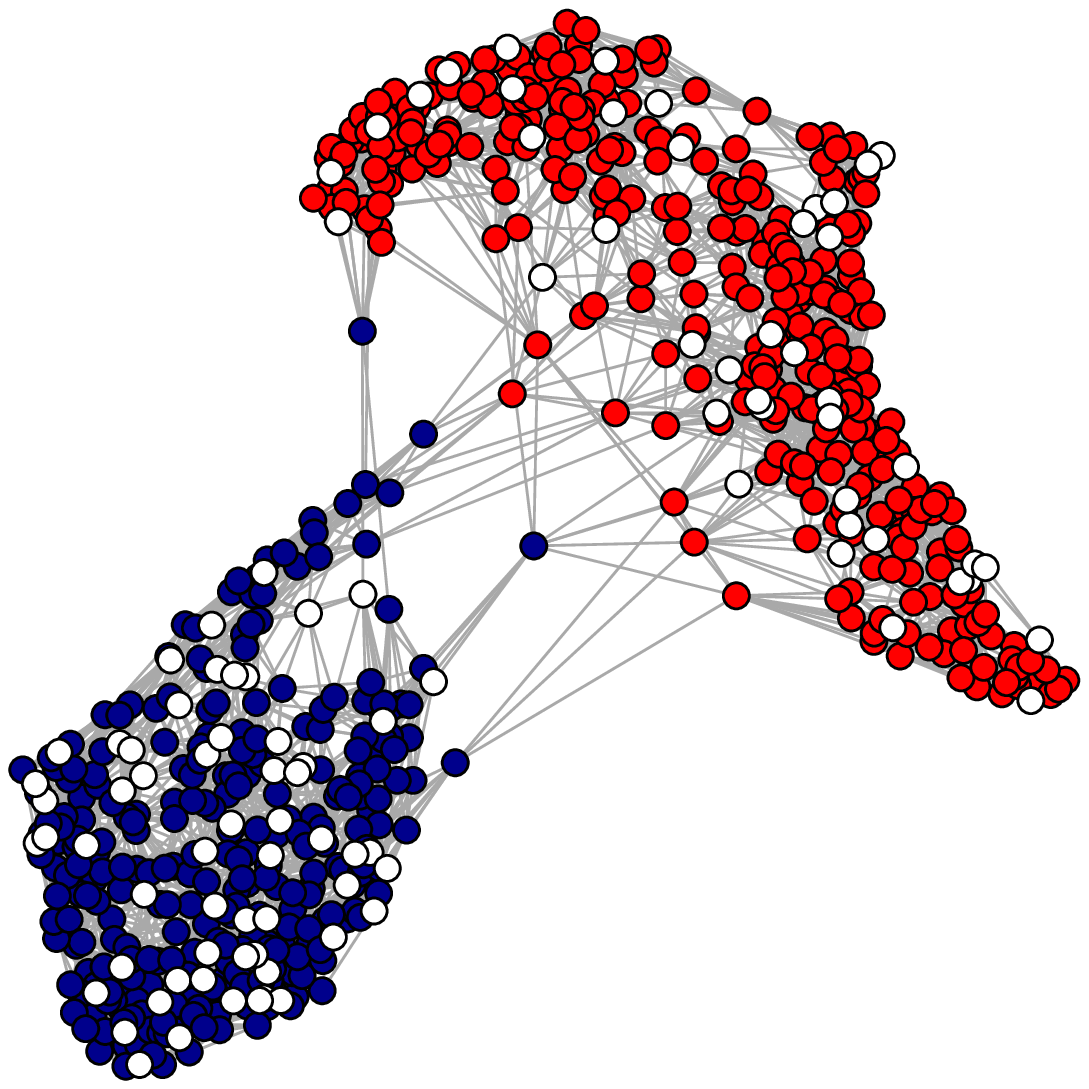}
\end{minipage}
\begin{minipage}{0.49\textwidth}
\includegraphics[width = \textwidth]{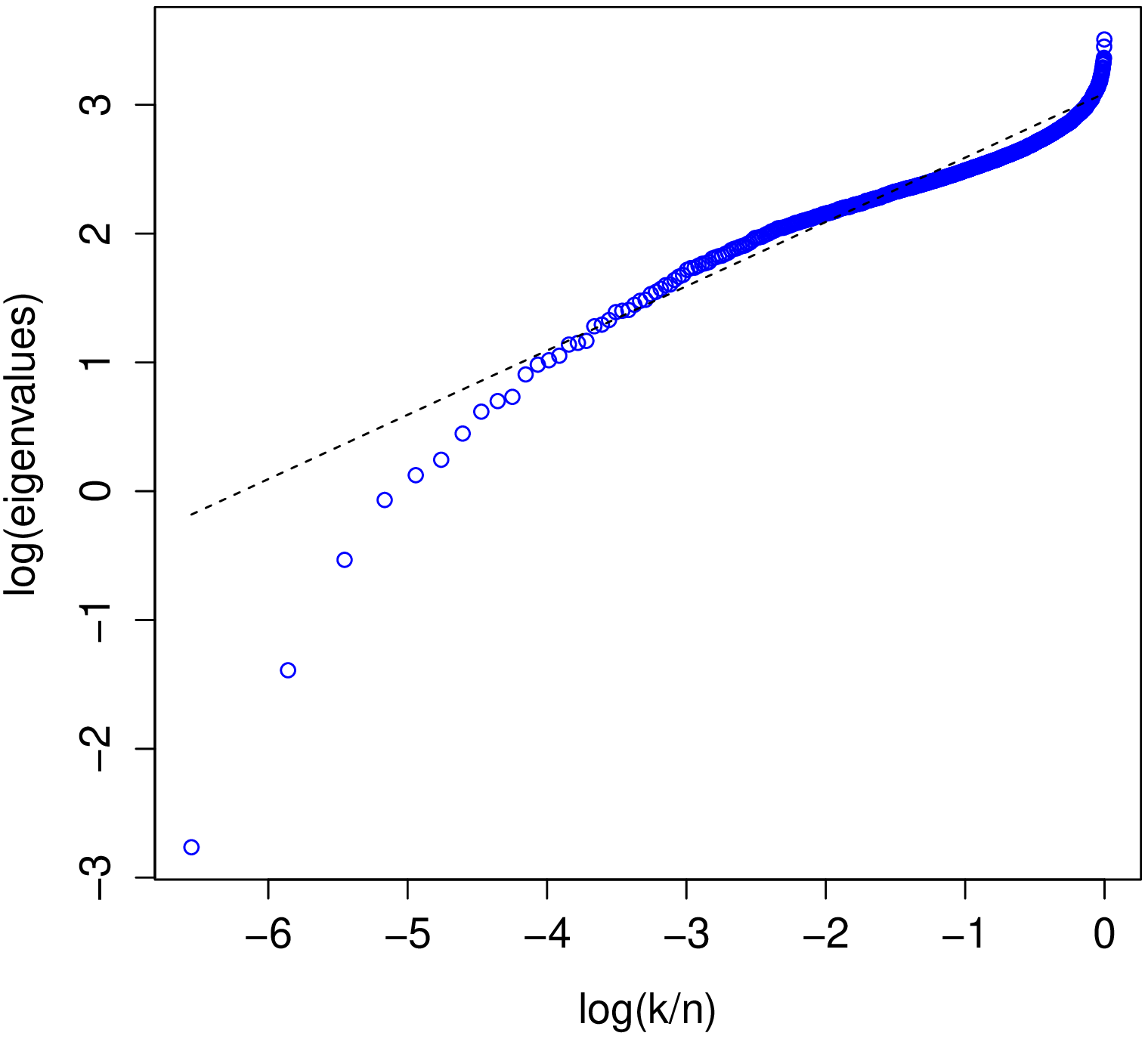}
\end{minipage}
\caption{Left: constructed MNIST graph. Digits $0$ are labeled blue, digits $1$ are labeled red and the missing labels are white. Right: Laplacian eigenvalues.}\label{fig: mnist}
\end{figure}

The MNIST dataset of handwritten digits has a training set of $60$ thousand examples and a test set of $10$ thousand. The digits are size-normalized and centered in a fixed-size image. The dataset is publicly available at \texttt{http://yann.lecun.com/exbd/mnist}. We have randomly selected a subsample of $700$ consisting of only the digits $0$ and $1$ of which $600$ in are the training set and $100$ are from the test set. Our goal is to classify the $100$ images from the test set. 
To turn this into a label prediction problem on a graph we construct a graph with $700$
nodes, corresponding to the images. For each image we determined the $10$ closest
images in terms of  pixel-distance and connected the corresponding nodes in the graph by 
an edge. The resulting graph {is} shown in Figure \ref{fig: mnist}.
The eigenvalue plot suggests that the graph does not satisfy the geometry condition
to the extent that the preceding graphs did.

\begin{figure}[H]
\begin{minipage}{0.49\textwidth}
\includegraphics[width = \textwidth]{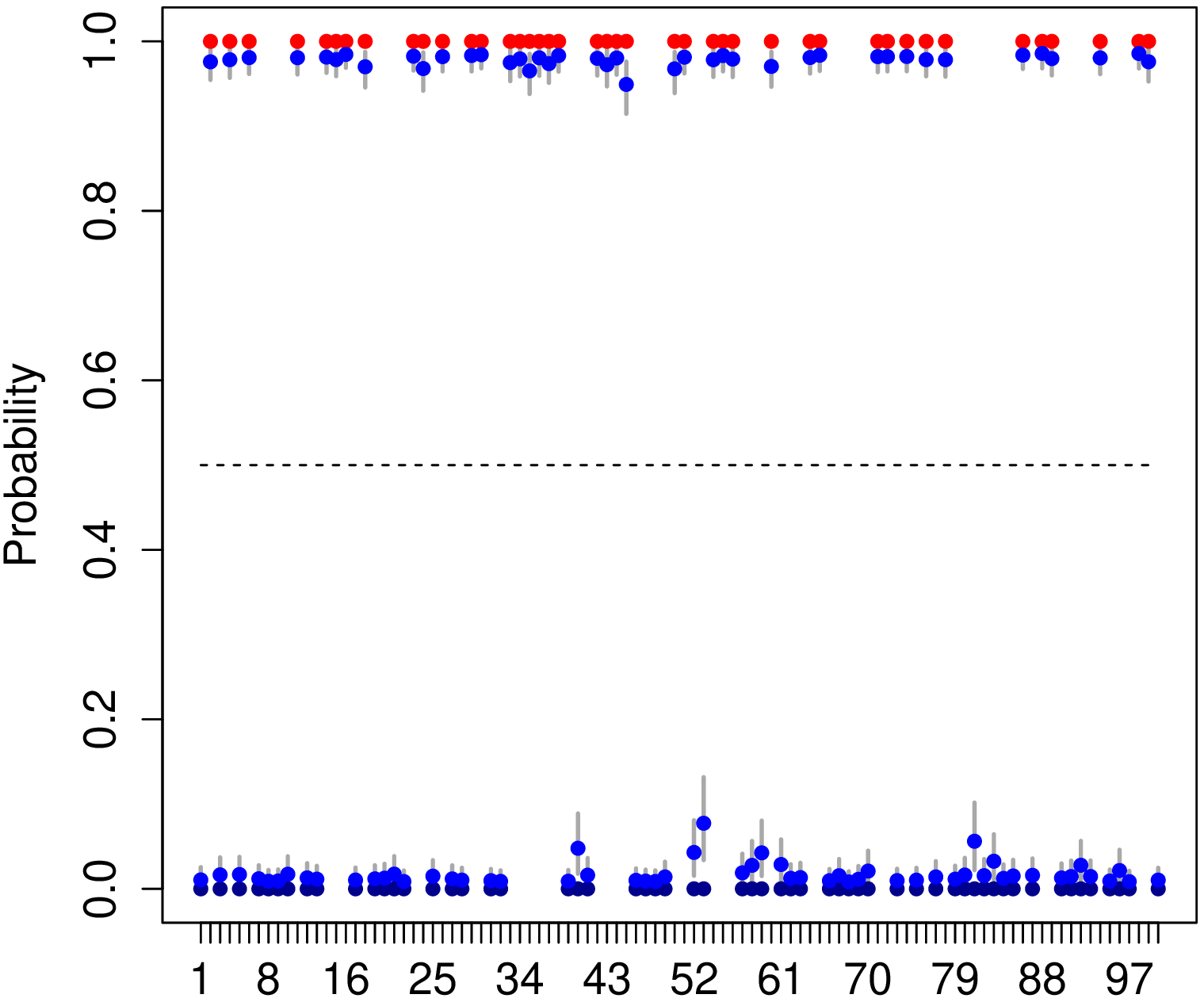}
\end{minipage}
\begin{minipage}{0.49\textwidth}
\includegraphics[width = \textwidth]{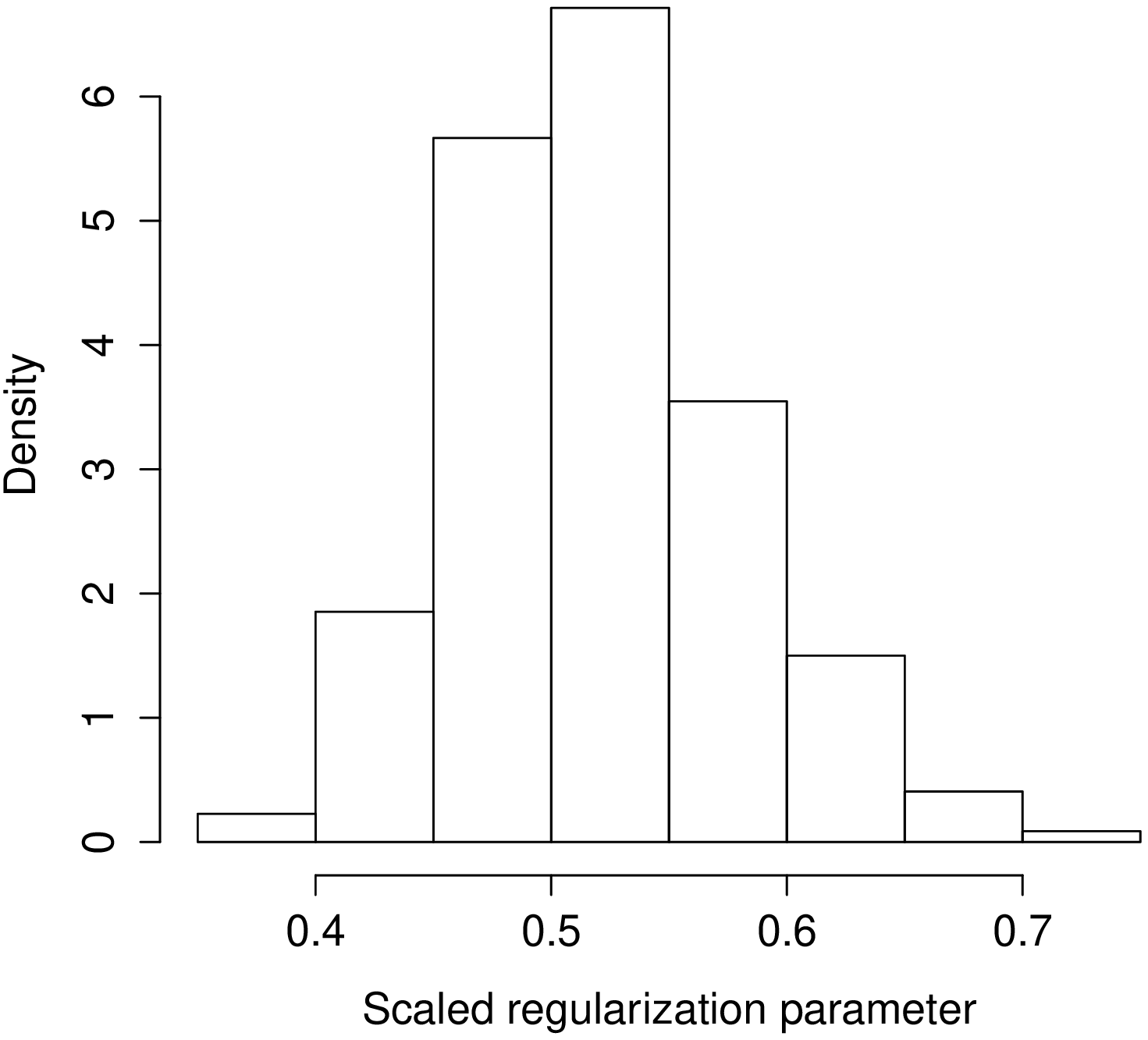}
\end{minipage}
\caption{Left: credible intervals for the soft threshold function at the nodes with missing labels. 
Blue dots are the posterior means, black and red dots are the true labels.
Right: posterior for tuning parameter $c$.}
\label{fig: res2}
\end{figure}

The picture indicates that predicting  the missing labels in this graph is not a very hard 
problem. And indeed, our procedure performs well in this case. We classify all missing labels
correctly, with very high certainties. See Figure \ref{fig: res2}.

\section{Concluding remarks}

We have described a nonparametric Bayes procedure to perform  binary classification on 
graphs. We have considered a hierarchical Bayesian approach with a randomly scaled 
Gaussian prior as in the theoretical framework in \cite{kirichenko2017}. We have 
implemented the procedure with the theoretically optimal prior from \cite{kirichenko2017} and a 
variant with a different prior on the scale, which exploits partial conjugacy 
and has some more flexibility. 

Our numerical experiments suggest that good results are obtained when using 
Algorithm \ref{alg: mcmc2}, i.e.\ using the ordinary gamma prior on the scale.
Suggested choices for the hyper parameters are $a=b=0$ and $q = r/2 + \alpha$. 
Here $r$ is the geometry parameter appearing in the geometry condition 
\eqref{eq: geom} and can
be determined numerically from the spectrum of the graph Laplacian. 
The parameter $\alpha$ reflects prior smoothness and should not be set too high
(e.g.\ $\alpha=1$ or $2$), to avoid oversmoothing.

In view of computational complexity it might be more advantageous  to consider other methods to adaptively find the tuning parameter, such as empirical Bayes methods. 
Also, it might be sensible to modify the prior by truncating the $n$-dimensional
Gaussian prior on $f$ to a lower dimensional one by writing a series expansion for 
$f$ and truncating the sum at a random point,  similar to the approach 
in \cite{liang2007} for instance. It is conceivable that in this way the procedure 
becomes both more flexible in terms of adaptation to smoothness and 
will also computationally scale better to large sample size $n$. 
We intend to investigate this in future work.

%\section*{References}

\bibliography{paper}

\end{document}